\documentclass{article}

\usepackage{arxiv}
\usepackage{comment}
\usepackage[title]{appendix}
\usepackage[utf8]{inputenc} 
\usepackage[T1]{fontenc}    
\usepackage{url}            
\usepackage{booktabs}       
\usepackage{amsfonts}       
\usepackage{amsmath}        
\usepackage{nicefrac}       
\usepackage{microtype}
\usepackage[numbers,sort&compress]{natbib}
\usepackage{lipsum}
\usepackage{graphicx}
\graphicspath{ {./plots/} }
\usepackage{subcaption}
\usepackage{caption}
\usepackage{float} 
\usepackage[section]{placeins}
\usepackage{hyperref}
\hypersetup{
    colorlinks=true,
    citecolor=blue,
    linkcolor=blue,
    urlcolor=black
}
\title{Deep Learning of Solver-Aware Turbulence Closures from Nudged LES Dynamics}

\author{
 Ashwin Suriyanarayanan \\
  School of Mechanical Engineering\\
  Purdue University\\
  West Lafayette, IN 47907 \\
  \texttt{asuriyan@purdue.edu} \\
  \And
 Dibyajyoti Chakraborty \\
  College of Information Sciences and Technology\\
  Pennsylvania State University\\
  University Park, PA 16802 \\
  \texttt{d.chakraborty@psu.edu} \\
  \And
 Romit Maulik \\
  School of Mechanical Engineering\\
  Purdue University\\
  West Lafayette, IN 47907 \\
  \texttt{rmaulik@purdue.edu} \\
}

\begin{document}
\maketitle
\begin{abstract}
The differentiable physics paradigm may be leveraged as an \textit{a-posteriori} approach for discovering turbulence closure models by embedding a neural network parameterization directly inside the solver and optimizing it given potentially sparse target data. This addresses a key limitation of \textit{a-priori} learning where direct numerical simulation (DNS) data is used to approximate the subgrid stress with the assumption of a low-pass filter. Closures trained in this \textit{a-priori} manner frequently lead to unstable deployments due to the mismatch between the assumed filter and the effect of numerical discretizations and coarse-graining. In comparison, while typically stable during deployment, \textit{a-posteriori} learning incurs high computational costs due to the need to backpropagate through a large eddy simulation (LES) solver. Furthermore, \textit{a-posteriori} methods are challenging to apply broadly since they require significant modification of existing solvers. Finally, both approaches are limited when generalization is desired across different numerical schemes with their implicit filtering characteristics. In this work, we present a deep-learning approach for turbulence closure modeling built on the continuous data assimilation framework. Our approach enables the \textit{a-priori} training of closures using DNS data observed only at coarse-grid points without modifying or differentiating through the LES solver, while preserving stability during deployment for the recovery of long-term statistics. We focus on the model's ability to adapt to different discretizations by explicitly conditioning it on the numerical scheme. We use two- and three-dimensional canonical cases to test our framework and show that the learned correction systematically tracks the discretization error of the coarse solver. 
\end{abstract}

\section{Introduction}
\label{sec:introduction}

Turbulent flows arise in a wide range of engineering and geophysical settings, yet their predictive simulation remains a major computational challenge. Direct Numerical Simulation (DNS) resolves all dynamically relevant scales, but its cost becomes prohibitive, especially at high Reynolds numbers. Reynolds-averaged Navier-Stokes (RANS) is considerably cheaper, but often lacks fidelity for strongly unsteady flows. Large-eddy simulation (LES) occupies an intermediate position by resolving the larger, energy-containing structures while modeling the effect of the unresolved \textit{subgrid scales} (SGS) through a \emph{turbulence closure model}. Its predictive accuracy, therefore, depends critically on the effectiveness of the closure model.

Since the purpose of the closure term is to represent the effect of the unresolved scales on the resolved dynamics, it is often modeled through dissipative terms in the filtered momentum equations, based on the assumption that the net effect of the unresolved motions is predominantly dissipative. The most common approach is the eddy-viscosity framework, in which the SGS stresses are represented using an effective viscosity, typically expressed as a function of the filtered strain rate. Such models remain widely used because of their simplicity, robustness and practical effectiveness. The Smagorinsky model \citep{smagorinsky1963general} is a popular choice for such modeling, in which the SGS eddy viscosity is defined in terms of a characteristic grid length scale, a Smagorinsky coefficient and the magnitude of the filtered strain-rate tensor. However, such models are purely dissipative, and are limited in their ability to represent the full complexity of inter-scale transfer, particularly backscatter \citep{Piomelli1991SubgridscaleBI}, in which energy is transferred from the unresolved scales back to the larger structures. The dynamic Smagorinsky model \citep{germano1991dynamic, lilly1992germano} provides a significant improvement by dynamically calculating the eddy-viscosity coefficient using self-similarity analysis leading to a more accurate SGS representation. This model can capture the backscatter by obtaining negative eddy viscosity, but also leads to numerical instability \citep{zang1993dynamic} and requires averaging across homogeneous directions or clipping.

In recent years, deep-learning-based approaches \citep{Duraisamy_2019, sanderse2024scientificmachinelearningclosure}, have been increasingly explored for LES turbulence closure modeling. In these methods, the closure term is learned in a data-driven manner by approximating the mapping between the evolved flow field and the corresponding corrective term. A common distinction in data-driven closure modeling is between \textit{a-priori} and \textit{a-posteriori} learning. In \textit{a-priori} approaches \citep{Guan_2022,beck2019deep, maulik2019subgrid, maulik2018data}, the model is trained offline against closure targets extracted from DNS. These methods are computationally attractive, but the learned correction often degrades when deployed within an evolving solver because the offline target does not account for discretization error \citep{PhysRevFluids.6.050504}. In \textit{a-posteriori} approaches \citep{List_Chen_Thuerey_2022, shankar2023differentiable, shankar2025differentiable, Sirignano_MacArt_2023, kim2026generalizable}, the closure is optimized within the solver itself, so the learned correction naturally reflects the numerical dynamics of the training configuration. Such an approach is termed ``online'' learning and comes under the broader umbrella of ``differentiable physics'' \citep{um2020solver}. This approach improves \textit{a-posteriori} deployment but incurs a higher compute cost due to the need for backpropagation through the solver. This requires that the solver be differentiable or the gradients be available through the discrete adjoint method, both of which need significant modifications to many existing solvers. 

Another difficulty is that differentiable-physics-based closures do not readily generalize across numerical discretizations. Because the learned correction is optimized within a particular solver and training configuration, it may become strongly coupled to the corresponding numerical errors. {As discussed by \citet{mohan2024you}, \citet{chakraborty2024note} and \citet{beck2023toward}, such approaches can effectively overfit to the discretization used during training, so that transfer to a different numerical scheme becomes non-trivial.} This limitation is particularly important in turbulence modeling~\citep{ghosal1996analysis,chow2003further}, where different numerical schemes may introduce different effective filtering characteristics and thereby alter the closure requirements for stable and accurate coarse grid evolution. These considerations motivate the need for an alternative strategy that retains the efficiency of offline learning while still incorporating the numerical artifacts of various schemes.

One such framework is continuous data assimilation (CDA) \citep{Azouani_2013}, or nudging, in which observational information is incorporated directly into the governing equations through a feedback term proportional to the discrepancy between the assimilated and reference states. This feedback acts as a corrective forcing that drives the coarse-grid solution (with associated numerical artifacts) towards the observed dynamics. Several works \citep{wu2025interpolateddiscrepancydataassimilation, cao2022continuous} have established convergence of CDA under suitable assumptions, and subsequent extensions have considered settings of practical relevance involving sparse observations and noisy measurements \citep{Bessaih_2015}.

Recent works have also explored the application of ideas from data assimilation for various problems involving turbulent flows. {\citet{Zauner_Mons_Marquet_Leclaire_2022} used nudging with sparse velocity observations to reconstruct the turbulent wake behind a square cylinder in an unsteady RANS solver. In homogeneous isotropic turbulence, nudging has been applied with Eulerian, Fourier and Lagrangian observations \citep{clark2020synchronization}, and also to infer unknown flow parameters and reconstruct large-scale turbulent structure from sparse measurements \citep{Clark_Di_Leoni_2018}.} In the context of LES closure modeling, \citet{wang2023ensemble} proposed an ensemble-Kalman-filter-based mixed SGS model, and \citet{ephrati2025continuous} developed a 3D variational data assimilation approach for stochastic spectral closure for coarse LES. {\citet{buzzicotti2020synchronizing} analyzed the relation between the degree of synchronization and the subgrid model when nudging an LES towards DNS data. \citet{li2022synchronizing} similarly reported that synchronizing an LES to DNS requires a minimum amount of assimilated large-scale data, with this threshold depending on the choice of subgrid-scale model}. A related study by \citet{ling2025numerically}, used a \emph{statistical} nudging procedure to assimilate mean profiles averaged over the homogeneous directions and time to generate numerically consistent training data for wall-modeled LES subgrid-scale closures in turbulent channel flows. {\citet{hoekstra2024reduced} reduced the closure to a few domain-integrated quantities of interest, measured the correction required in these after a single coarse step, and resampled it from the measured distribution. \citet{ephrati2025probabilistic} instead measured the subgrid force field by advancing the coarse and reference solutions one step from a common filtered state, and used it to define a stochastic forcing, with the solution ensemble corrected towards target statistics.} In contrast, the present work learns the \emph{instantaneous} coarse-grid nudging discrepancy directly  {as a deterministic function of the evolved state} and examines how the required correction depends on the numerical discretization of the coarse solver.

Here, nudging applied to an LES acts as an effective correction for the combined effect of unresolved scales and solver-dependent error, since the feedback term acts directly to synchronize the coarse-grid system with the available reference data, obtained for example from DNS or experiments. Specifically, we first perform synchronized coarse-grid simulations in which the LES is driven towards reference observations through nudging, and then \emph{record} the corresponding forcing required for that synchronization with the goal of using this as training data. A neural network is subsequently trained \textit{a-priori} to approximate this correction from the evolving coarse-grid state. In this manner, the learned closure inherits information not only about the missing subgrid contribution, but also about the numerical discretization error of the target solver, while avoiding the need for adjoints or backpropagation through the solver. {Thus, the primary contribution of this work is a single, scheme-conditioned closure model that adapts across multiple numerical discretizations, trained \emph{a-priori} using the recorded \emph{solver-aware} forcing term.} We assess this framework on forced two-dimensional turbulence with Kolmogorov forcing and three-dimensional homogeneous isotropic turbulence, and further examine its dependence on the underlying numerical discretization by considering different training data from numerical schemes within a unified learning framework.

{The remainder of this paper is organized as follows. The nudging-based methodology is described in \S~\ref{sec:methodology}. Section~\ref{sec:2d_results} and \S~\ref{sec:3d_results} present results for the two- and three-dimensional cases, followed by an analysis of how the learned correction depends on the numerical discretization error and a comparison of its forcing contribution against classical closures and against the forcing required by the coarse solver itself. The adaptation of the scheme-conditioned model to a discretization not seen during training is examined in \S~\ref{sec:generalization_fine_tune}. Concluding remarks follow. A table of notation, together with further details on dataset generation, the choice of nudging strength, training, temporal-scheme generalization, {extension to higher Reynolds number,} and nudging using a subset of fields and sparser observations, is given in Appendices~\ref{sec:appendix_A}-\ref{sec:appendix_E}.}

\section{Methodology}\label{sec:methodology}
We begin by recalling the continuous data assimilation (CDA) framework in a generic form. Let $\boldsymbol{u}$ denote a reference solution of a dynamical system on a domain $\Omega$, governed by
\begin{equation}
\partial_t \boldsymbol{u}=\mathcal{N}[\boldsymbol{u}],
\end{equation}
where $\mathcal{N}[\cdot]$ denotes the underlying PDE operator. Let $\boldsymbol v$ denote the assimilated solution. In CDA, observations of the reference state are incorporated into the evolution of $\boldsymbol{v}$ through a feedback term proportional to the discrepancy between the assimilated and reference states, so that
\begin{equation}
{\partial_t \boldsymbol{v}=\mathcal{N}[\boldsymbol{v}]+\boldsymbol{f}_{nudge}(\boldsymbol{d}),}
\end{equation}
\begin{equation}\label{eq:nudge_force}
{\boldsymbol{f}_{nudge}(\boldsymbol{d}) = \mu \, \boldsymbol{d},}
\end{equation}
{where $\boldsymbol{d}$ is the observation-induced discrepancy field, $\mu>0$ the nudging coefficient (or nudging strength) and $\boldsymbol{f}_{nudge}(\boldsymbol{d})$ the corresponding feedback (nudging) term.} Under suitable assumptions on the observation resolution and the nudging coefficient, CDA drives the assimilated state towards the reference dynamics. Hence, such a forcing can be used for constructing a correction term for the coarse-grid LES. Our implementation specializes this general setting to the case in which the reference observations are available on the coarse computational grid. Let $\boldsymbol{u}_{obs}$ denote the ground truth restricted to the coarse grid, and let $\boldsymbol v$ denote the corresponding coarse-grid LES solution. The discrepancy is then defined pointwise on the coarse grid as
\begin{equation}\label{eq:discrepancy}
{\boldsymbol{d}=\boldsymbol{u}_{obs}-\boldsymbol{v}.}
\end{equation}

This full-state coarse-grid form is the version used in all of the numerical experiments reported here. Theoretical CDA formulations treat observations through an interpolation operator to reconstruct the discrepancy in the continuous space. Here, we consider a case where the observations are available directly on the coarse grid using a subsampling operation.

Since the assimilated system here is itself a coarse-grid LES equipped with a baseline closure, exact synchronization with the reference field is not generally expected; rather, consistent with prior work on {data assimilation in such reduced-order and LES-type systems \citep{larios2025data, zerfas2019continuous, buzzicotti2020synchronizing}}, one expects convergence only to a residual error floor. It should also be noted that a non-zero residual may still persist even without an explicit SGS closure, due to the effect of high-frequency structures from ground truth which cannot be represented on the coarse grid.

For the present problem, the reference dynamics (DNS) are governed by the Navier-Stokes equations solved on a grid that resolves all dynamically relevant
scales up to its numerical cutoff, while the assimilated system is evolved on a coarse mesh in an LES-type setting. Formally, the filtered representation of the DNS may be written as \begin{equation}
\bar{\boldsymbol{u}}(\boldsymbol{x},t)=G_h \star \boldsymbol{u}(\boldsymbol{x},t),
\end{equation}
where $G_h$ denotes a filter of characteristic width $h$. Applying this filtering operation to the governing equations introduces an unresolved contribution, written here as $\boldsymbol{f}_{sgs}=\nabla\cdot\boldsymbol{\tau}_{sgs}$, in the coarse-grid evolution. In practice, however, the exact nature of the filter is impossible to ascertain, and the performance of LES closure heavily depends on the assumptions used to prescribe this. {Also, applying such a filter presupposes access to the full-fidelity field and in practical settings, the full-fidelity solution is not typically available.}  Hence, the approach of this article is to assume that sparse observations of the DNS must correspond to (and synchronize with) the LES predictions, framing the learning procedure as an inverse problem. Accordingly, we use $\bar{\boldsymbol{u}}$, or equivalently $\boldsymbol{u}_{obs}$, to denote the coarse-grid representation of the reference state {obtained through subsampling. Moreover, such a coarse-graining intuitively translates closest to what measurements provide by acting as an implicit filter with no specific form.}

{An important clarification here is regarding the use of the term DNS for the source of the ground truth observations. The reference state used in the study should be understood as an \emph{agnostic} higher-fidelity state, whose role is to supply information that the assimilated system does not otherwise possess. The reference enters the formulation only through $\boldsymbol{u}_{obs}$, so neither the full-fidelity field nor the numerics used to generate it are required by the procedure. This ground truth observation could equally be drawn from a finer-fidelity LES or from sparse experimental measurements. Although nudging itself is not constrained by the source of the observations, provided these are of higher fidelity than the coarse grid; their quality, such as the spacing of the measurements and the presence of noise, does affect the degree of synchronization attained (see Appendix~\ref{sec:appendix_E}). In the present study, the observations are taken to correspond to all points of the coarse grid, for ease of analysis.}

The baseline coarse-grid model is taken to be an LES equipped with a Smagorinsky closure. However, since the Smagorinsky term does not fully capture all subgrid-scale interactions, the nudging term is introduced as an additional correction. The resulting coarse-grid evolution may therefore be written schematically as
\begin{equation}\label{eq:navier_stokes_nudging_equation}
{\partial_t \boldsymbol{v}=\mathcal{N}[\boldsymbol{v}]+\mathcal{S}[\boldsymbol{v}]+\boldsymbol{f}_{nudge}(\boldsymbol{d}),}
\end{equation}
where $\mathcal{S}[\cdot]$ denotes the baseline Smagorinsky closure and $\boldsymbol{f}_{nudge}$ is given by~\eqref{eq:nudge_force}, with the discrepancy defined in~\eqref{eq:discrepancy}.

The above nudging equation~\eqref{eq:navier_stokes_nudging_equation} is formulated at the continuous level, in which the feedback enters through an interpolation operator that reconstructs the discrepancy field in continuous space. In practice, however, the assimilating system is a numerical solver advancing a discrete state on a coarse grid. The continuous PDE operators are thereby discretized using a chosen numerical scheme, which introduces its own discretization error. The feedback term compensates not only for the unresolved subgrid contribution but also for the discretization error of the coarse solver. Hence this nudging term is, in general, an unknown function of both the SGS deficit and the numerical error,
\begin{equation}\label{eq:fnudge_function}
\boldsymbol{f}_{nudge} \approx
\mathcal{G}\!\left(\text{SGS deficit},\ \text{numerical error}\right).
\end{equation}

\begin{figure}
    \centering
    \includegraphics[width=0.75\textwidth]{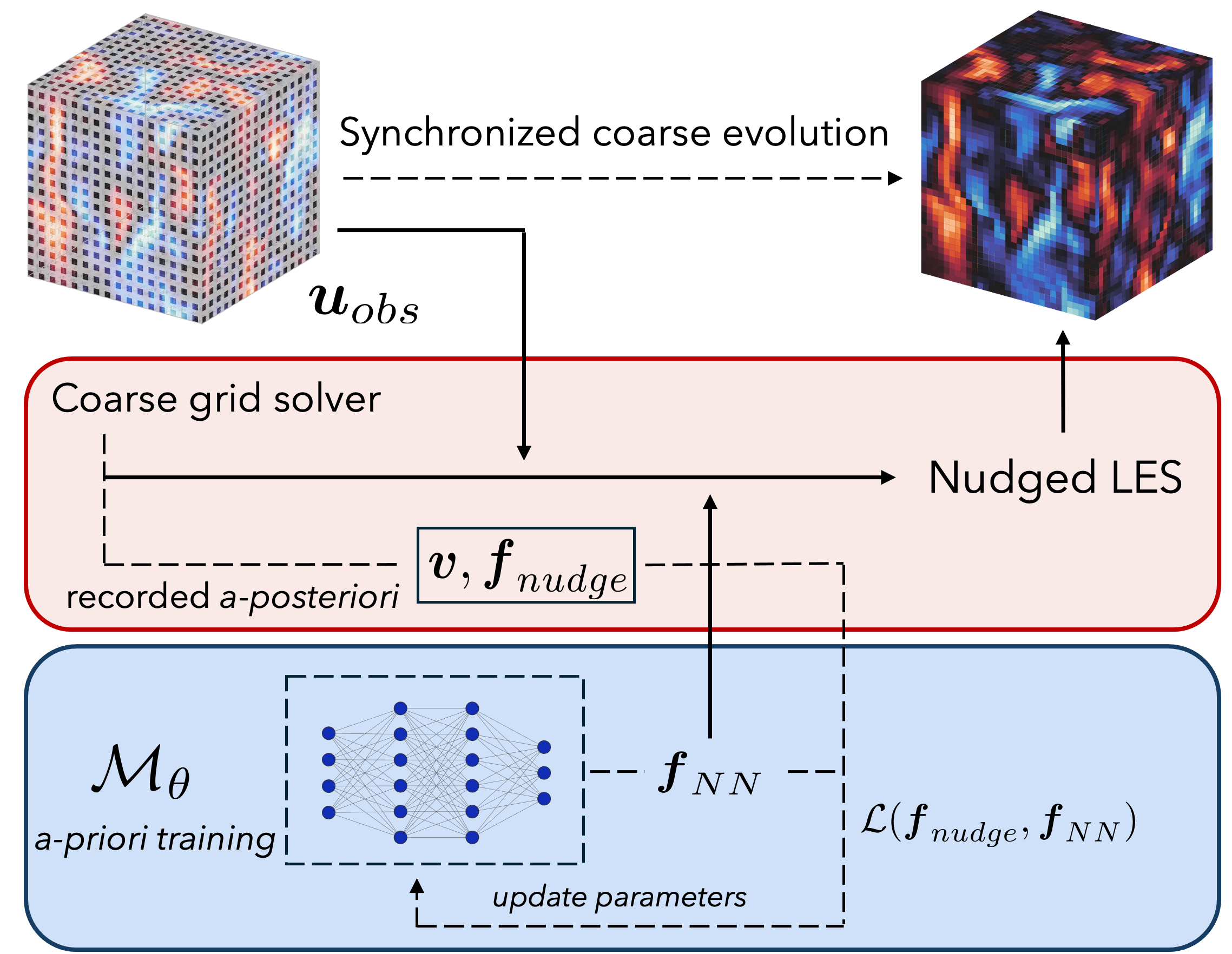}
    \caption{Schematic of the proposed data-driven closure modeling using nudging. $\boldsymbol{u}_{obs}$, $\boldsymbol{v}$ represent the ground truth observations and the assimilated LES field. {The observations $\boldsymbol{u}_{obs}$ enter the coarse-grid solver as a feedback term that synchronizes the LES state with the ground truth. The resulting field state and the corresponding correction $\boldsymbol{f}_{nudge}$ are recorded and used as input-target pairs to train the model $\mathcal{M}_\theta$ \textit{a-priori} on the mapping from the coarse-grid state to the required correction. At deployment, the observation-based feedback is removed and replaced by the network prediction $\boldsymbol{f}_{NN}$.}}
\label{fig:schematic}
\end{figure}

The main idea of the present work is to approximate the discrepancy between the coarse reference observations and the evolving LES state using a deep neural network trained on data obtained from synchronized coarse-grid nudging simulations (figure~\ref{fig:schematic}). Specifically, for each synchronized trajectory, we record the discrepancy/error field $\boldsymbol{d}=(\boldsymbol{u}_{obs}-\boldsymbol{v})$, which is also the unscaled nudge forcing,
and use this quantity as the supervised target for training. The neural network, denoted by $\mathcal{M}_\theta$ with parameters $\theta$, is therefore trained to approximate the map
\begin{equation}
\mathcal{M}_\theta(\boldsymbol{v}) \approx \boldsymbol{d},
\end{equation}
and the corresponding corrective forcing is then recovered as
\begin{equation}
\boldsymbol{f}_{nudge} \approx \mu\,\mathcal{M}_\theta(\boldsymbol{v}).
\end{equation}
Training is performed \textit{a-priori} using recorded trajectories from synchronized nudged simulations, and therefore does not require {any modification of}, or gradients to be propagated through a solver. After training, the explicit observation-driven nudging term is removed and replaced by the neural network prediction, which provides the corrective forcing required to sustain the reference dynamics. {The nudging term can in principle account for all subgrid interactions, and synchronization is attained regardless of the baseline; without the Smagorinsky contribution the recorded correction simply absorbs the full subgrid component together with the discretization error. The baseline is nonetheless retained because it provides a stable, physics-based dissipation, so that stability at deployment does not rest solely on the network. It also allows the network to learn a residual rather than the entire closure, the Smagorinsky term acting as the primary contribution and the learned correction supplying only what it leaves.}

To account for different numerical discretizations, we introduce a scheme encoding $s$ and condition the network through FiLM layers \citep{perez2018film}, which apply a feature-wise affine transformation to
the network activations. {From each encoding $s$, a small embedding generator produces a per-channel scale $\gamma^{(s)}_{c}$ and shift $\beta^{(s)}_{c}$ for each conditioned stage of the network, and is shared by the transformer blocks within that stage.} Hence, the channel features $\boldsymbol{h}_c$ are modified as
\begin{equation}
\mathrm{FiLM}\!\left(\boldsymbol{h}_{c}\mid\gamma^{(s)}_{c},\beta^{(s)}_{c}\right)
=\gamma^{(s)}_{c}\,\boldsymbol{h}_{c}+\beta^{(s)}_{c},
\end{equation}
so that the encoding scales and shifts each feature channel independently. In this way a single set of network weights is shared across different scheme encodings, with the scheme dependence captured entirely by the learned, per-scheme parameters $(\gamma^{(s)},\beta^{(s)})$. {The encoding $s$ is supplied explicitly as an additional input at both training and inference; the network does not infer the discretization from the coarse-grid state itself.}

This scheme-dependent model evolving the coarse-grid field $\boldsymbol{v}^{(s)}$ on numerical scheme $s$ is then written as
\begin{equation}
\mathcal{M}_\theta(\boldsymbol{v},s) \approx (\boldsymbol{u}_{obs}-\boldsymbol{v}^{(s)}),
\end{equation}
and the corresponding corrective forcing is given by
\begin{equation}
\boldsymbol{f}_{nudge}^{(s)}=\mu\,\mathcal{M}_\theta(\boldsymbol{v},s).
\end{equation}

In this way, the learned model provides a scheme-aware estimate of the coarse-grid discrepancy field, from which the corrective forcing is constructed during rollout. The reference observations are kept fixed on the coarse grid, while the numerical discretization of the coarse solver is varied to examine how the required correction depends on numerical discretization error. {The proposed framework is assessed on two-dimensional turbulence with Kolmogorov forcing and three-dimensional homogeneous isotropic turbulence; the corresponding flow configurations, dataset generation and training details are summarized in the Appendix.}

\section{Results}
\subsection{Case 1: Two-dimensional turbulence with Kolmogorov forcing}\label{sec:2d_results}

We consider forced two-dimensional turbulence with Kolmogorov forcing on a periodic square domain. Such flows are widely used as test cases for developing data-driven turbulence closures because they are computationally inexpensive while still retaining key nonlinear and multiscale features of turbulent dynamics relevant to closure modeling \citep{Guan_2023, Jakhar_2024, Subel_2023}. The reference data are obtained from DNS, while the corresponding coarse-grid simulations are evolved on a substantially lower-resolution mesh using an LES formulation with a Smagorinsky baseline closure (Smagorinsky coefficient, $C_s = 0.2$). The learned model is assessed by replacing the explicit nudging correction during rollout and comparing the resulting dynamics with the subsampled DNS, the baseline Smagorinsky LES (SMAG) and the nudged LES. We also compare against a dynamic Smagorinsky benchmark based on the formulation reviewed by \citet{meneveau2000scale}. In the present implementation, the test filter is taken to be a spectral cutoff with test-to-grid filter scale ratio equal to two, and the local eddy viscosity is constrained to remain positive, following \citet{zang1993dynamic}. Since the coarse-grid reference state is obtained by direct subsampling rather than by applying an explicit filter, this dynamic Smagorinsky implementation is not identical to the standard filtered formulation; it is nevertheless retained as a useful benchmark for comparison with the learned model. Full details of the governing equations, forcing, numerical setup and dataset generation are provided in Appendix~\ref{sec:appendix_data_2d_turbulence}.

\begin{figure}
  \centering
  \begin{tabular}{@{}c@{\hspace{0.02\textwidth}}c@{}}
    \includegraphics[width=0.45\textwidth]{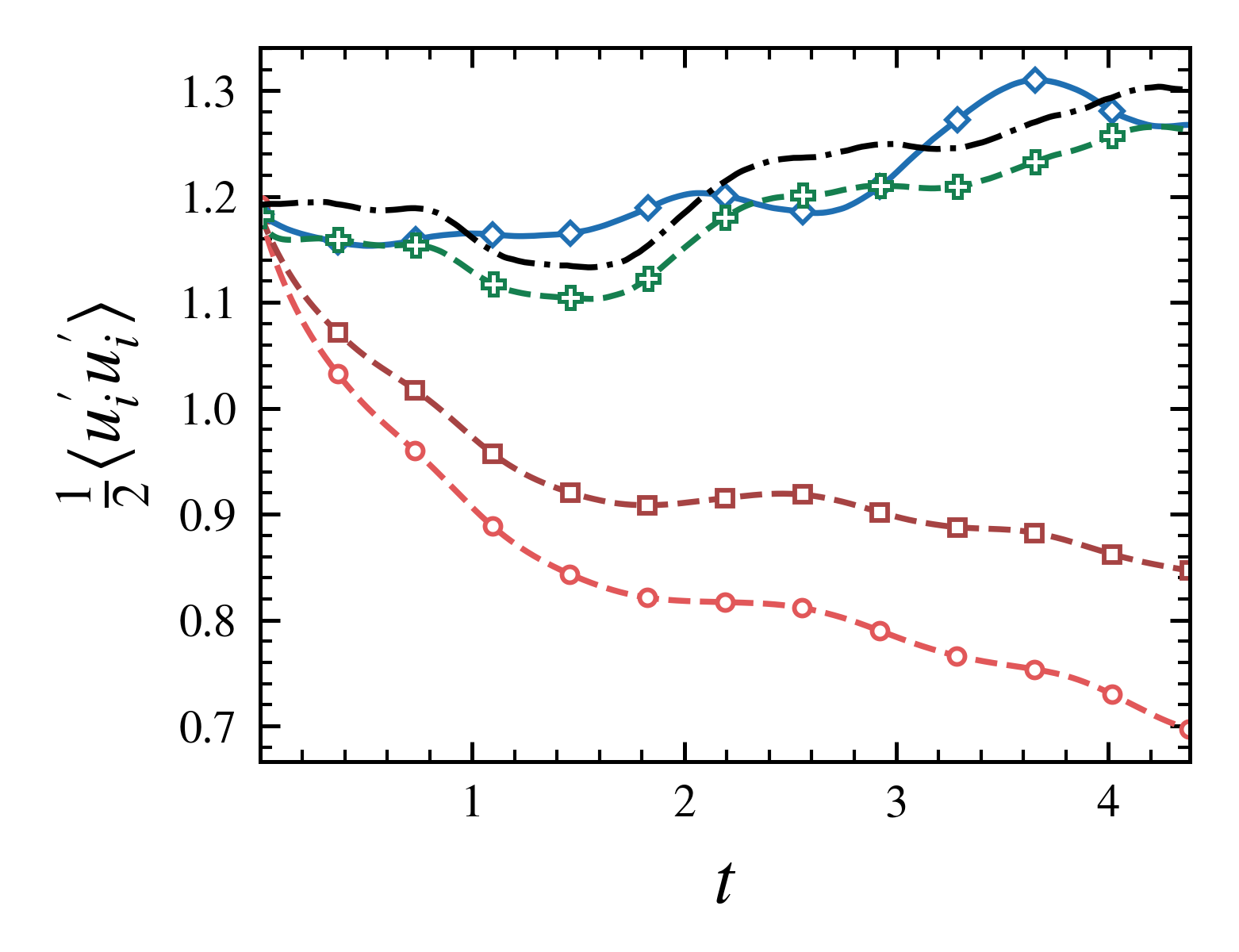} &
    \includegraphics[width=0.45\textwidth]{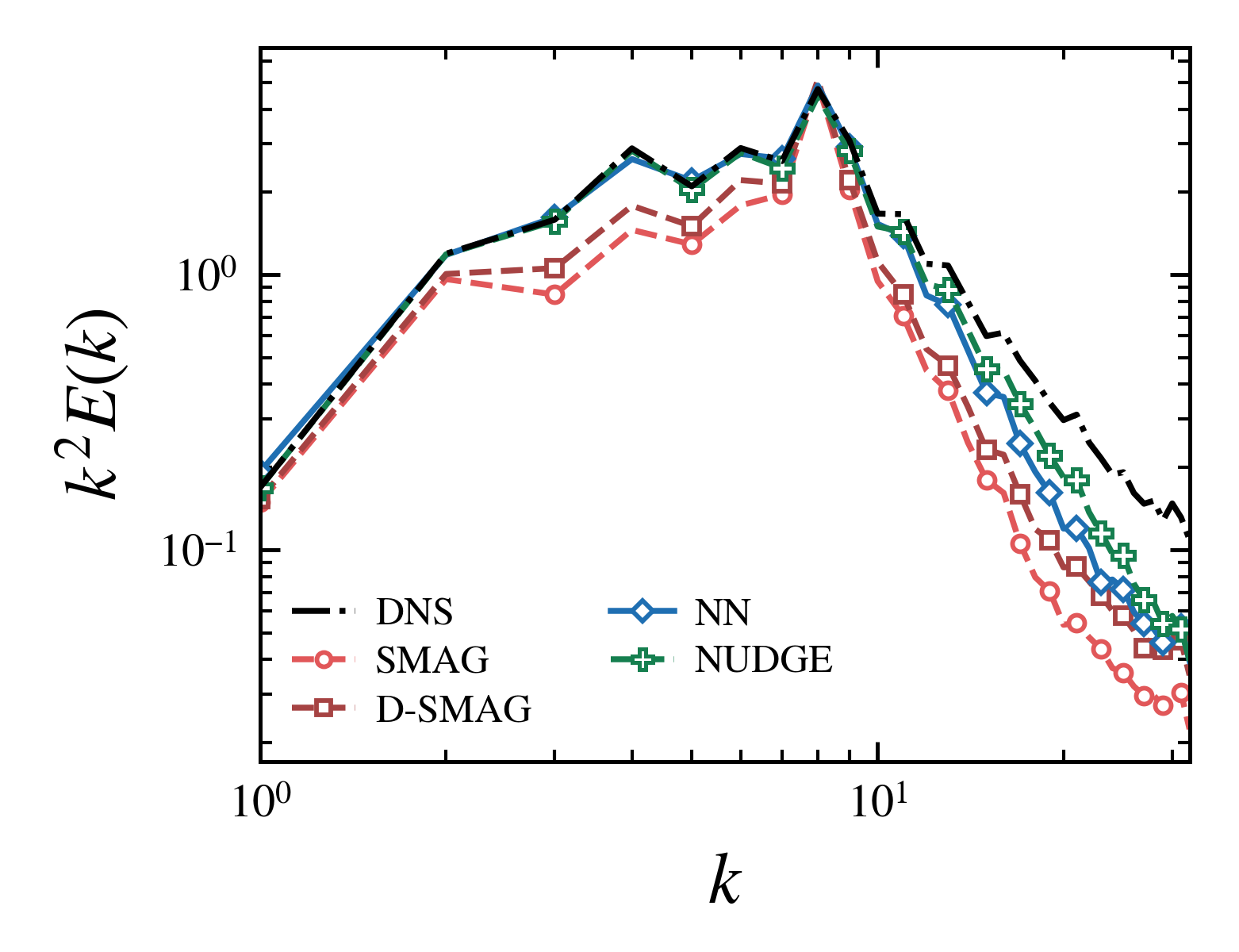} \\
    (a) & (b) \\[0.5em]
    \includegraphics[width=0.45\textwidth]{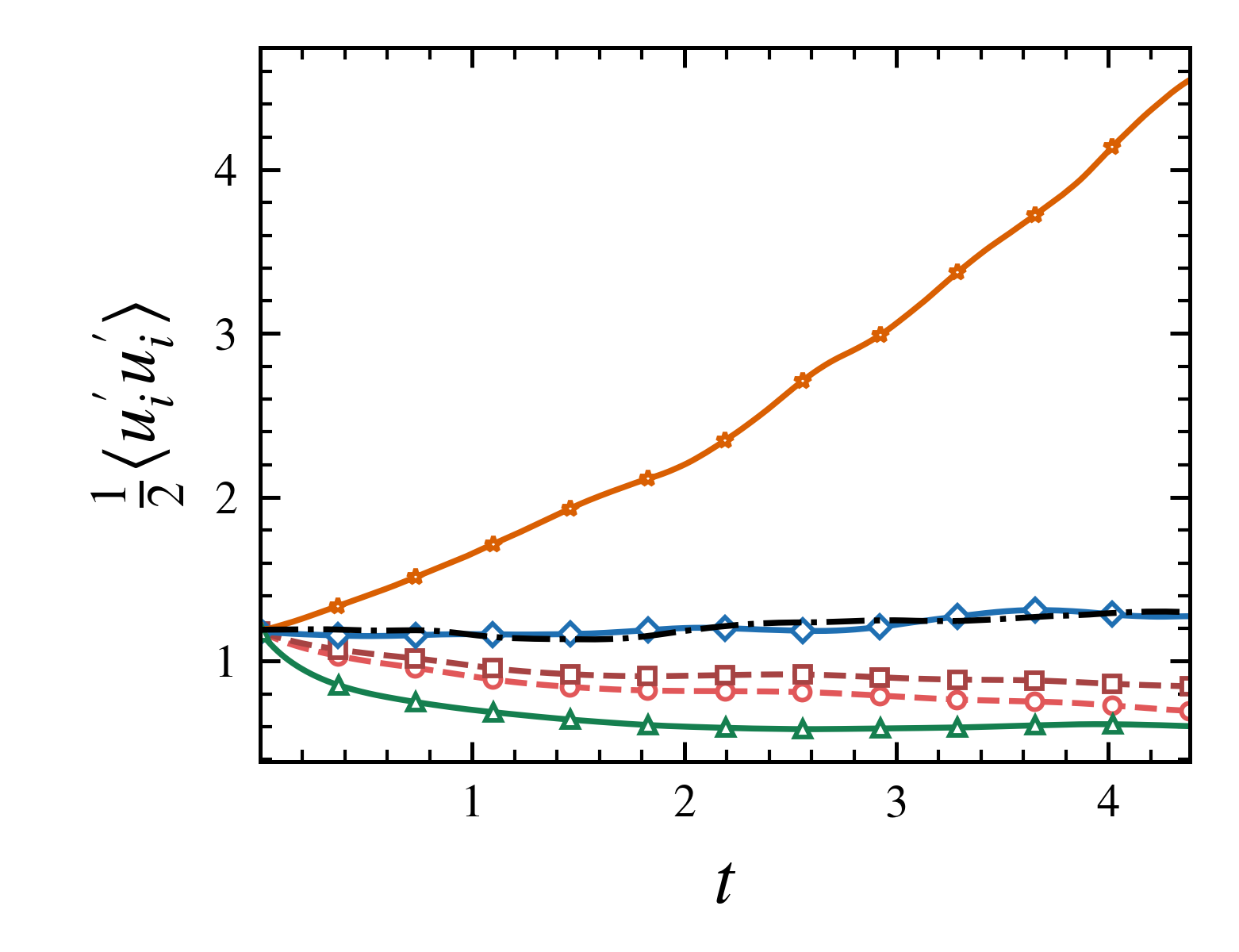} &
    \includegraphics[width=0.45\textwidth]{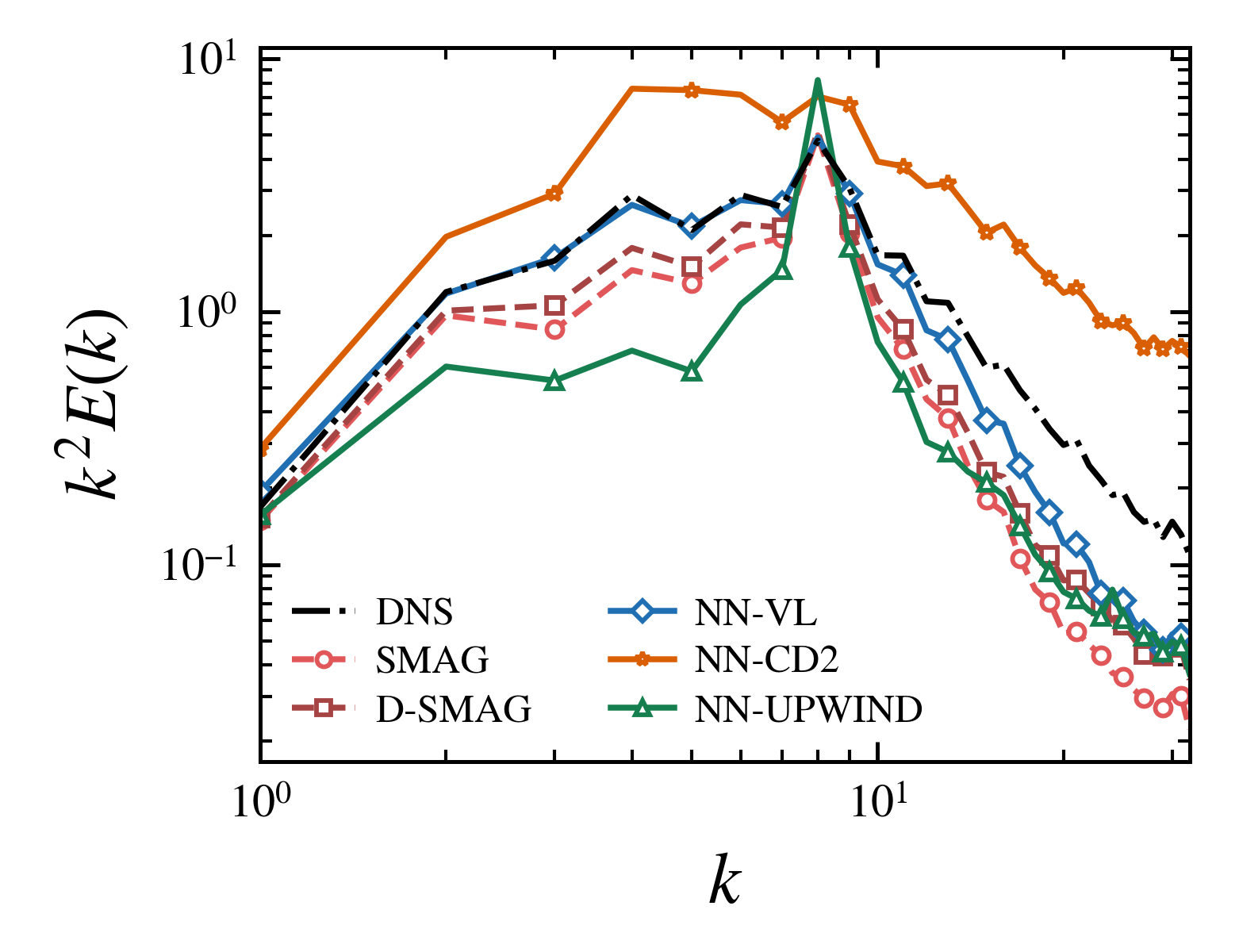} \\
    (c) & (d)
  \end{tabular}
  \caption{
{Results for model trained using a single numerical scheme (Van Leer convective scheme) for the two-dimensional turbulence case. (a) Turbulent kinetic energy evolution (TKE), $\tfrac{1}{2}\langle u_i' u_i'\rangle$ versus $t$, and (b) scaled time-averaged energy spectrum $k^2 E(k)$, comparing the subsampled DNS, the baseline Smagorinsky LES (SMAG), the dynamic Smagorinsky LES (D-SMAG), the recorded nudged simulation (NUDGE) and the trained model (NN). (c,d) Corresponding TKE evolution and spectra when the same NN model is evaluated with different convective discretizations: Van Leer (NN-VL), second-order central differencing (NN-CD2) and first-order upwind (NN-UPWIND), together with the subsampled DNS, SMAG and D-SMAG references; SMAG and D-SMAG use the VL scheme. NN and NUDGE show substantially closer agreement with the reference DNS statistics than SMAG and D-SMAG. {Here NUDGE denotes the synchronized coarse-grid solution obtained with the observational feedback active at all coarse-grid points, from which the training targets are recorded.}}}
  \label{fig:single_scheme_2d}
\end{figure}

We first assess the learned closure on the same numerical scheme used for training. In this setting, both training and inference are performed using the Van Leer discretization for the convective term. Figure~\ref{fig:single_scheme_2d} compares the turbulent kinetic energy (TKE) evolution and time-averaged energy spectrum for the subsampled DNS, the Smagorinsky LES (SMAG), dynamic Smagorinsky (D-SMAG), the nudged LES (NUDGE) and the learned model (NN). SMAG is seen to be excessively dissipative, leading to a gradual loss of energy relative to the reference dynamics. As expected, D-SMAG performs better than SMAG in tracking the TKE of the system.  The nudged simulation outperforms both D-SMAG and SMAG, by acting as the additional corrective forcing required to maintain the reference statistics, and the learned model reproduces this behavior closely. This agreement is evident in both the TKE evolution and the time-averaged spectrum, indicating that the learned forcing (or the closure) provides an effective surrogate for the corrective term required by the coarse-grid LES.

We next examine whether a model trained on a single numerical scheme transfers to other discretization settings. As expected, the results indicate that the learned correction is not universal across schemes, but instead retains a strong dependence on the numerical dissipation associated with the training discretization. To illustrate this, we consider first-order upwind (UPWIND), which is highly dissipative, Van Leer (VL), which is less dissipative than UPWIND but still damps the flow in regions of strong gradients, and second-order central differencing (CD2), which introduces substantially less damping than the other two. In all cases, only the discretization of the convective term is varied. As shown in figure~\ref{fig:single_scheme_2d}, the model trained on the VL scheme fails to compensate for the stronger numerical dissipation of UPWIND, while over-correcting in the less dissipative CD2 case, leading to excessive energy injection in the latter.

{To represent this dependence on numerical discretization, we next train the model on multiple schemes and condition it explicitly on the target discretization by encoding the numerical schemes as input labels using FiLM encodings (see \S~\ref{sec:methodology}).}
For this case, training is performed using two convective term discretizations, namely UPWIND and VL, encoded by the scheme labels $0$ and $1$, respectively. The dataset is constructed in the same manner as in the single-scheme setting, except that the LES-nudging simulations are performed for both schemes and the corresponding encoding is supplied as an additional input to the model. Since the two schemes introduce different levels of numerical dissipation, the learned correction must adapt accordingly. In particular, for the same synchronized coarse-grid input, the correction associated with UPWIND is expected to inject more energy than that associated with VL in order to compensate for the stronger numerical damping.

\begin{figure}
  \centering
  \begin{tabular}{@{}c@{\hspace{0.02\textwidth}}c@{}}
    \includegraphics[width=0.45\textwidth]{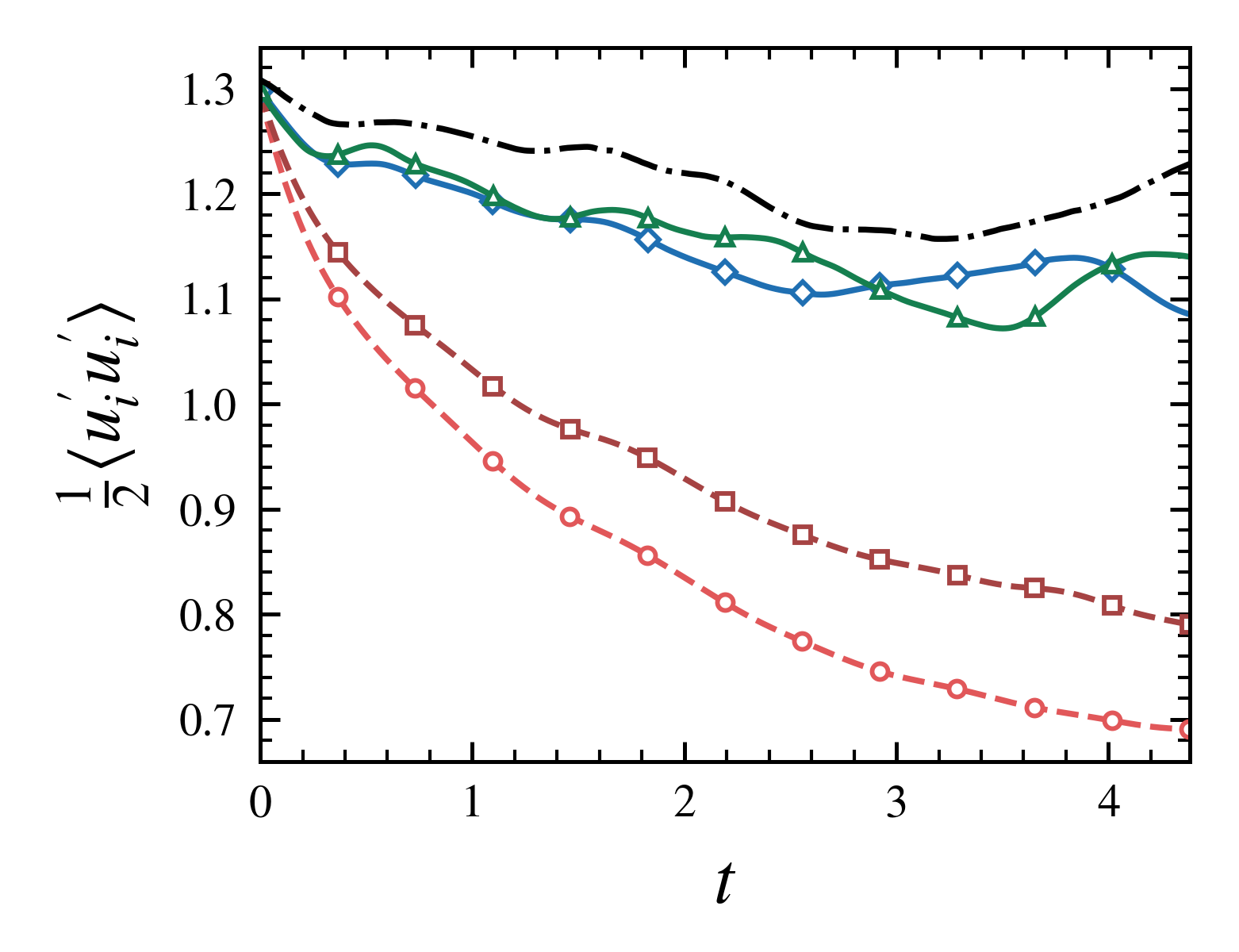} &
    \includegraphics[width=0.45\textwidth]{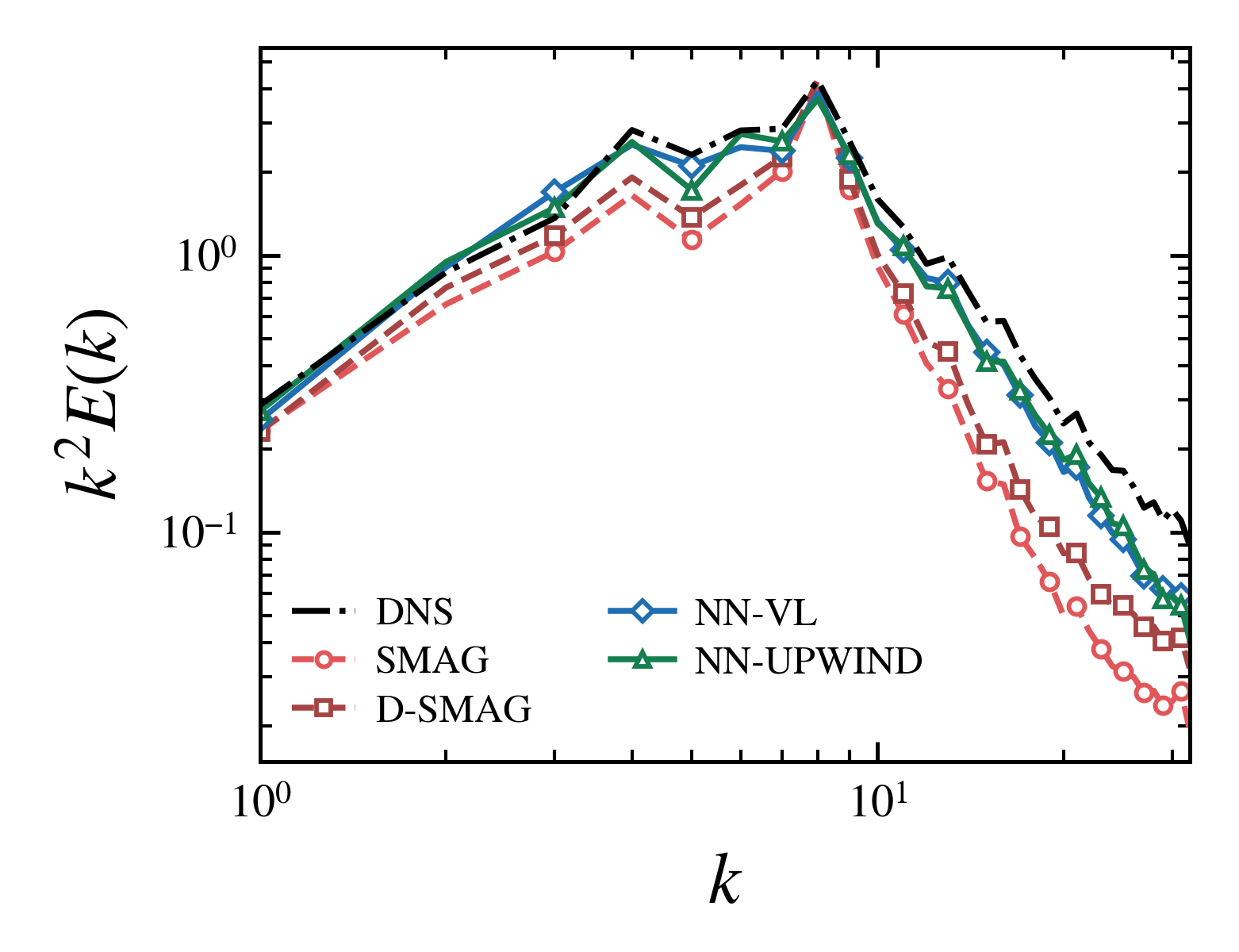} \\
    (a) & (b) \\[0.5em]
    \includegraphics[width=0.45\textwidth]{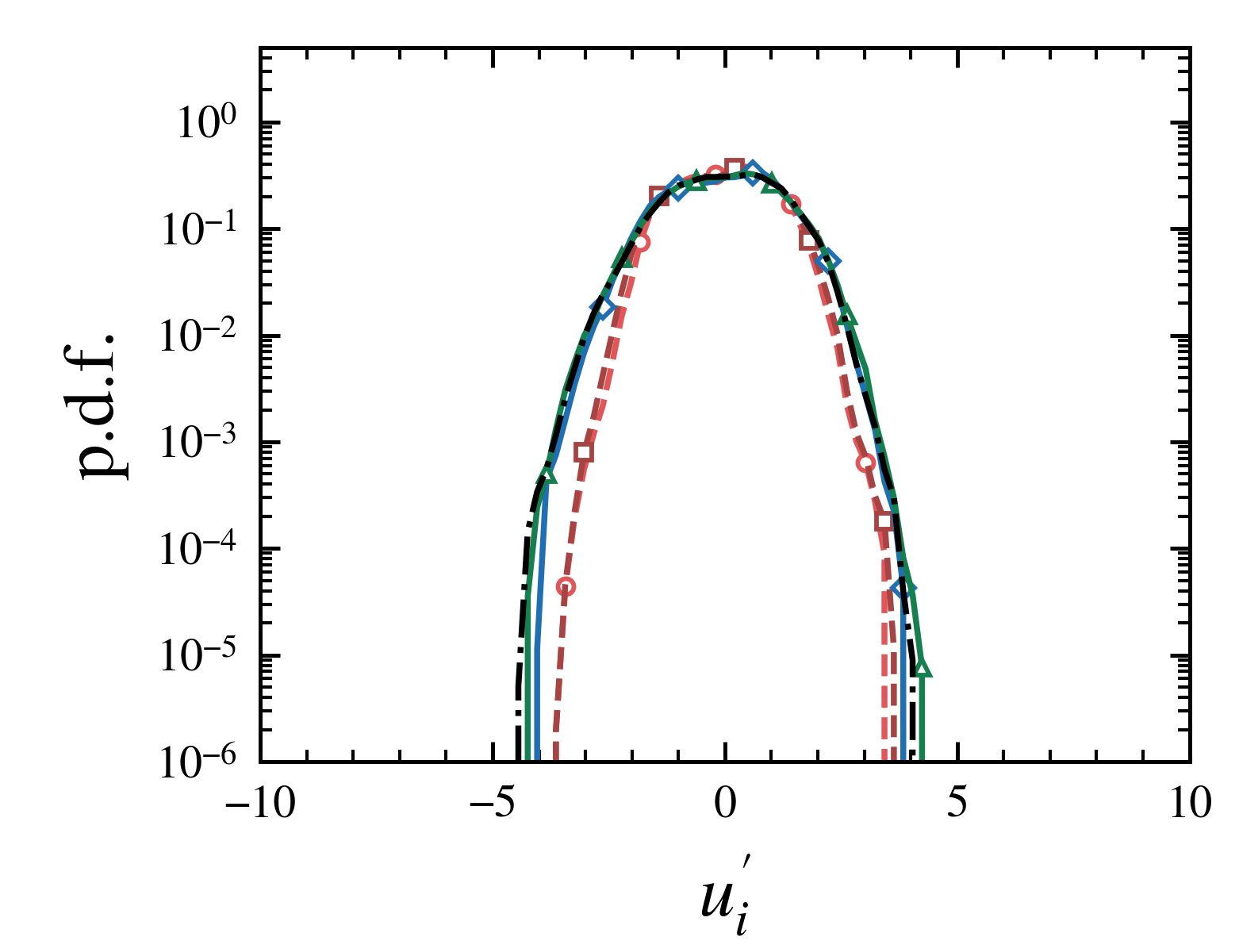} &
    \includegraphics[width=0.45\textwidth]{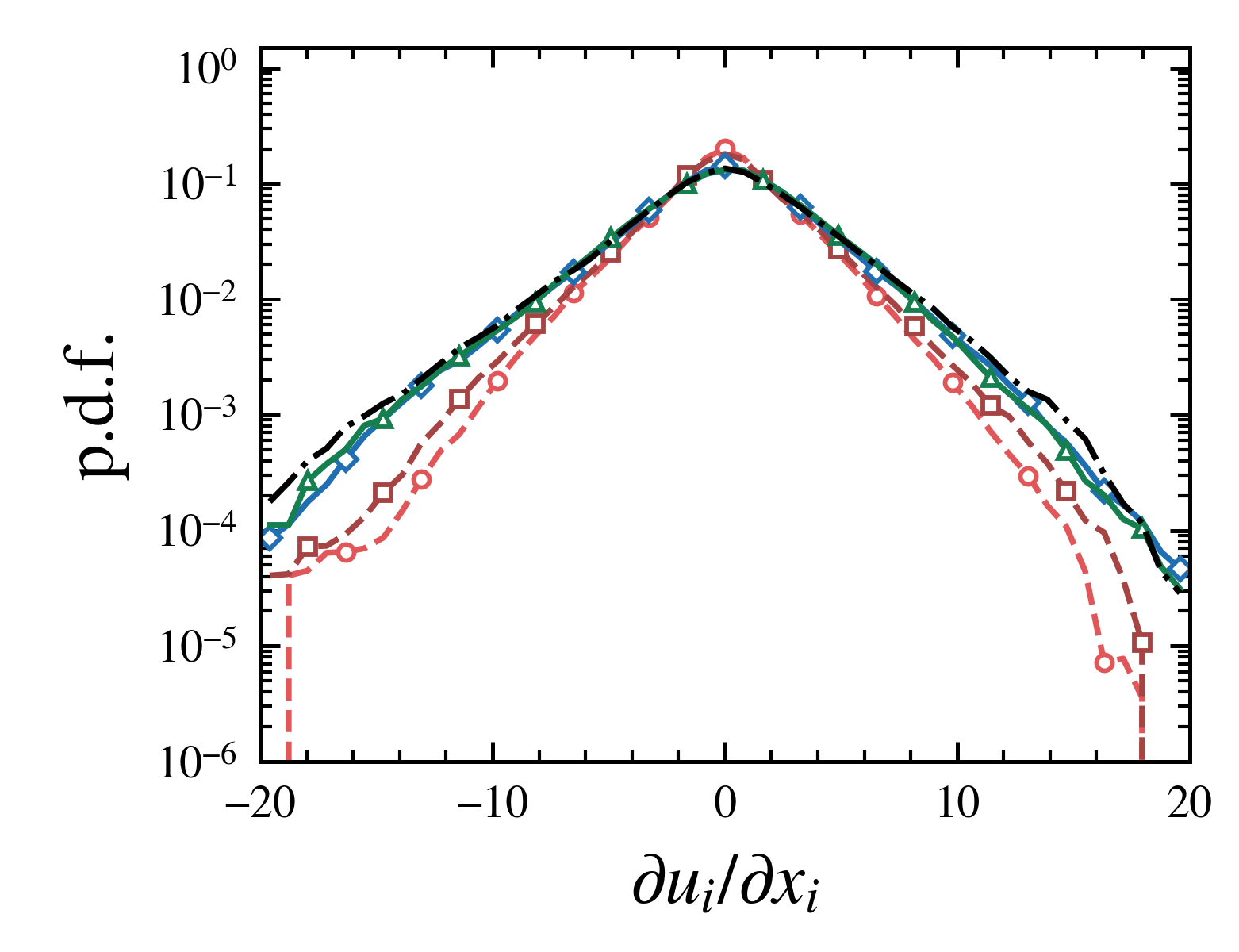} \\
    (c) & (d)
  \end{tabular}
  \caption
{{Results of the same scheme-conditioned model trained on the Van Leer (VL) and upwind (UPWIND) schemes deployed using different discretization encodings. (a) TKE, $\tfrac{1}{2}\langle u_i' u_i'\rangle$ versus $t$. (b) Scaled time-averaged energy spectrum $k^2 E(k)$. (c,d) Probability distribution functions of the velocity fluctuations $u_i'$ and the longitudinal velocity gradients $\partial u_i/ \partial x_i$. All panels compare DNS, SMAG, D-SMAG, and the scheme-conditioned model evaluated on the two trained discretizations, NN-VL and NN-UPWIND. The scheme-conditioned model adapts to both schemes and shows closer agreement with the reference statistics than SMAG and D-SMAG.}}
  \label{fig:case1_multischeme_stats}
\end{figure}

\begin{table}
\centering
\begin{tabular}{lccc}
\hline
Method & rel. $L_2$ & TKE & rel. TKE error \\
\hline
DNS & -- & 1.216$\times 10^{0}$ & -- \\
SMAG & 7.054$\times 10^{-1}$ & 5.221$\times 10^{-1}$ & 5.706$\times 10^{-1}$ \\
D\mbox{-}SMAG & 6.463$\times 10^{-1}$ & 9.411$\times 10^{-1}$ & 2.260$\times 10^{-1}$ \\
NN\mbox{-}VL & 7.291$\times 10^{-1}$ & 1.112$\times 10^{0}$ & \textbf{8.577$\times 10^{-2}$} \\
NN\mbox{-}UPWIND & 7.525$\times 10^{-1}$ & 1.102$\times 10^{0}$ & \textbf{9.399$\times 10^{-2}$} \\
\hline
\end{tabular}
\vspace{3mm}
\caption{{Comparison of relative $L_2$ norm error, time-averaged TKE and relative error in TKE with coarse DNS. SMAG, D-SMAG and NN do not retain pointwise synchronization with the reference (unlike nudging), and their relative errors are of the same order; however, the TKE (and the statistics) of the model rollouts remains closer to the ground truth than those of the baseline SMAG and D-SMAG.}}
\label{tab:nudging-summary-2d}
\end{table}

\begin{figure}
  \centering
  \includegraphics[width=1.00\textwidth]{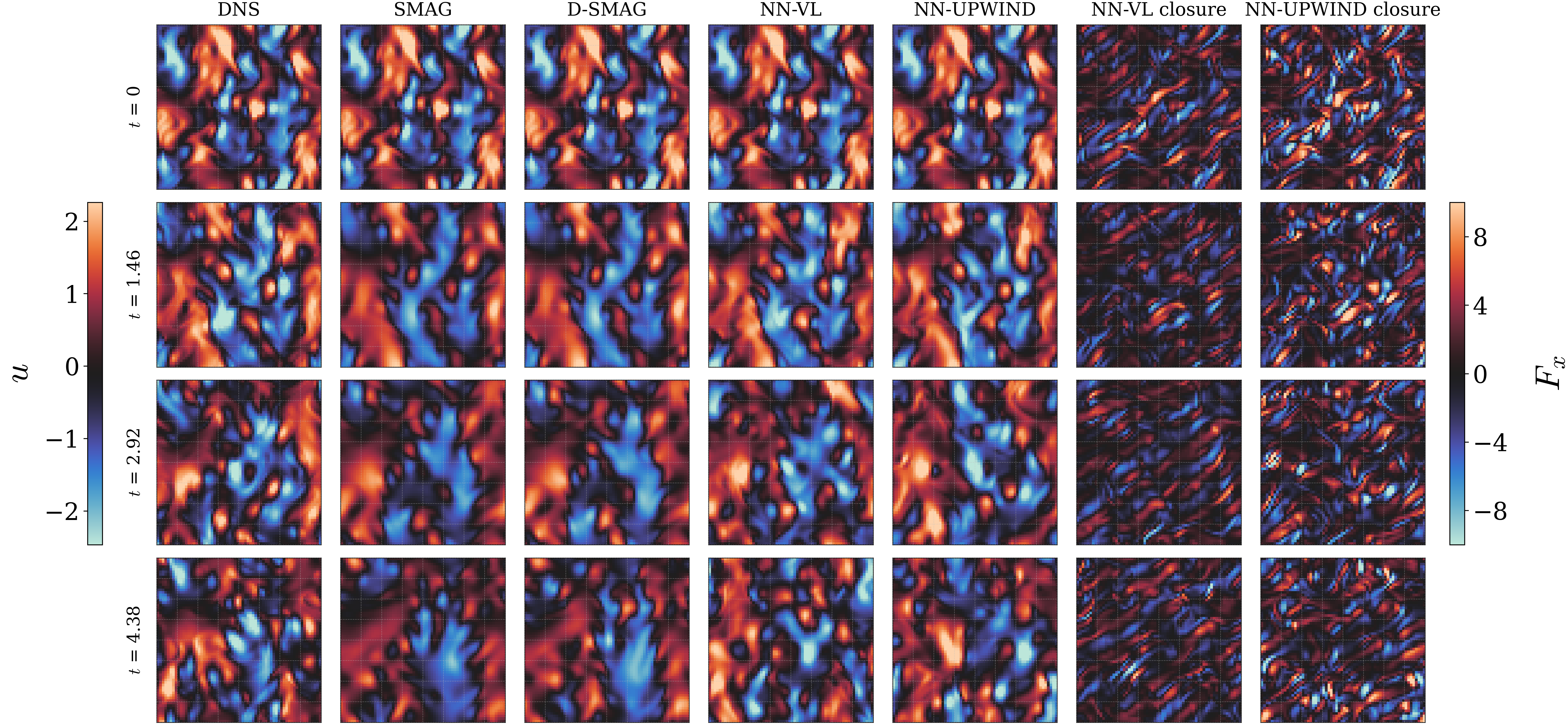}
  \caption{
{Representative coarse-grid $x$-component velocity fields at different times. The first five columns show the velocity field $u$ for the subsampled DNS, SMAG, D-SMAG and model deployed in the two schemes, NN-VL and NN-UPWIND. The last two columns show the corresponding predicted forcing fields $F_x$ (after scaling model output with the nudging coefficient, $\mu$) for NN-VL and NN-UPWIND. SMAG and D-SMAG are visibly more dissipative than the NN rollouts. The predicted forcing fields for the two schemes also differ, reflecting the different levels of numerical dissipation associated with the underlying discretizations.}}
  \label{fig:case1_fields}
\end{figure}

\begin{figure}
  \centering
  \begin{tabular}{@{}c@{\hspace{0.02\textwidth}}c@{}}
    \includegraphics[width=0.47\textwidth]{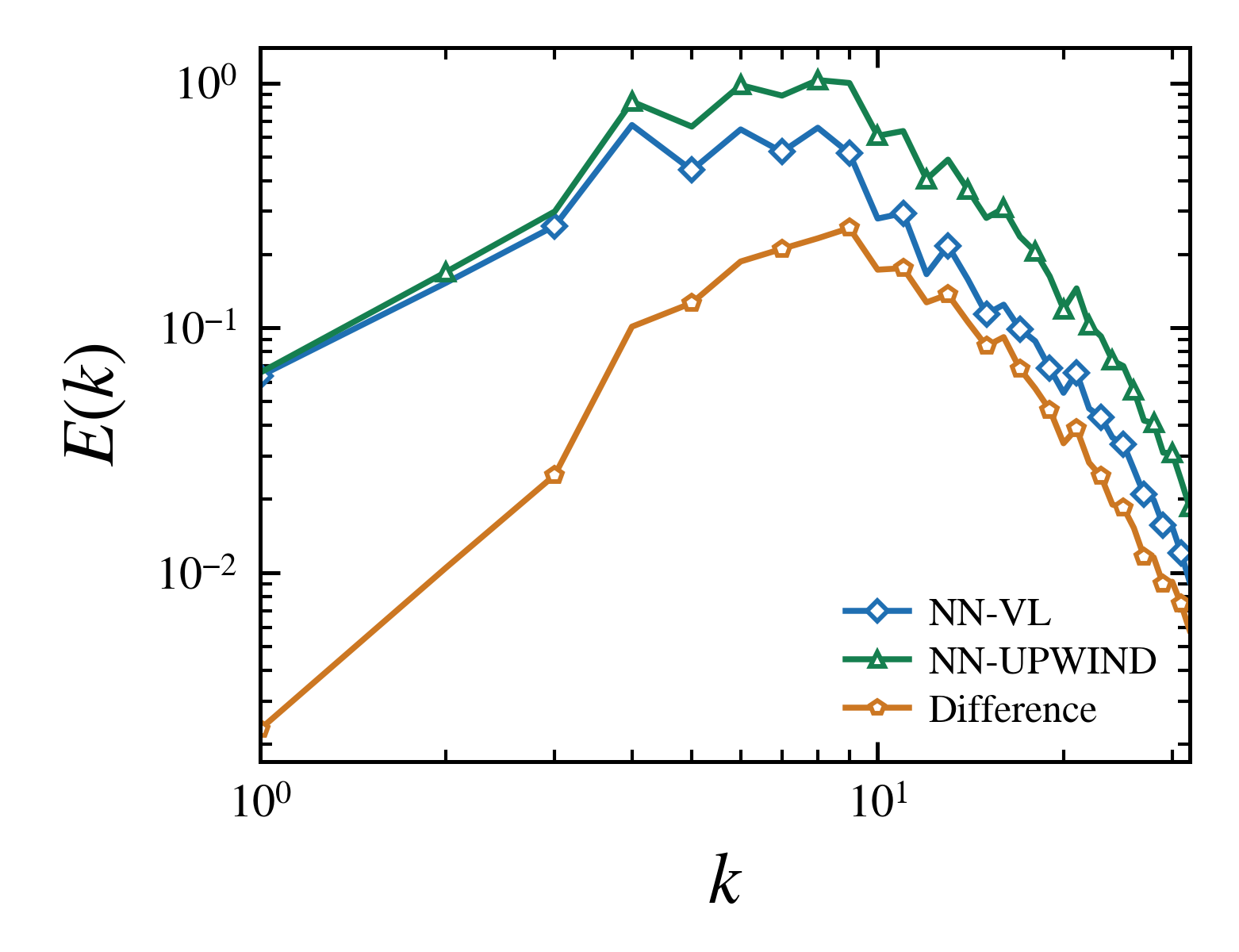} &
    \includegraphics[width=0.47\textwidth]{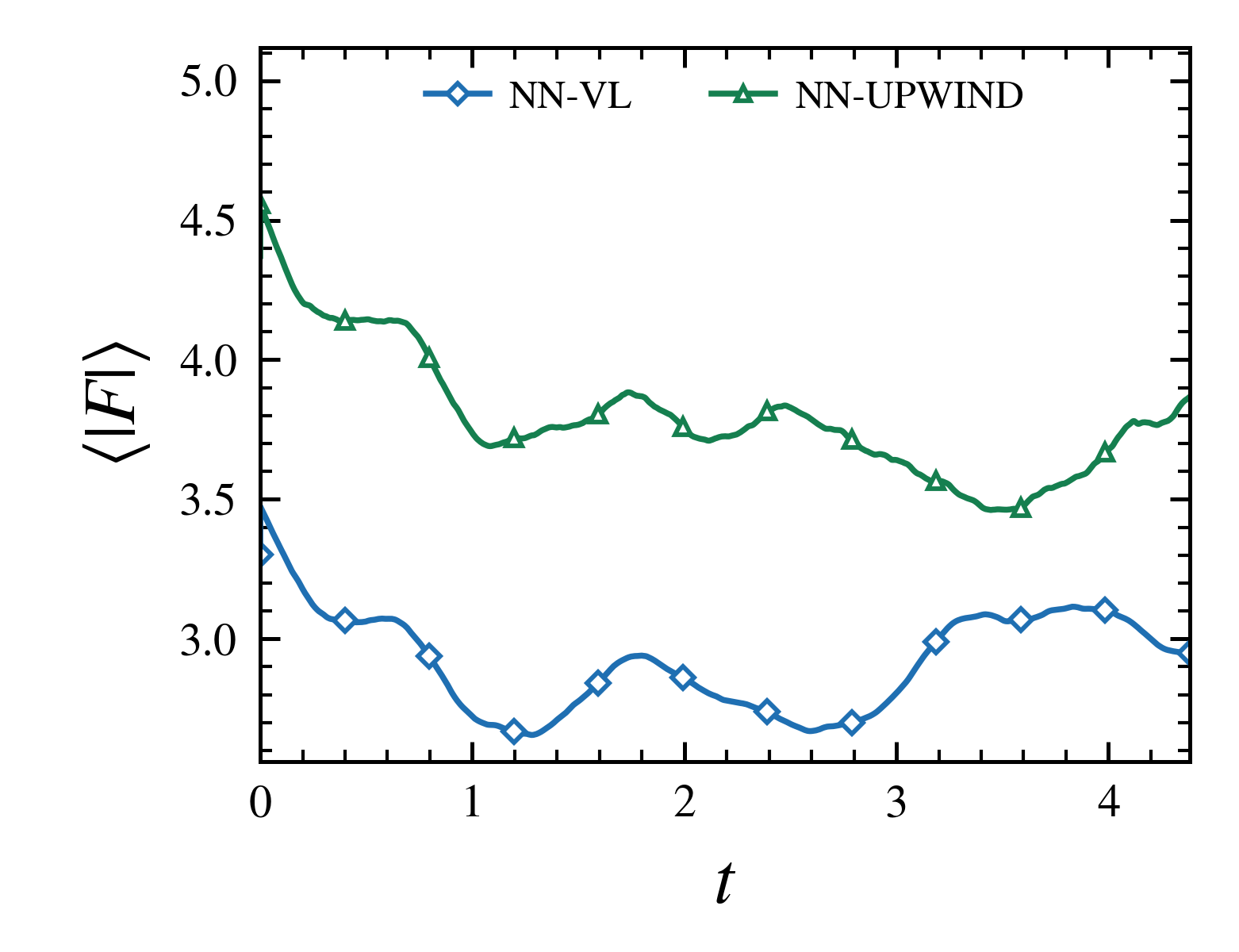} \\
    (a) & (b)
  \end{tabular}

  \vspace{0.5em}

  \begin{tabular}{@{}c@{}}
    \includegraphics[width=1.0\textwidth]{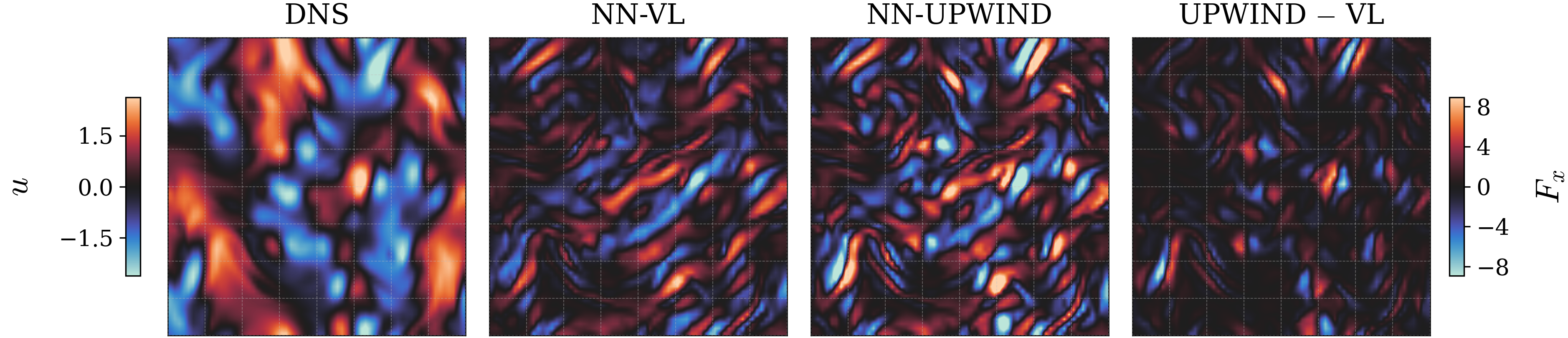} \\
    (c)
  \end{tabular}
  \caption{
{Comparison of the predicted forcing associated with the VL and UPWIND schemes for the same synchronized coarse-grid input. (a) Energy spectrum $E(k)$ of the predicted forcing for VL and UPWIND, together with their difference UPWIND$-$VL. (b) Time evolution of the spatially averaged forcing magnitude $\langle|F|\rangle$ during rollout for VL and UPWIND. (c) Representative $x$-component fields: the DNS velocity $u$, and the predicted forcing $F_x$ for NN-VL, NN-UPWIND and their difference UPWIND$-$VL.}}
  \label{fig:closure_discretization_2d}
\end{figure}

Figure~\ref{fig:case1_multischeme_stats} shows the inference results for the model trained on both VL and UPWIND using FiLM conditioning. The TKE evolution, time-averaged energy spectra and probability distribution functions indicate that the scheme-conditioned model recovers the required correction for both discretizations while retaining close agreement with the reference statistics. In contrast to the single-scheme model, the present model adapts its predicted correction according to the target scheme, thereby accommodating the different levels of numerical dissipation introduced by VL and UPWIND. These results show that explicit conditioning on the numerical discretization substantially improves the ability of the learned model to reproduce the statistical behavior of the reference flow across schemes.

The quantitative comparison is summarized in table~\ref{tab:nudging-summary-2d}. Here the relative $L_2$ error (averaged in time) is defined as $\varepsilon=\langle\|\boldsymbol{v}-\boldsymbol{u}_{obs}\|_2/\|\boldsymbol{u}_0\|_2\rangle_t$, where $\boldsymbol{v}$ is the coarse-grid field produced by the method under test (SMAG, D-SMAG or NN), $\boldsymbol{u}_{obs}$ is the subsampled DNS reference, and $\boldsymbol{u}_{0}$ is the reference field at the initial time. Similarly, the relative TKE error is the discrepancy in the time-averaged TKE between the method and the DNS. 

To further examine the scheme-dependent behavior of the learned correction, figure~\ref{fig:case1_fields} presents representative coarse-grid fields for the DNS, SMAG, D-SMAG and model inference on the VL and UPWIND schemes, together with the corresponding predicted closures. Consistent with the statistical results above, the SMAG appears overly dissipative relative to the reference field, in comparison to D-SMAG and the trained model. The predicted closure fields for VL and UPWIND are also visibly distinct, reflecting the different levels of numerical damping introduced by the two schemes. 

Figure~\ref{fig:closure_discretization_2d} compares the predicted correction associated with the VL and UPWIND schemes for the same synchronized coarse-grid input. The corresponding spectra indicate good agreement at the largest and smallest resolved scales, with the most pronounced differences occurring in the inertial range, consistent with the stronger numerical dissipation of UPWIND relative to VL.

Overall, the two-dimensional results show that the proposed framework recovers the correction required to reproduce the reference coarse-grid statistics on the training scheme, but that this correction depends strongly on the underlying discretization. Explicit scheme conditioning allows a single model to adapt across schemes while maintaining close agreement with the reference statistics.

\subsection{Case 2: Three-dimensional homogeneous isotropic turbulence}\label{sec:3d_results}
We next assess the proposed framework on forced three-dimensional homogeneous isotropic turbulence. In contrast to the two-dimensional case, the data are generated and evolved using a compressible finite-volume solver operated in a low-Mach-number regime, so that the dynamics remain close to the incompressible limit. The nudging correction is introduced in conservative form through momentum and energy source terms, and the full details of the numerical setup and how our nudging approach generalizes to the compressible representation are given in Appendix~\ref{sec:appendix_data_3d_turbulence}.

Let $\rho_{obs}$ and $\boldsymbol{u}_{obs}$ denote the observed DNS density and velocity fields on the coarse grid, and let $\rho$ and $\boldsymbol{v}$ denote the corresponding LES density and velocity fields. The nudging source is written in conservative form, with momentum source,
\begin{equation}
\boldsymbol{F}_{p}=\mu\left(\rho_{obs}\boldsymbol{u}_{obs}-\rho\boldsymbol{v}\right),
\end{equation}
and its corresponding energy source, 
\begin{equation}
F_{e}=\boldsymbol{v}\cdot \boldsymbol{F}_{p}.
\end{equation}
Under the present low-Mach conditions, density variations remain small, and no additional density forcing is applied in this case. For the coarse-grid nudging simulations the convective term is varied across second-, fourth- and sixth-order central schemes, denoted C2, C4 and C6, respectively, with the net effect of C2 {along with the Smagorinsky term} being the most dissipative and that of C6 the least.

\begin{figure}
  \centering
  \begin{tabular}{@{}c@{\hspace{0.02\textwidth}}c@{}}
    \includegraphics[width=0.48\textwidth]{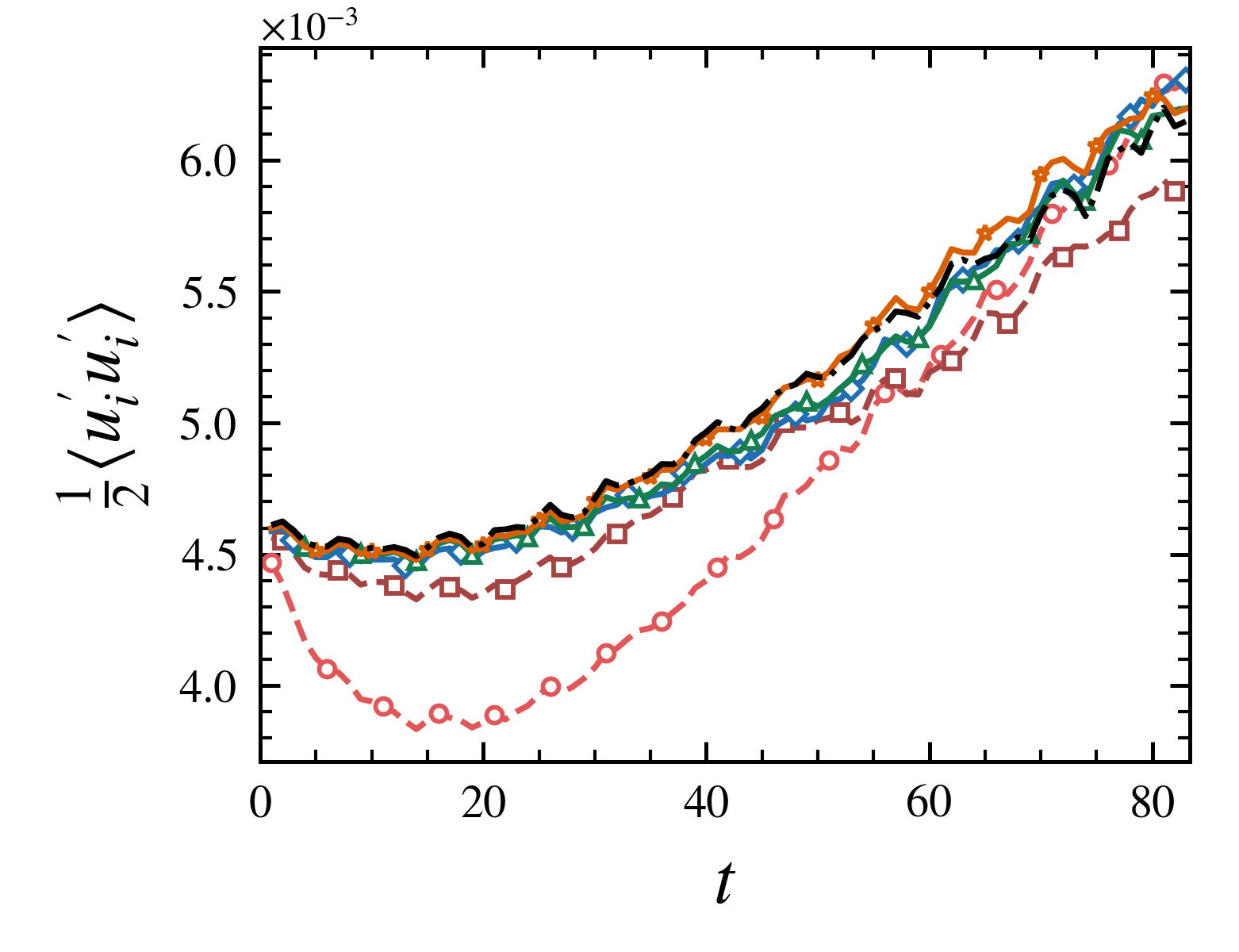} &
    \includegraphics[width=0.48\textwidth]{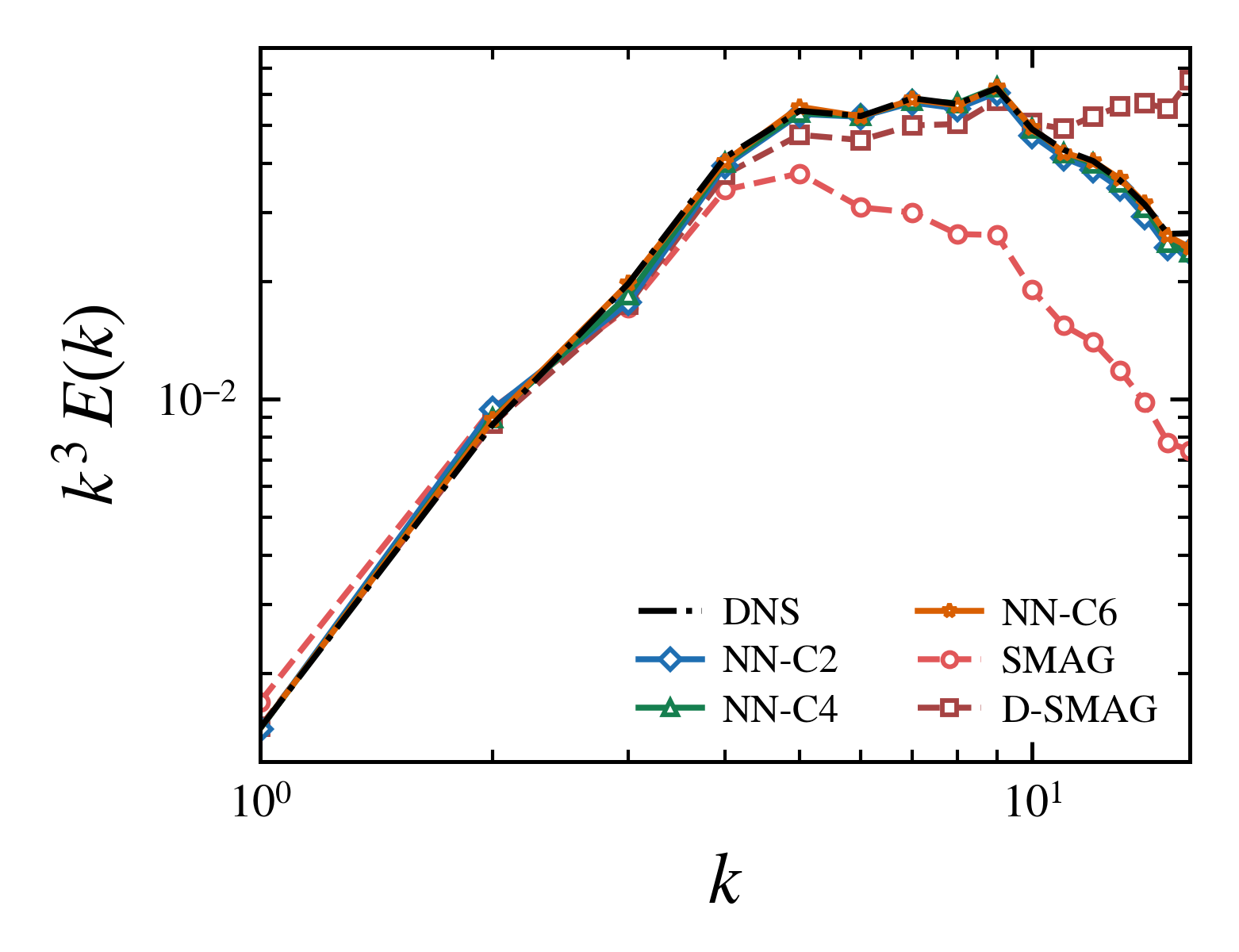} \\
    (a) & (b) \\[0.5em]
    \includegraphics[width=0.48\textwidth]{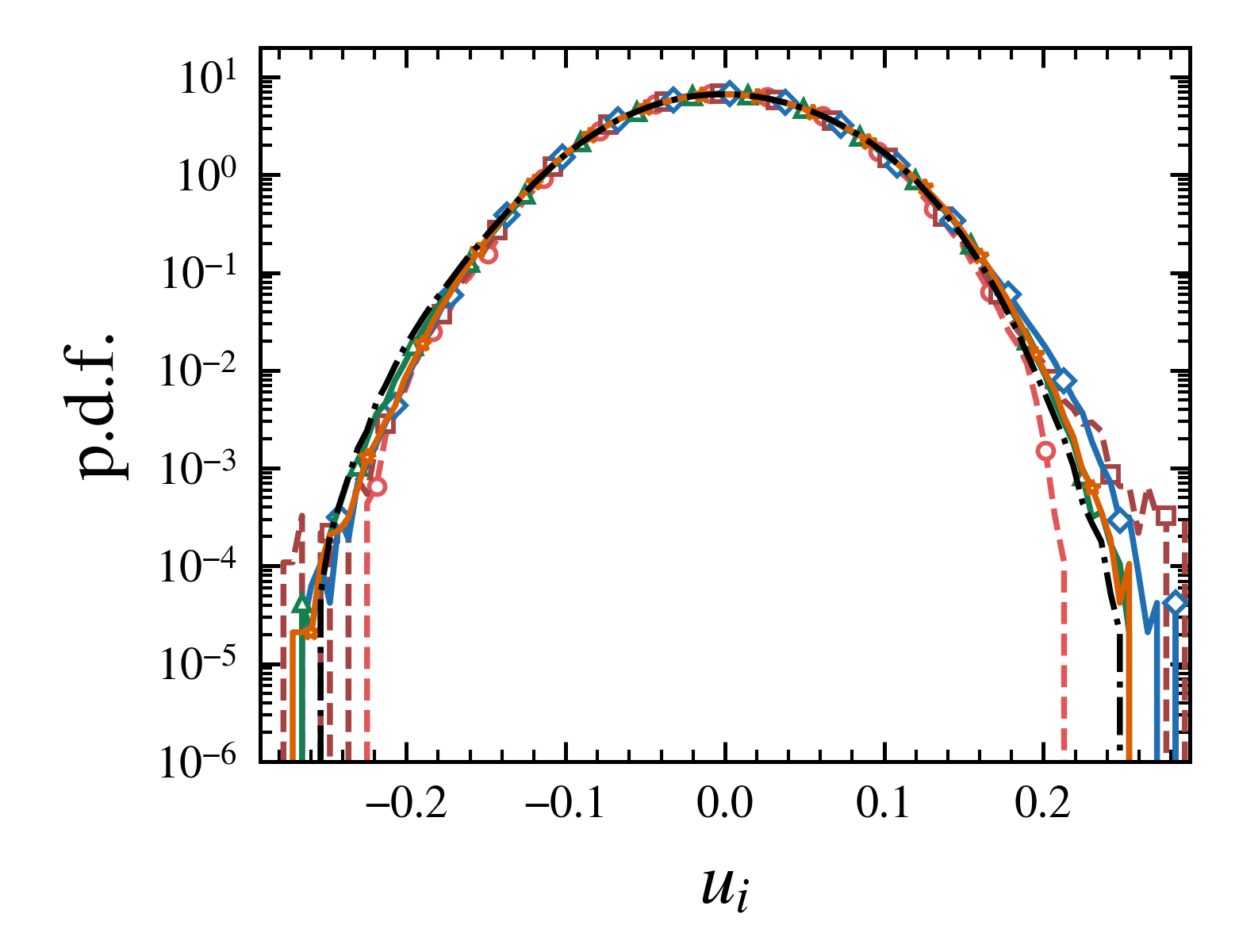} &
    \includegraphics[width=0.48\textwidth]{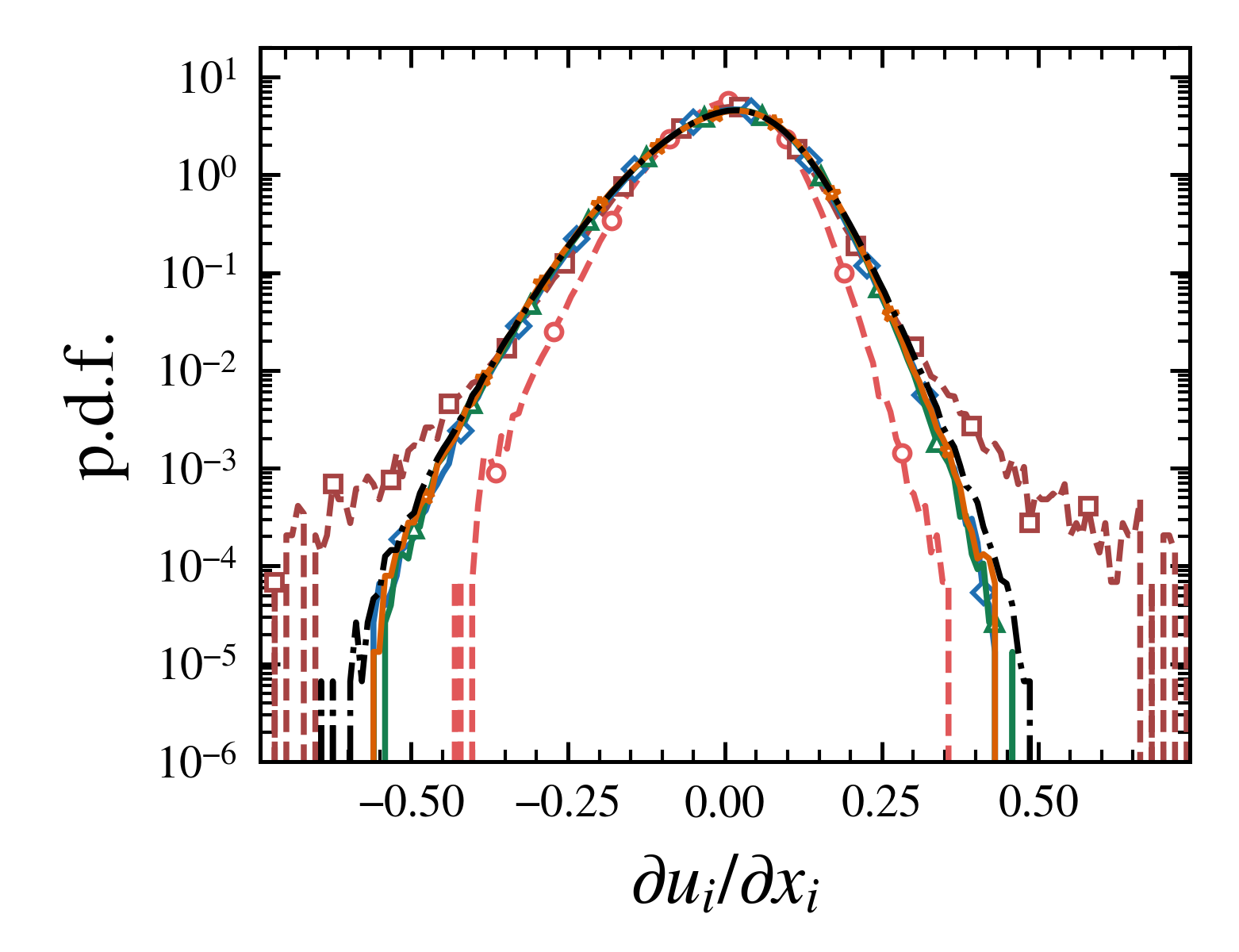} \\
    (c) & (d)
  \end{tabular}
  
\caption{
{Results for the model trained on the C2, C4 and C6 schemes as applied to the three-dimensional test case. (a) TKE evolution, $\tfrac{1}{2}\langle u_i' u_i'\rangle$ versus $t$. (b) Scaled time-averaged energy spectrum $k^3 E(k)$. (c,d) Probability distribution functions of the velocity fluctuations $u_i'$ and the longitudinal velocity gradients $\partial u_i/\partial x_i$. {The rollout spans around five eddy-turnover times~\citep{rosales2005linear}.} All panels compare DNS, SMAG, D-SMAG and the scheme-conditioned model evaluated on the three trained discretizations, NN-C2, NN-C4 and NN-C6. Here C2, C4 and C6 denote the second-, fourth- and sixth-order central convective schemes. As in the two-dimensional case, the NN rollouts retain the reference statistics more accurately than SMAG and D-SMAG.}}
  \label{fig:inference_stats_3d}
\end{figure}

\begin{figure}
  \centering
  \includegraphics[width=0.98\textwidth]{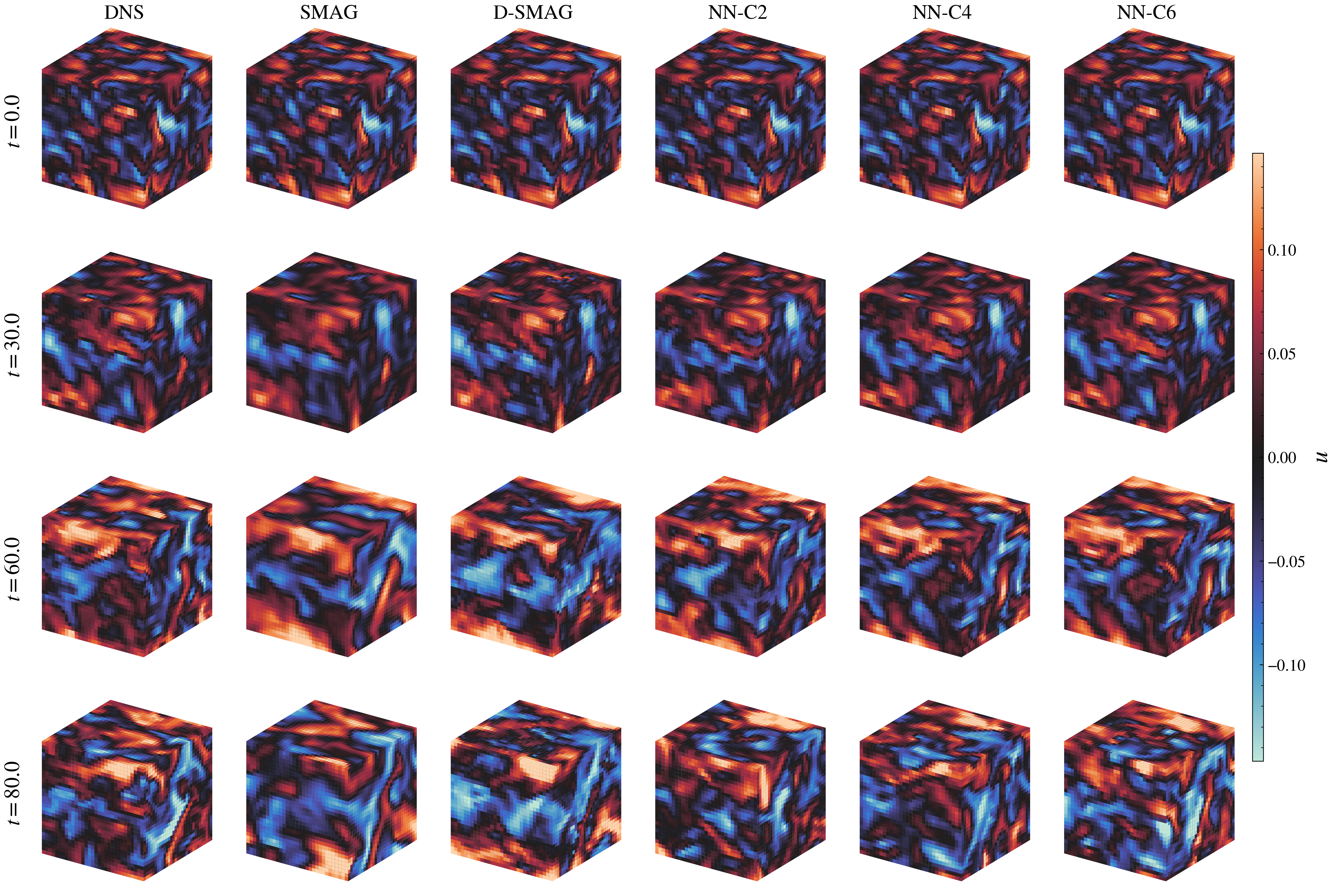}
  \caption{Representative coarse-grid $x$-component velocity fields at different times for the three-dimensional test case. The columns represent the velocity field $u$ for subsampled DNS, SMAG, D-SMAG and model rollouts in the three schemes: NN-C2, NN-C4 and NN-C6.}
  \label{fig:3d_fields}
\end{figure}

{Figure~\ref{fig:inference_stats_3d} presents the statistical comparison for the ground truth, SMAG, D-SMAG and model inference on the three schemes. As in the two-dimensional case, SMAG remains overly dissipative at the higher wavenumbers, but also accumulates some energy at the larger scales due to the linear forcing used to sustain the turbulence (see Appendix~\ref{sec:appendix_B}) acting on the flow. This can also be seen from the initial rapid TKE decay of SMAG and then a gradual recovery with time. The excessive damping of SMAG is also evident from the probability distribution functions and the flow field being visibly smoother (see figure~\ref{fig:3d_fields}) which is associated with loss of sharper structures.} In contrast to the two-dimensional case, D-SMAG performs comparatively better; however, the learned model still shows closer agreement with the reference statistics, whereas D-SMAG exhibits excess small-scale content, as reflected by the elevated energy at high wavenumbers and the mismatch between its PDFs and those of the ground truth. {One key difference between NUDGE and other coarse grid runs (SMAG, D-SMAG and NN) is that, unlike nudging, which is constructed directly from the instantaneous discrepancy with the reference observations, the other rollouts do not remain synchronized in the pointwise sense with the ground truth due to deterministic chaos. The corresponding metrics, defined as in \S~\ref{sec:2d_results}, are reported in table~\ref{tab:nudging-summary-3d}. {For SMAG, the TKE error is reduced by its excess energy at the larger scales, so the energy spectrum provides the fairer comparison.} In both the two- and three-dimensional cases, the relative $L_2$ errors of all methods are of the same order of magnitude; hence, we do not infer any ordering from $\varepsilon$, and rely on the statistical comparisons instead.}

\begin{table}
\centering
\begin{tabular}{lccc}
\hline
Method & rel. $L_2$ & TKE & rel. TKE error \\
\hline
DNS & -- & {5.107$\times 10^{-3}$} & -- \\
SMAG & {7.320$\times 10^{-1}$} & {4.715$\times 10^{-3}$} & {7.668$\times 10^{-2}$} \\
D\mbox{-}SMAG & {8.181$\times 10^{-1}$} & {4.922$\times 10^{-3}$} & {3.628$\times 10^{-2}$} \\
NN\mbox{-}C2 & {7.326$\times 10^{-1}$} & {5.056$\times 10^{-3}$} & \textbf{{9.942$\times 10^{-3}$}} \\
NN\mbox{-}C4 & {5.801$\times 10^{-1}$} & {5.066$\times 10^{-3}$} & \textbf{{8.047$\times 10^{-3}$}} \\
NN\mbox{-}C6 & {5.326$\times 10^{-1}$} & {5.135$\times 10^{-3}$} & \textbf{{5.501$\times 10^{-3}$}} \\
\hline
\end{tabular}
\vspace{3mm}
\caption{Comparison of relative $L_2$ norm error, time-averaged TKE, and relative error in TKE with coarse DNS. {As seen in table~\ref{tab:nudging-summary-2d}, the errors of the model rollouts are comparable with the baseline SMAG and D-SMAG; however, the model tracks the reference TKE closely.} {Note that the TKE of SMAG benefits from its excess energy at the larger scales due to the linear forcing, which compensates for its deficit at the smaller scales (see figure~\ref{fig:inference_stats_3d}).}}
\label{tab:nudging-summary-3d}
\end{table}

\begin{figure}
  \centering
  \begin{tabular}{@{}c@{\hspace{0.02\textwidth}}c@{}}
    \includegraphics[width=0.5\textwidth]{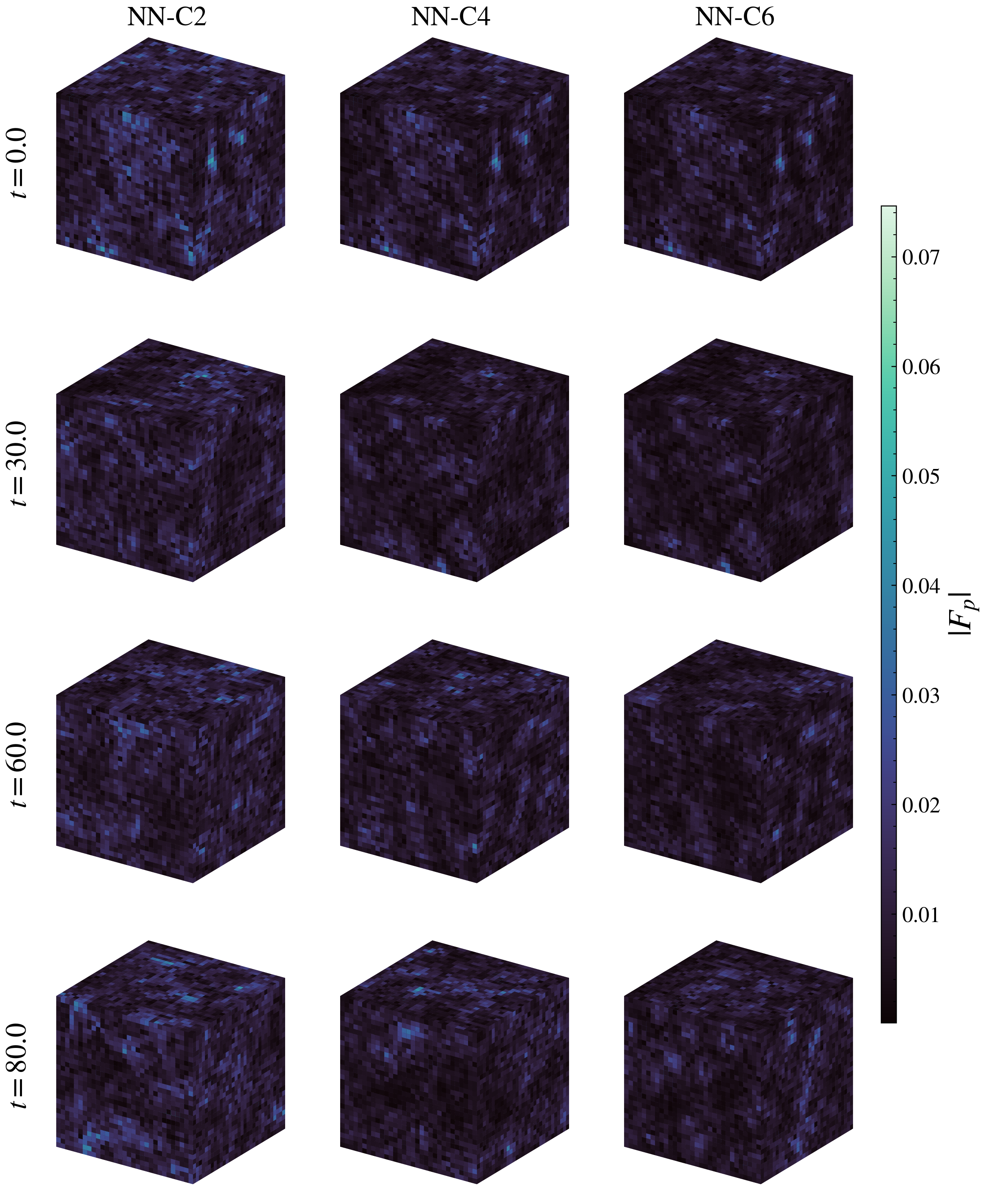} &
    \includegraphics[width=0.42\textwidth]{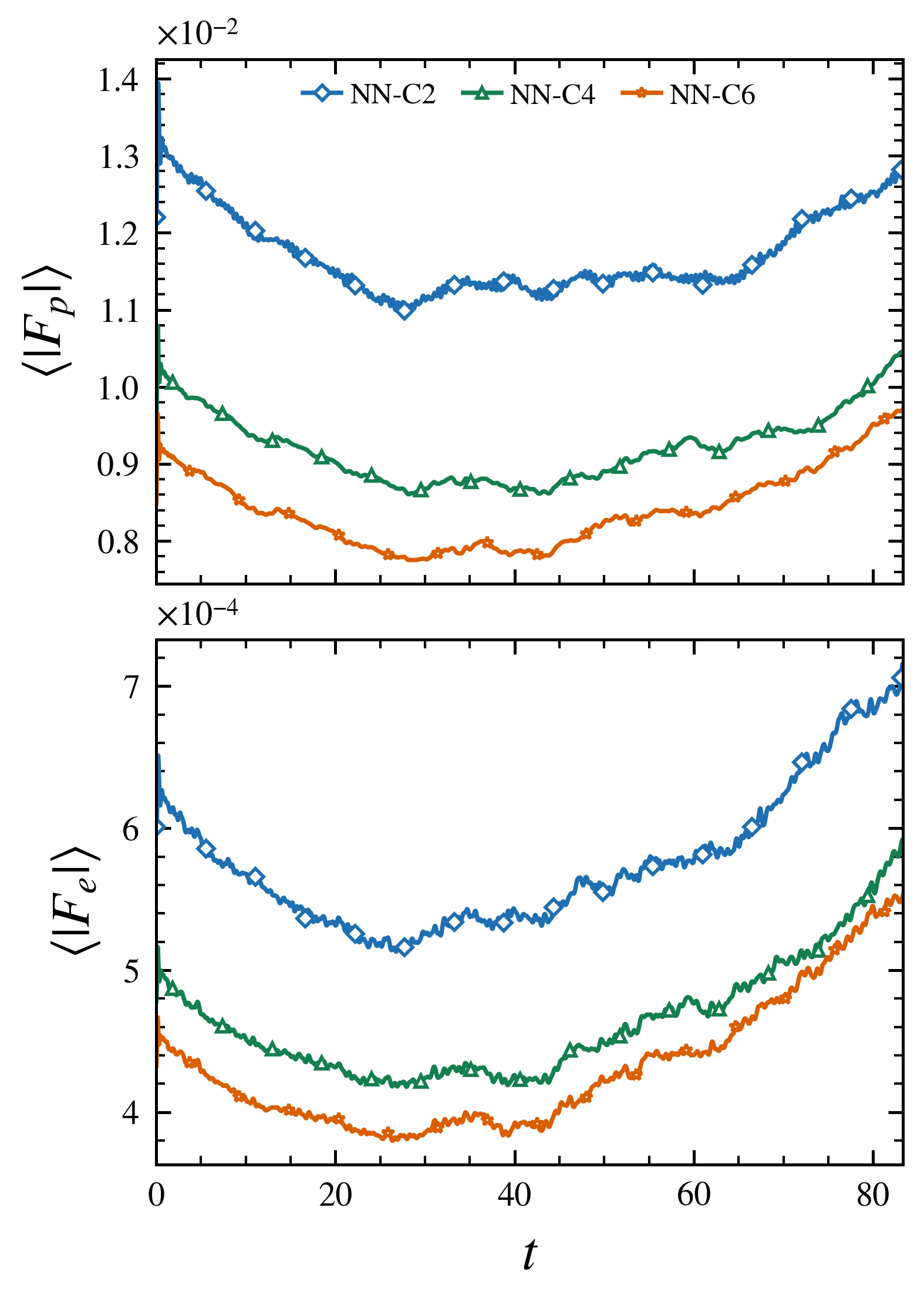} \\
    (a) & (b)
  \end{tabular}

  \vspace{0.5em}

  \begin{tabular}{@{}c@{}}
    \includegraphics[width=1.0\textwidth]{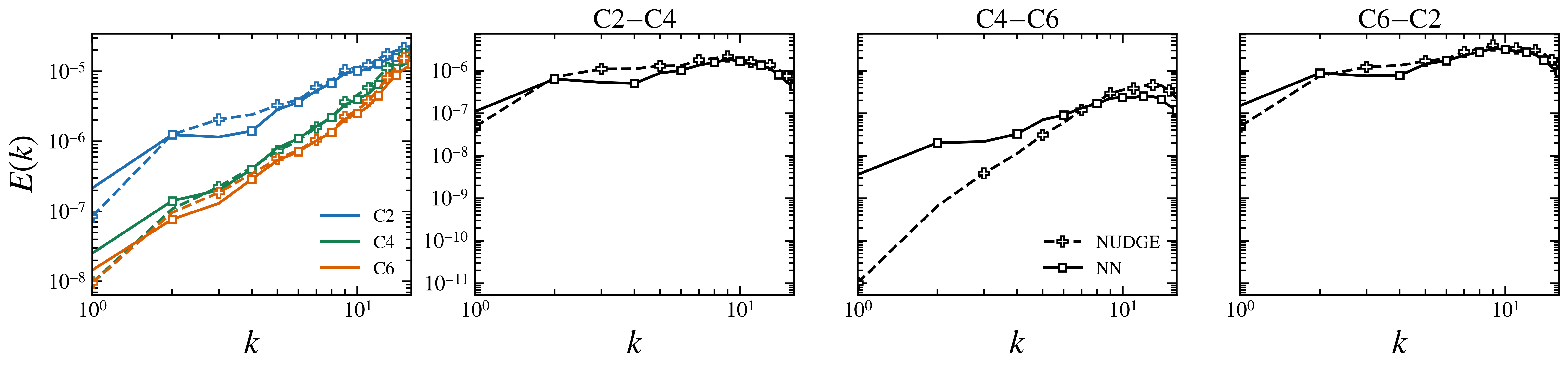} \\
    (c)
  \end{tabular}
\caption{
{Comparison of the predicted correction when applied for the three-dimensional turbulence test case. (a) Representative momentum-forcing magnitude fields $|F_p|$ at different times for the C2, C4 and C6 schemes. (b) Time evolution of the spatially averaged momentum and energy forcing magnitudes, denoted by $\langle|F_p|\rangle$ and $\langle|F_e|\rangle$, for NN-C2, NN-C4 and NN-C6. (c) Energy spectrum $E(k)$ of the predicted and actual nudging momentum corrections: the leftmost panel shows the actual (NUDGE) and predicted (NN) correction spectra per scheme, C2, C4 and C6, with the NUDGE drawn using {dashed} and NN using {solid}; the remaining three panels show the scheme-to-scheme difference spectra C2$-$C4, C4$-$C6 and C6$-$C2.}}
  \label{fig:closure_magnitude_3d}
\end{figure}

Figure~\ref{fig:closure_magnitude_3d} also shows that the model predicts a non-zero correction at \(t=0\), even when the initial condition is taken directly from the reference field. This reflects the construction of the training target: the recorded nudging is taken after the initial transient associated with the onset of nudging, once the correction has attained an approximately steady magnitude. In practice, this transient is short-lived, and in the experiments conducted for this study, the discrepancy settles within approximately 500 time steps (simulation time, $t \approx 0.5$) after the onset of nudging. Despite this initial mismatch, the model rollout remains in good agreement with the reference statistics. The results further show that the learned correction adapts to the different levels of numerical dissipation across C2, C4 and C6, yielding comparable statistical agreement with the reference flow for all three schemes.

\subsection{Numerical discretization-based error analysis}\label{sec:numerical_discretization_error}

The preceding results indicate that the learned correction retains a clear dependence on the numerical discretization. To examine this dependence more directly, we compare the variation of the predicted correction with the corresponding change in the numerical scheme. {In both the two- and three-dimensional cases we vary only the discretization of the convective term. Its nonlinearity governs the transfer of energy across scales, so changing this discretization makes the interaction between the numerical error and the subgrid contribution more pronounced; whereas varying the linear terms would instead be expected to produce a weaker and more uniform damping. Therefore, the differences in the predicted correction between two scheme encodings may be used as a proxy for the change in numerical error induced by the underlying discretization.} The objective of this section is therefore to assess whether the learned scheme dependence is consistent with the corresponding differences in the discretized operators.

For the two-dimensional incompressible case, we compare the discretized convective operator
\begin{equation}
\mathcal{C}(\boldsymbol{v})=-(\boldsymbol{v}\cdot\nabla)\boldsymbol{v}.
\end{equation}
For the three-dimensional compressible case, we compare the scheme-dependent contribution of the advective momentum flux, written here schematically as
\begin{equation}
\mathcal{C}(\rho,\boldsymbol{v},p)=-\nabla\cdot(\rho\boldsymbol{v}\otimes\boldsymbol{v}+p\mathbf{I}),
\end{equation}
where the latter expression includes the pressure-flux contribution associated with the compressible solver. For two numerical schemes, $s_1$ and $s_2$, we then compare the difference of these fluxes with the corresponding difference in the forcing prediction obtained for the same synchronized input but with different scheme labels.

In this sense, we examine whether
\begin{equation}
\mathcal{C}^{s_1}(\boldsymbol{v})-\mathcal{C}^{s_2}(\boldsymbol{v}) \approx \mu (\mathcal{M}_\theta(\boldsymbol{v},s_1)-\mathcal{M}_\theta(\boldsymbol{v},s_2))
\end{equation}
in two dimensions, and the corresponding conservative-form difference in three dimensions, exhibit similar spatial and spectral trends.

\begin{figure}
  \centering
  \begin{tabular}{@{}c@{\hspace{0.02\textwidth}}c@{}}
    \includegraphics[width=0.6\textwidth]{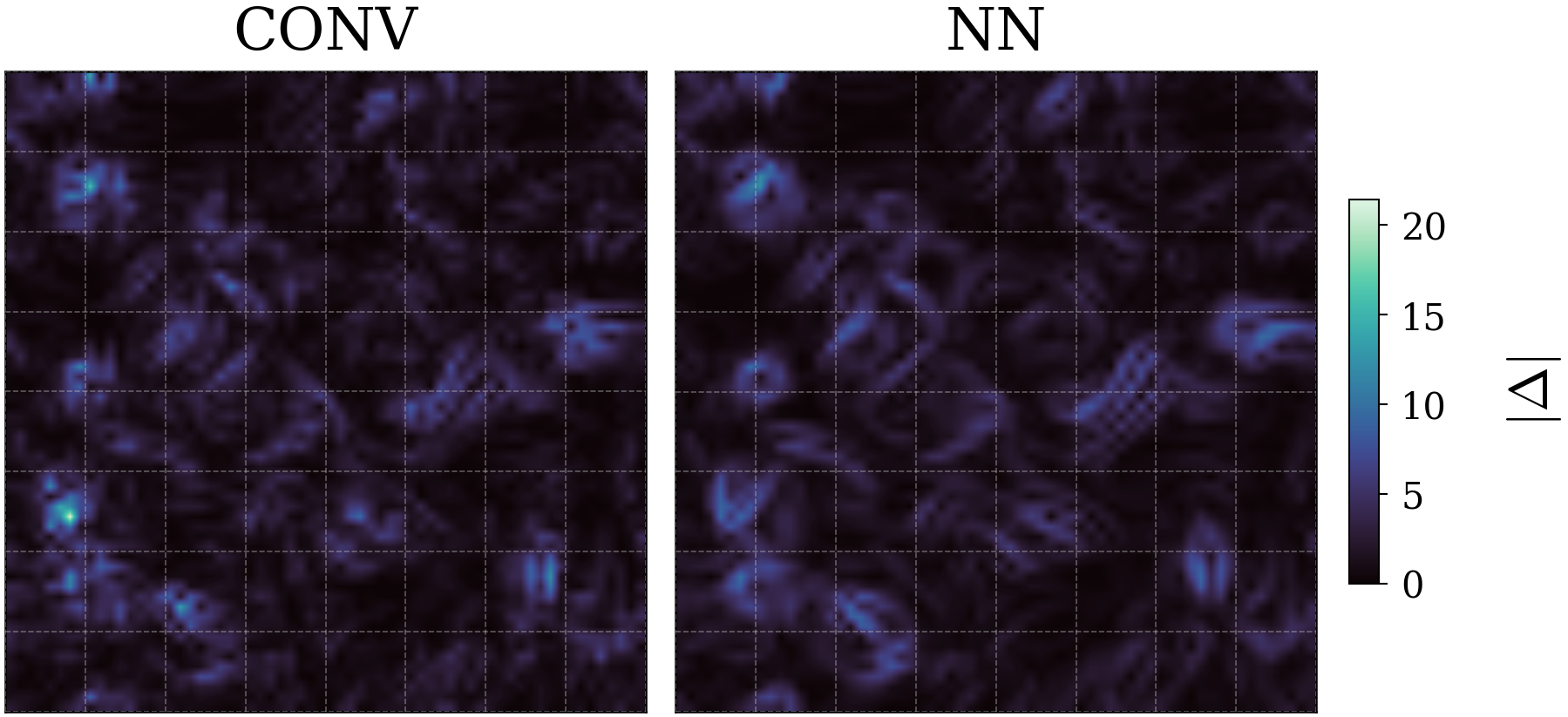} &
    \includegraphics[width=0.37\textwidth]{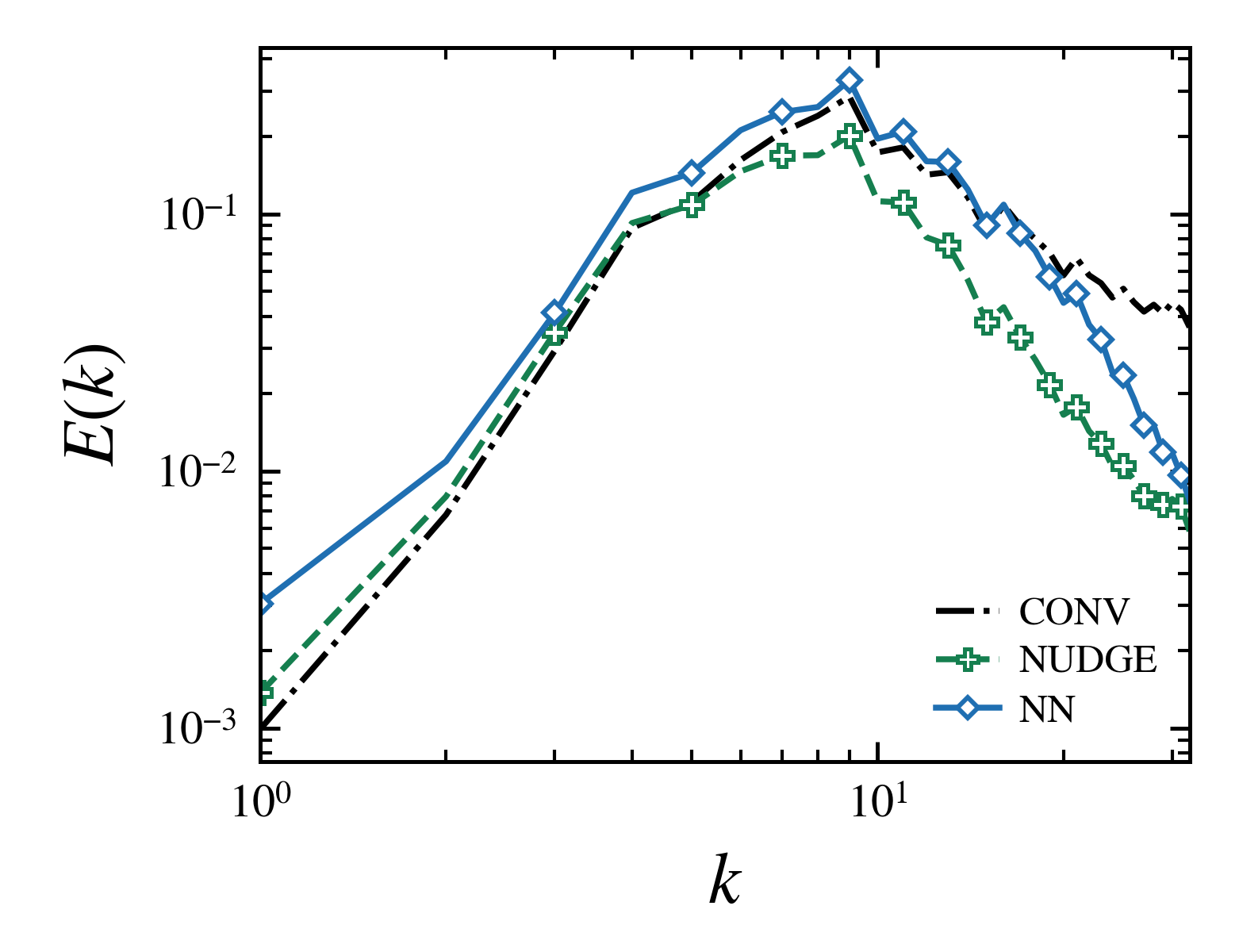} \\
    (a) & (b)
  \end{tabular}
  \caption{
{Comparison of convective-term and model-prediction differences for the two-dimensional case. (a) Magnitude of the UPWIND$-$VL difference field for the convective term (CONV) and for the corresponding model prediction (NN), both evaluated from the same synchronized coarse-grid input. (b) Energy spectrum $E(k)$ of the convective-term difference, CONV, the corresponding model-prediction difference (NN) and nudging term (NUDGE). The spatial trend of the nudging term is excluded for comparison due to the nature of NUDGE being collected in an \textit{a-posteriori} manner; however, the average spectral trends of NUDGE and NN still remain consistent. Similar spatial and spectral trends are observed for the convective-term, nudging and predicted-forcing differences.}}
  \label{fig:convective_flux_combined_2D}
\end{figure}

\begin{figure}
  \centering
  \begin{tabular}{@{}c@{}}
    \includegraphics[width=0.8\textwidth]{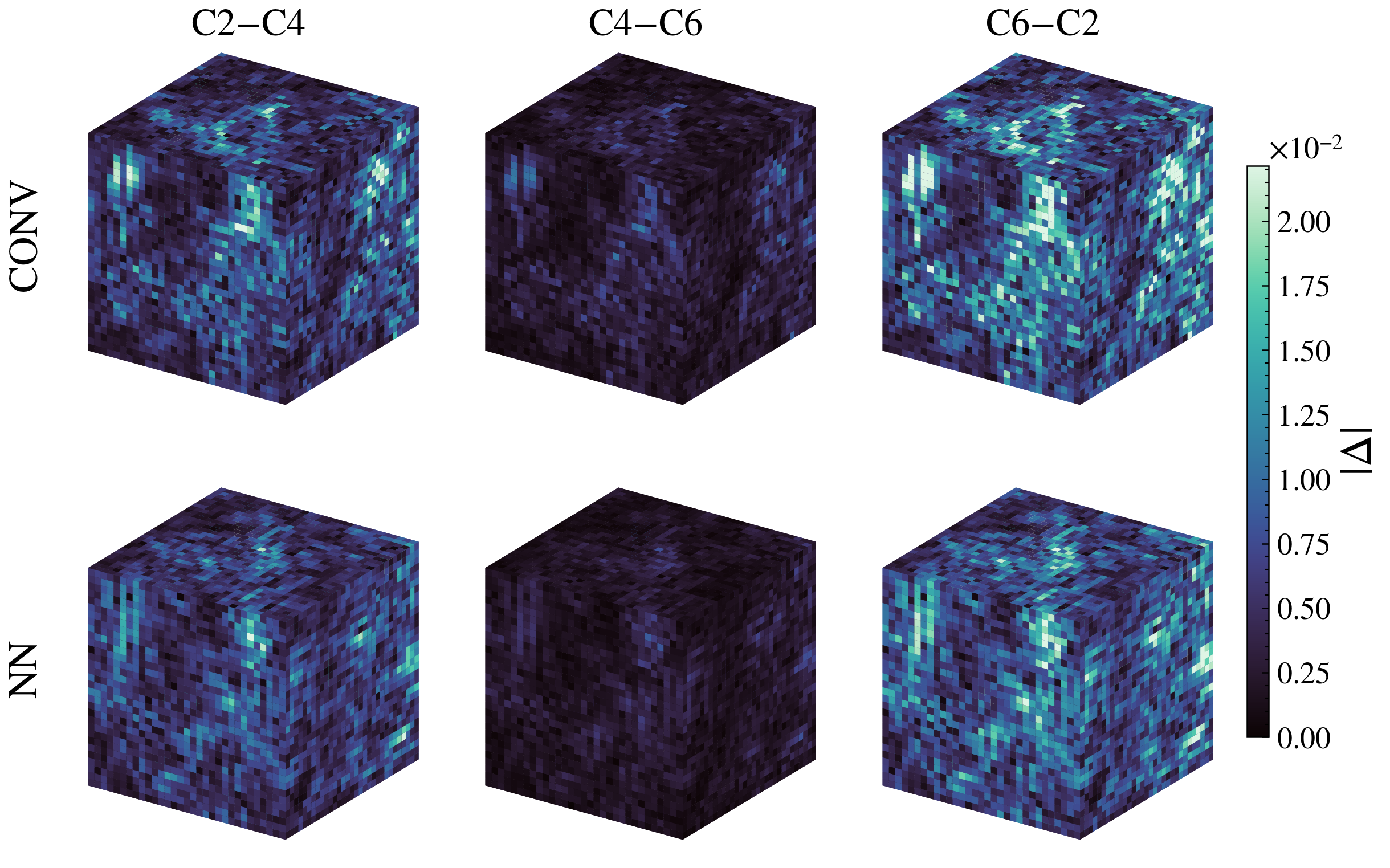} \\
    (a)
  \end{tabular}

  \vspace{0.5em}

  \begin{tabular}{@{}c@{}}
    \includegraphics[width=0.96\textwidth]{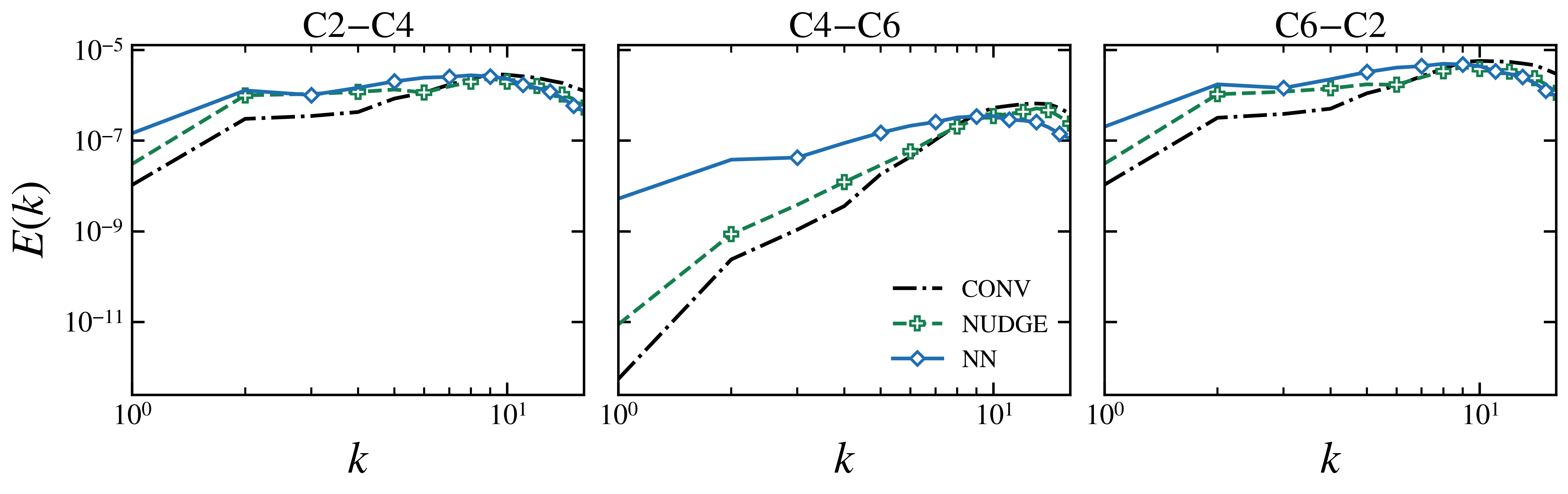} \\
    (b)
  \end{tabular}
  \caption{
{Comparison of convective-term and model-prediction differences for the three-dimensional case. (a) Magnitude of the scheme-to-scheme difference field for the momentum convective term (CONV) and for the corresponding model prediction (NN), shown for the three scheme pairs C2$-$C4, C4$-$C6 and C6$-$C2 (columns), each evaluated from the same synchronized coarse-grid input. (b) Energy spectrum $E(k)$ of the difference fields: the convective-term difference, CONV, the actual nudging difference, NUDGE and the model-prediction difference, NN. Here C2$-$C4 denotes the difference between the C2 and C4 schemes, and the remaining labels are defined similarly.}}
  \label{fig:convective_flux_combined_3d}
\end{figure}

\begin{table}
\centering
\begin{tabular}{lccccc}
\hline
 & $E(k{=}1)$ & $E(k{=}3)$ & $E(k{=}5)$ & $E(k{=}7)$ & cum.\ $k\le7$ [\%] \\
\hline
NUDGE, C2 & $8.43\times10^{-8}$ & $2.06\times10^{-6}$ & $3.26\times10^{-6}$ & $5.87\times10^{-6}$ & $11.99$ \\
NN, C2    & $2.15\times10^{-7}$ & $1.15\times10^{-6}$ & $2.83\times10^{-6}$ & $5.27\times10^{-6}$ & $12.32$ \\
NUDGE, C4 & $9.64\times10^{-9}$ & $2.24\times10^{-7}$ & $7.25\times10^{-7}$ & $1.58\times10^{-6}$ & $4.40$ \\
NN, C4    & $2.52\times10^{-8}$ & $2.00\times10^{-7}$ & $8.08\times10^{-7}$ & $1.63\times10^{-6}$ & $6.03$ \\
NUDGE, C6 & $9.48\times10^{-9}$ & $1.85\times10^{-7}$ & $5.63\times10^{-7}$ & $1.05\times10^{-6}$ & $4.09$ \\
NN, C6    & $1.45\times10^{-8}$ & $1.30\times10^{-7}$ & $5.34\times10^{-7}$ & $1.02\times10^{-6}$ & $5.05$ \\
\hline
\end{tabular}
\vspace{3mm}
\caption{{Energy spectrum $E(k)$ of the momentum forcing $\boldsymbol{F}_{p}$ at low wavenumbers, for the recorded nudging correction (NUDGE) and the model prediction (NN), on the C2, C4 and C6 schemes. The final column gives the cumulative contribution of $k\le7$ as a percentage of the total forcing energy.}}
\label{tab:lowk_closure_spectra}
\end{table}

To perform this comparison, we evaluate the convective flux term and the model forcing for the same synchronized input under different schemes. Figures~\ref{fig:convective_flux_combined_2D} and \ref{fig:convective_flux_combined_3d} show that the resulting differences display similar spatial and spectral trends. This agreement supports the interpretation that the learned scheme dependence is closely connected to the numerical error introduced by the coarse-grid discretization scheme.  The NN C4-C6 spectral difference seems to deviate from the trends displayed by NUDGE and CONV. The apparent deviation of the NN C4-C6 spectrum difference should be viewed in the context of the near-overlap of the C4 and C6 spectra themselves (see figure~\ref{fig:closure_magnitude_3d}), which makes their difference intrinsically small. {This deviation is, in part, a limitation of the model itself. Table~\ref{tab:lowk_closure_spectra} lists the forcing spectrum at low wavenumbers, together with the fraction of the total forcing energy carried by these bands, for the three schemes. The correction to be learned lies almost entirely at the smaller scales for all the three schemes, with the exception of C2, which carries a (relatively) higher proportion of low-$k$ forcing. The \textit{a-priori} accuracy of the model at low $k$ is therefore of limited importance during inference.} Consequently, the corresponding difference spectrum of the learned correction is more sensitive to small variations, even though the underlying discrepancy remains negligible. The nudging difference obtained from separate rollout simulations is less suitable for pointwise comparison, since the scheme-dependent trajectories do not remain identical, but its spectrum is still informative in an averaged sense and is therefore included as an additional reference.

\subsection{Forcing term analysis}
Since nudging and model predictions are added to the governing equations as forcing terms, we compare the forcing contributions associated with each of the closure corrections considered above. Smagorinsky-type closures (SMAG, D-SMAG) are added to the momentum equations as $\nabla \cdot \tau_{smag}$, which acts as a dissipative forcing to represent the forward cascade of energy to the unresolved subgrid scales. In this work, both NUDGE and NN forcing are added as an extra correction to this forcing, which corrects both the SGS effects and the numerical dissipation on the coarse grid. 

\begin{figure}
  \centering
  \begin{tabular}{@{}c@{\hspace{0.02\textwidth}}c@{}}
    \includegraphics[width=0.47\textwidth]{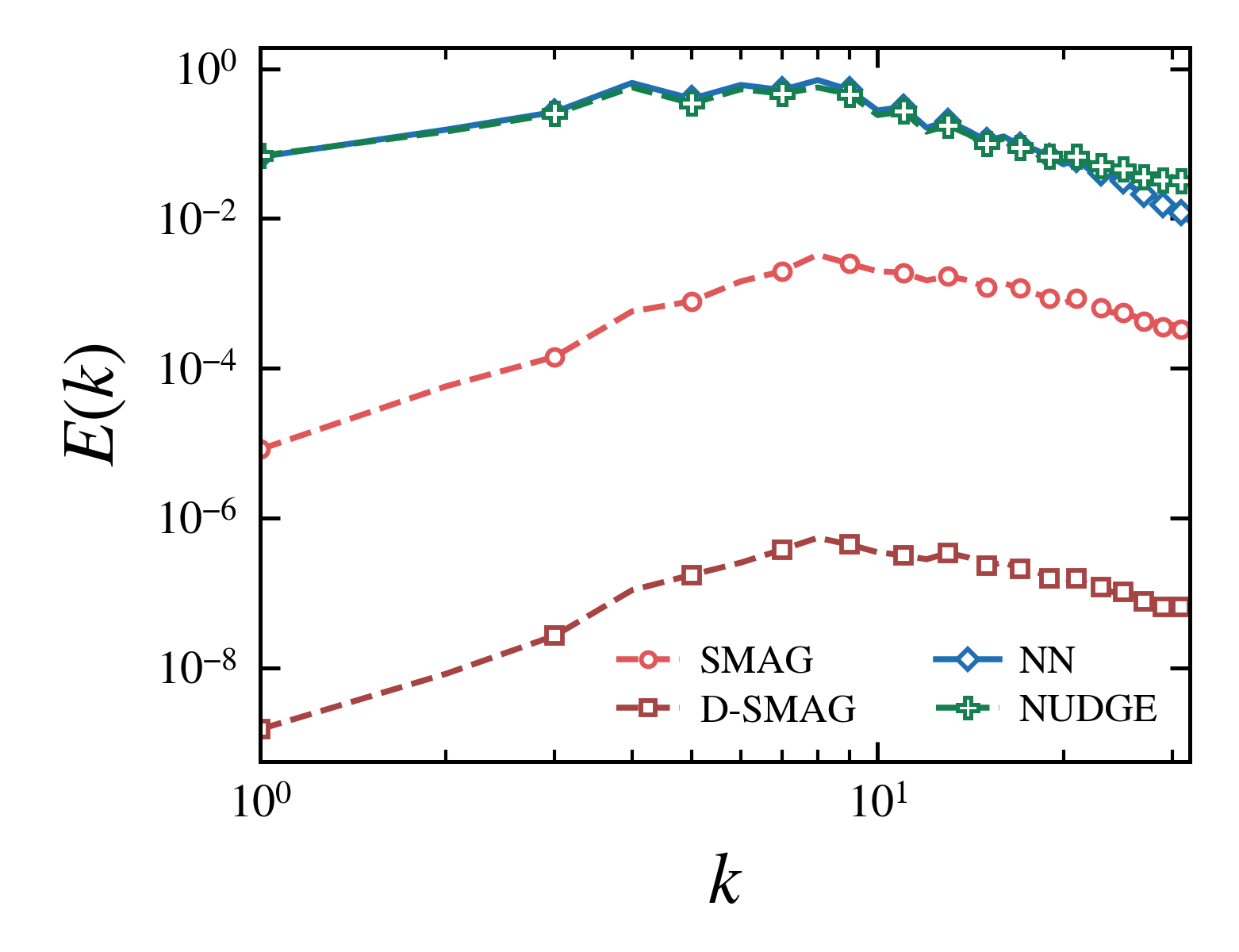} &
    \includegraphics[width=0.47\textwidth]{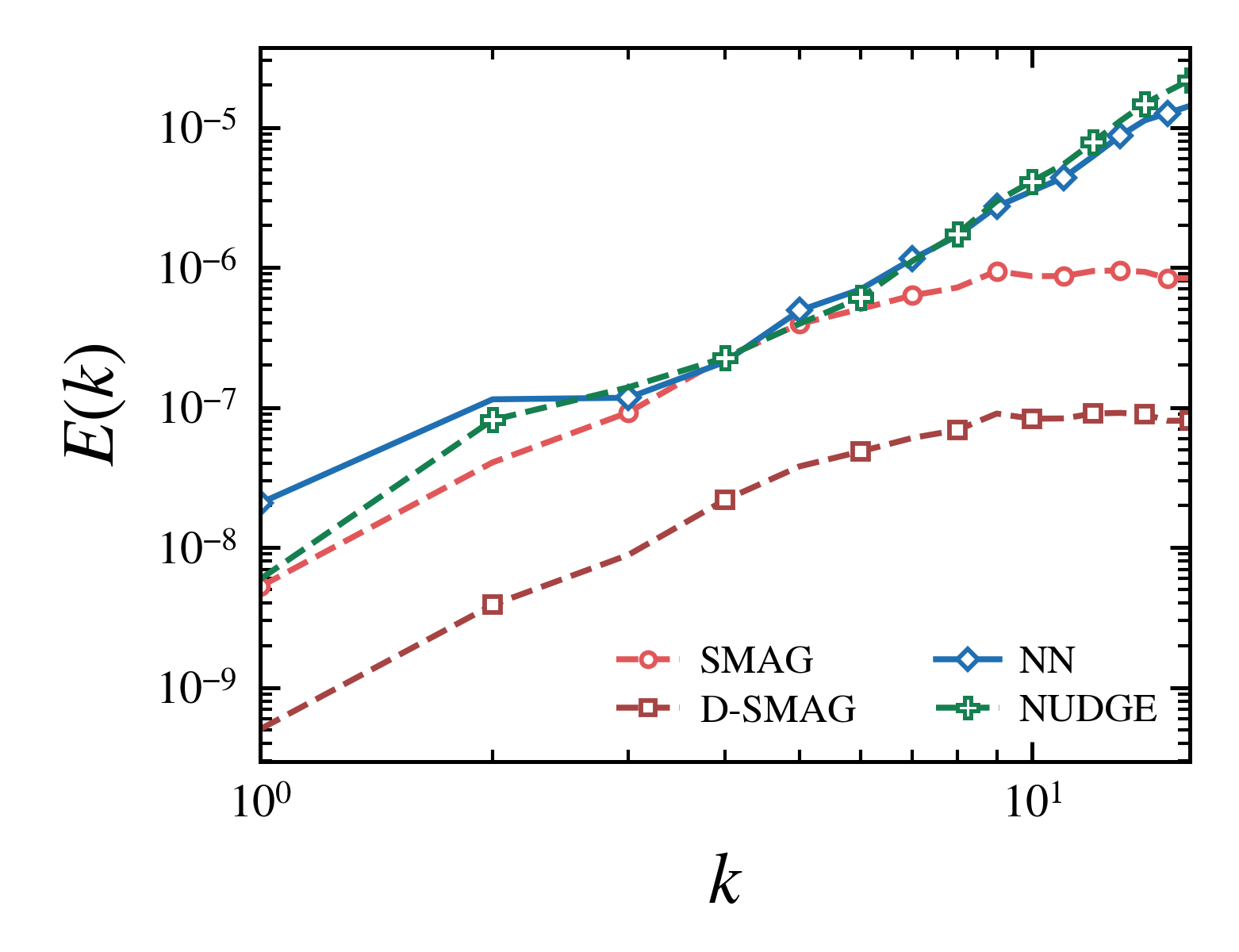} \\
    (a) & (b)
  \end{tabular}
  \caption{
{Comparison of the forcing-term spectrum $E(k)$: (a) two-dimensional case (VL) and (b) three-dimensional case (C4). The SMAG and D-SMAG forcings do not explicitly account for discretization-induced error. The D-SMAG spectrum is of smaller magnitude than that of SMAG, consistent with the adaptive eddy-viscosity formulation of D-SMAG and its reduced net dissipation relative to SMAG.}}
  \label{fig:forcing_spectrum}
\end{figure}

\begin{figure}
  \centering
  \begin{tabular}{@{}c@{\hspace{0.02\textwidth}}c@{}}
    \includegraphics[width=0.98\textwidth]{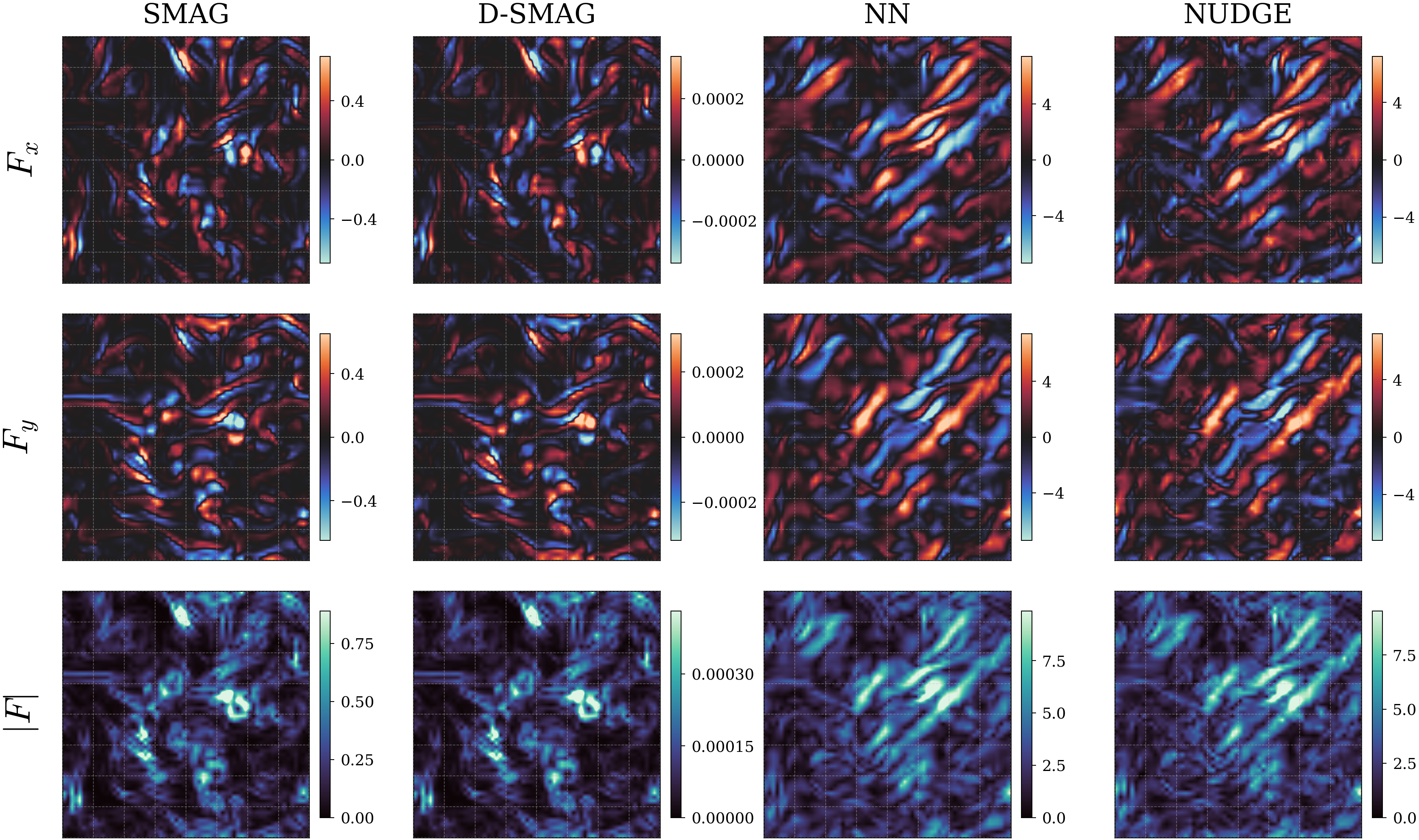}\\
  \end{tabular}
  \caption{Visualization of the component-wise net forcing for SMAG, D-SMAG, NN and NUDGE for the two-dimensional case. NN and NUDGE provide additive corrections to the baseline SMAG forcing and explicitly account for the coarse-grid dynamics.}
  \label{fig:cum_forcing_2d}
\end{figure}

\begin{figure}
  \centering
  \includegraphics[width=0.98\textwidth]{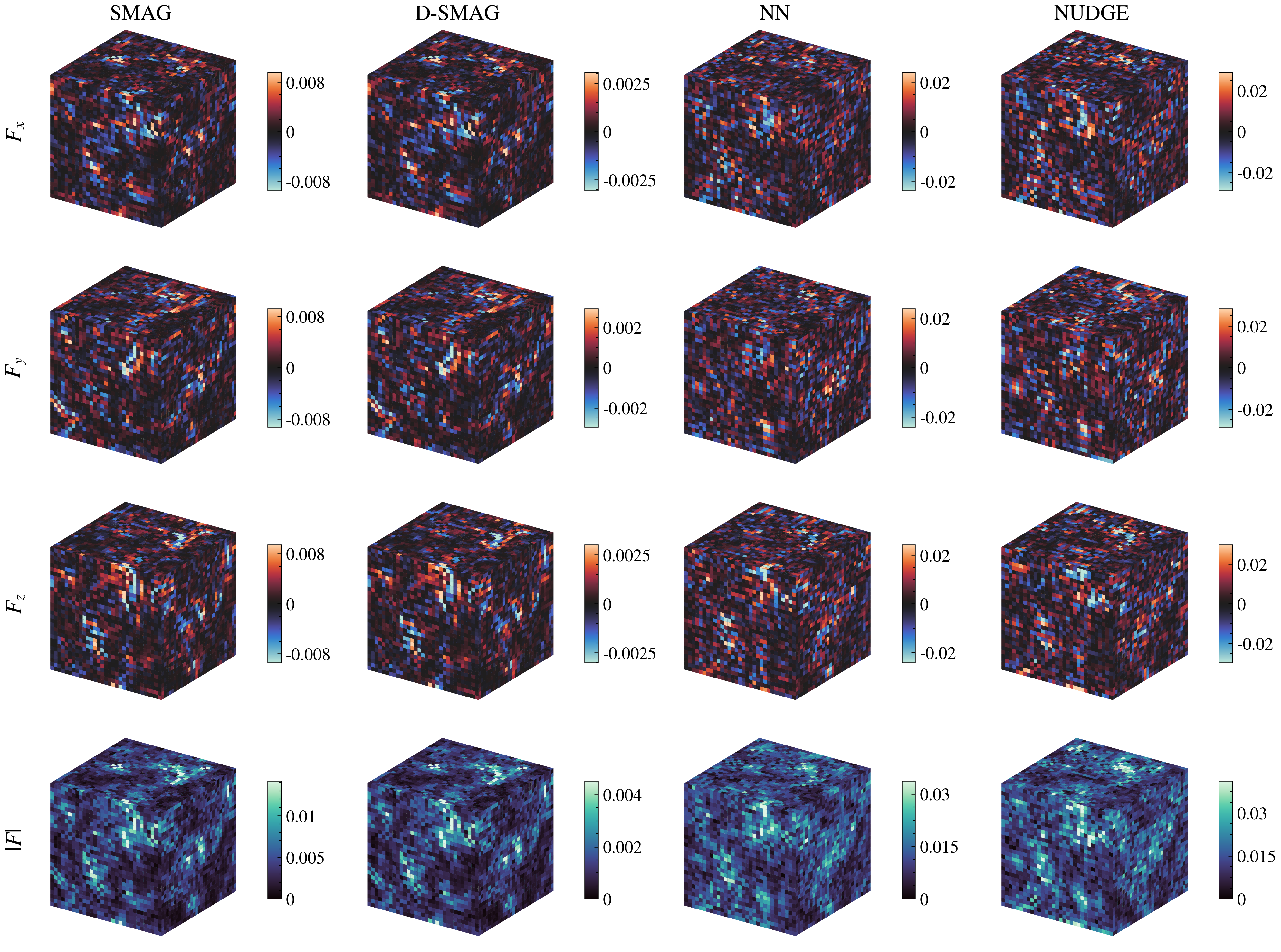}
  \caption{
{Visualization of the component-wise net forcing for the three-dimensional case. Labels are defined as in figure~\ref{fig:cum_forcing_2d}}.}
  \label{fig:cum_forcing_3d}
\end{figure}

The forcing spectra exhibit a consistent ordering (see figure~\ref{fig:forcing_spectrum}), with D-SMAG showing the smallest magnitude, NN and NUDGE the largest, and SMAG lying in between. This is qualitatively consistent with SMAG being overly dissipative, whereas D-SMAG adapts to the local flow and hence provides a more moderate correction. However, neither SMAG nor D-SMAG explicitly compensates for the numerical damping introduced by the LES discretization, while NN and NUDGE supply an additional correction that accounts for both the closure deficit and the discretization-induced damping. The corresponding component-wise net forcing fields are visualized in figure~\ref{fig:cum_forcing_2d} for the two-dimensional case and figure~\ref{fig:cum_forcing_3d} for the three-dimensional case.

A notable observation from the visual comparison in the three-dimensional case (see figure~\ref{fig:cum_forcing_3d}) is that, although the forcing magnitudes differ across models, their dominant spatial structure remains broadly similar. This suggests that the regions requiring the largest correction are associated with localized flow features characterized by large resolved gradients and strain rates. It further suggests that regions of strong numerical damping also coincide with regions of large strain rate, in which case the largest corrective forcing would reflect a combined influence of subgrid effects and discretization-induced dissipation.

{Since the reference observations here are subsampled DNS fields, a further comparison can be made between the net forcing considered above and the \emph{true forcing} required by the solver to retain the reference evolution. Let $\Phi_c$ denote the coarse-grid solver without any closure term, that is, without the Smagorinsky contribution. Given the reference field on the coarse grid at time step $n$, denoted $\boldsymbol{u}_{obs}^{\,n}$, the coarse evolution of this field is obtained as}
\begin{equation}\label{eq:coarse_evolution}
{\tilde{\boldsymbol{u}}^{\,n+1}=\Phi_c\!\left(\boldsymbol{u}_{obs}^{\,n}\right),}
\end{equation}
{and since the reference solution on the coarse grid is available at the next time step, the forcing required by the coarse-grid solver may be assigned as the discrepancy between the coarse-evolved field and the reference field,}
\begin{equation}\label{eq:true_forcing}
{\boldsymbol{F}_{true}=\left(\boldsymbol{u}_{obs}^{\,n+1}-\tilde{\boldsymbol{u}}^{\,n+1}\right)/\Delta t ,}
\end{equation}
{where $\Delta t$ is the time step, common to the reference and coarse-grid simulations. This is the forcing required to recover the reference pointwise over a single step, and therefore contains all of the content that the coarse grid cannot sustain. The comparison is made against the net closure forcing, namely the sum of the Smagorinsky term and the NUDGE or NN correction, evaluated on the nudged coarse-grid field $\boldsymbol{v}^{n}$, which is synchronized with $\boldsymbol{u}_{obs}^{n}$.}

\begin{figure}
  \centering
  \includegraphics[width=1.0\textwidth]{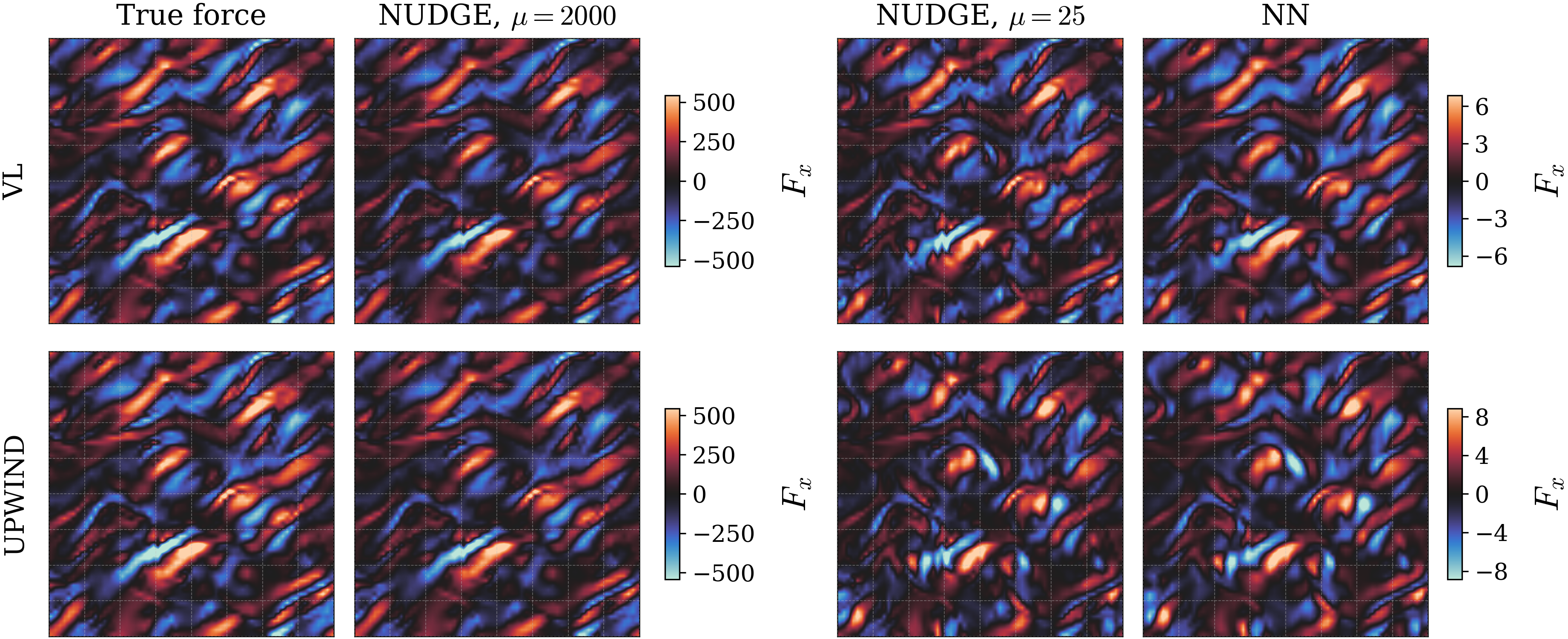}
  \caption{{{Comparison of the net coarse-grid forcing with the true forcing $\boldsymbol{F}_{true}$ of equation~\eqref{eq:true_forcing}, for the two-dimensional case on the VL and UPWIND schemes, shown for the $x$ component. The net forcing is the sum of the Smagorinsky and NUDGE or NN terms, evaluated for an input field $\boldsymbol{v}$ synchronized with $\boldsymbol{u}_{obs}$; the solver $\Phi_c$ used to obtain $\boldsymbol{F}_{true}$ carries no Smagorinsky term. The second column shows the nudging correction when the fields are nudged using $\mu=2000$.}}}
  \label{fig:true_forcing_fields_2d}
\end{figure}

{Figure~\ref{fig:true_forcing_fields_2d} compares the $x$ component of the forcing for the three cases, $\boldsymbol{F}_{true}$, NUDGE and NN. The forcings match closely in structure for both schemes, but the magnitude of the NUDGE and NN fields at $\mu=25$ is smaller than that of $\boldsymbol{F}_{true}$. This is because, although the coarse-grid field is synchronized, a residual remains whose magnitude is set by the nudging strength $\mu$. The value of $\mu$ also determines the amount of forcing injected into the system, larger values producing a more aggressive correction for the discrepancy between $\boldsymbol{v}$ and $\boldsymbol{u}_{obs}$. The second column is an illustration of how the nudging strength $\mu$ can be tuned so as to reproduce the true forcing.}

\begin{figure}
  \centering
  \includegraphics[width=0.8\textwidth]{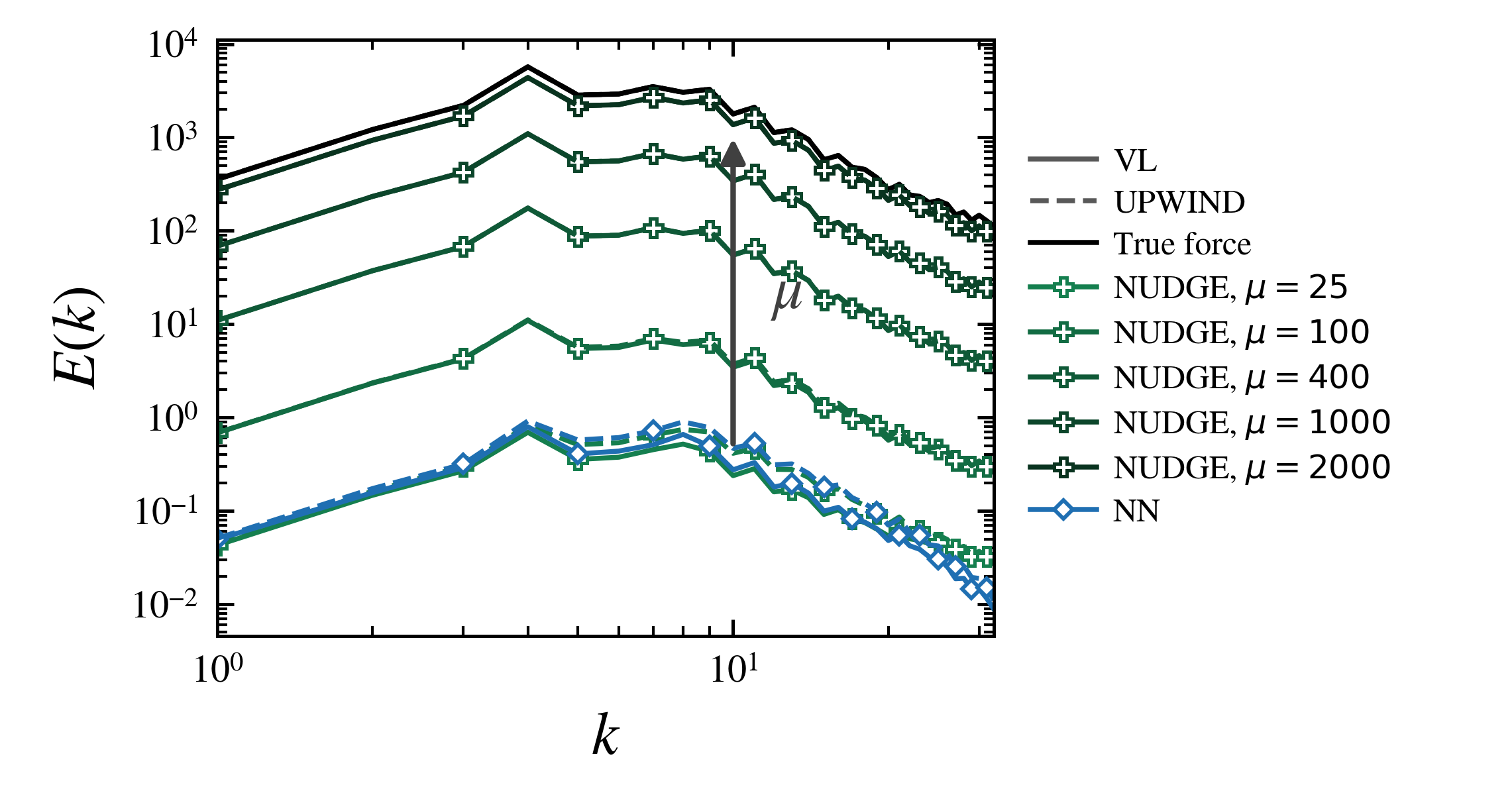}
  \caption{{{Spectrum $E(k)$ of the net forcing compared with that of the true forcing $\boldsymbol{F}_{true}$, for the VL and UPWIND schemes in the two-dimensional case. The nudging forcing is shown for a sweep in $\mu$, from the value $\mu=25$ used throughout to $\mu=2000$, at which the two nearly match.}}}
  \label{fig:true_forcing_spectrum_2d}
\end{figure}

{The forcing spectra in figure~\ref{fig:true_forcing_spectrum_2d} show the same behavior. At the value $\mu=25$ used throughout, the NUDGE and NN spectra follow the same spectral trend as that of $\boldsymbol{F}_{true}$ for both schemes, with the NN prediction lying essentially on top of the recorded nudging spectrum, both differing from the true forcing approximately by a constant factor in magnitude, rather than in their distribution across scales. Together with the pointwise comparison of figure~\ref{fig:true_forcing_fields_2d}, this indicates that the recorded correction reproduces the structure of the forcing required by the coarse solver, and differs from it principally in magnitude. Increasing the nudging strength displaces the spectrum upwards without altering its shape, the two nearly coinciding at $\mu=2000$. This agreement in magnitude by increasing the nudging strength is shown only as an illustration of similar behavior between the true forcing and the nudging (and NN) forcing. Table~\ref{tab:true_sgs_mu} reports the residual and the pointwise correlation with the reference over the same range of $\mu$, and both are essentially unchanged, so that recovering the magnitude of $\boldsymbol{F}_{true}$ brings minimal gain in synchronization. The value $\mu=25$ is therefore retained throughout; a more detailed explanation for the choice of this parameter for both two- and three-dimensional cases is given in Appendix~\ref{sec:appendix_nudge_strength}.}

\begin{table}
\centering
\begin{tabular}{ccc}
\hline
$\mu$ & $\varepsilon$ & $\mathrm{corr}(\boldsymbol{v},\boldsymbol{u}_{obs})$ \\
\hline
$25$   & $0.0818$ & $0.9965$ \\
$100$  & $0.0753$ & $0.9970$ \\
$400$  & $0.0745$ & $0.9971$ \\
$1000$ & $0.0744$ & $0.9971$ \\
$2000$ & $0.0744$ & $0.9971$ \\
\hline
\end{tabular}
\vspace{3mm}
\caption{{Relative $L_2$ error $\varepsilon$ and the Pearson correlation coefficient, $\mathrm{corr}(\boldsymbol{v},\boldsymbol{u}_{obs})$, evaluated over all grid points between the synchronized coarse-grid field and the reference, for the nudged two-dimensional runs at different values of $\mu$.}}
\label{tab:true_sgs_mu}
\end{table}

\begin{figure}
  \centering
  \includegraphics[width=1.0\textwidth]{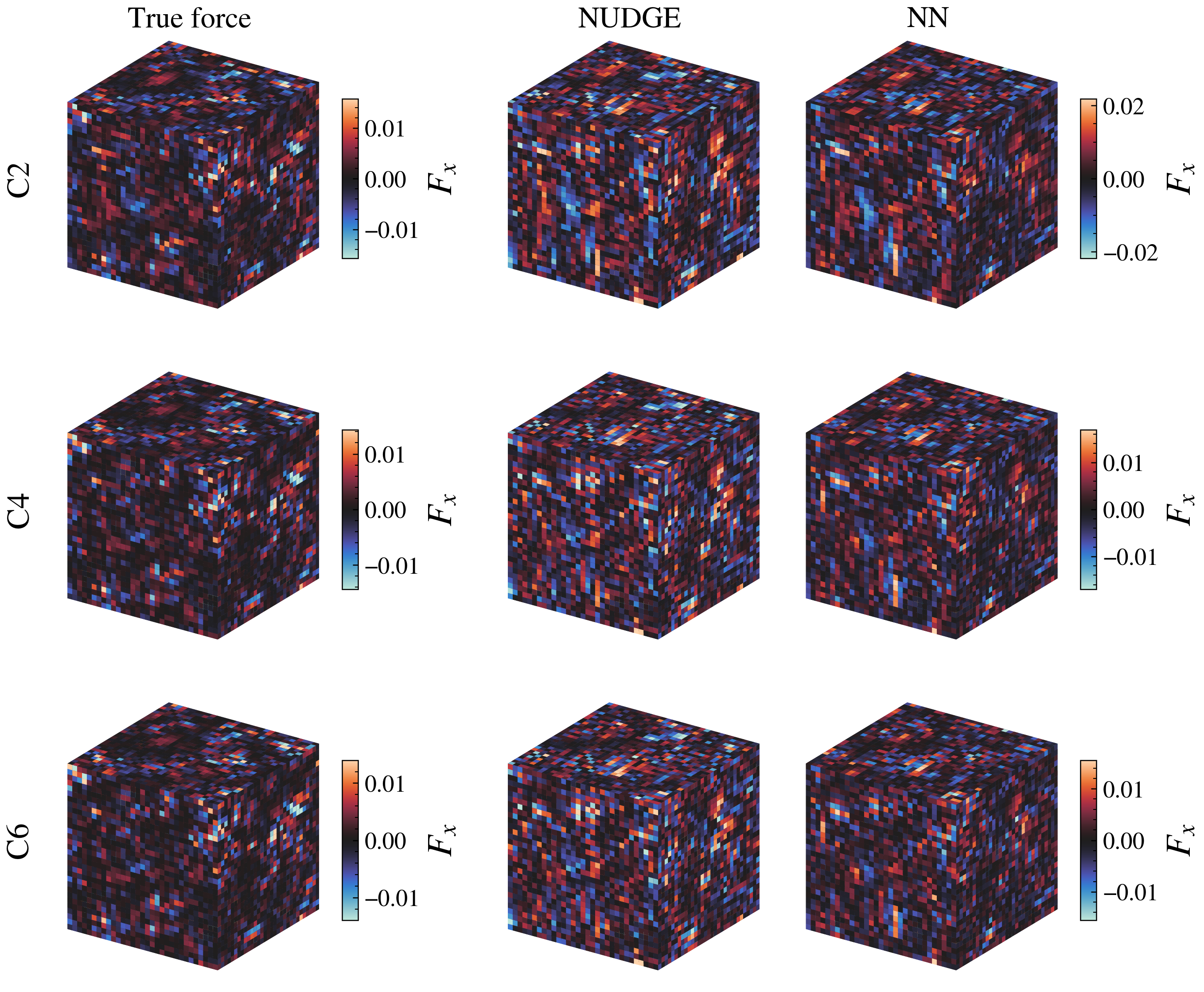}
  \caption{{{Comparison of the NUDGE and NN forcing with the true forcing $\boldsymbol{F}_{true}$ of equation~\eqref{eq:true_forcing_3d}, for the three-dimensional case, shown for the $x$ component of the momentum forcing. As in figure~\ref{fig:true_forcing_fields_2d}, the net forcing is the sum of the Smagorinsky and NUDGE or NN terms, evaluated on the nudged coarse-grid field $\boldsymbol{v}^{n}$ synchronized with $\boldsymbol{u}_{obs}^{n}$.}}}
  \label{fig:true_forcing_fields_3d}
\end{figure}

{In the three-dimensional case the pointwise comparison differs from that of the two-dimensional one. The same procedure is followed, with the coarse-grid reference fields $\rho_{obs}^{\,n}$ and $\boldsymbol{u}_{obs}^{\,n}$ advanced through the coarse solver, so that the true forcing is obtained in conservative form as}
\begin{equation}\label{eq:true_forcing_3d}
{\boldsymbol{F}_{true}=\left(\rho_{obs}^{\,n+1}\boldsymbol{u}_{obs}^{\,n+1}-\tilde{\rho}^{\,n+1}\tilde{\boldsymbol{u}}^{\,n+1}\right)/\Delta t ,}
\end{equation}
{where $\tilde{(\cdot)}$ again denotes a field obtained through the coarse evolution and $n+1$ the next time step. This case is slightly different from the previous one; here, the recorded correction lies predominantly in high wavenumber range. The nudging is also only partial, since the density is not itself corrected in the present runs, so a difference persists between the reference and coarse-grid density fields. For these reasons the true forcing, NUDGE and NN are not expected to show the same degree of visual correspondence (figure~\ref{fig:true_forcing_fields_3d}) as in two dimensions.}

\begin{figure}
  \centering
  \includegraphics[width=1.0\textwidth]{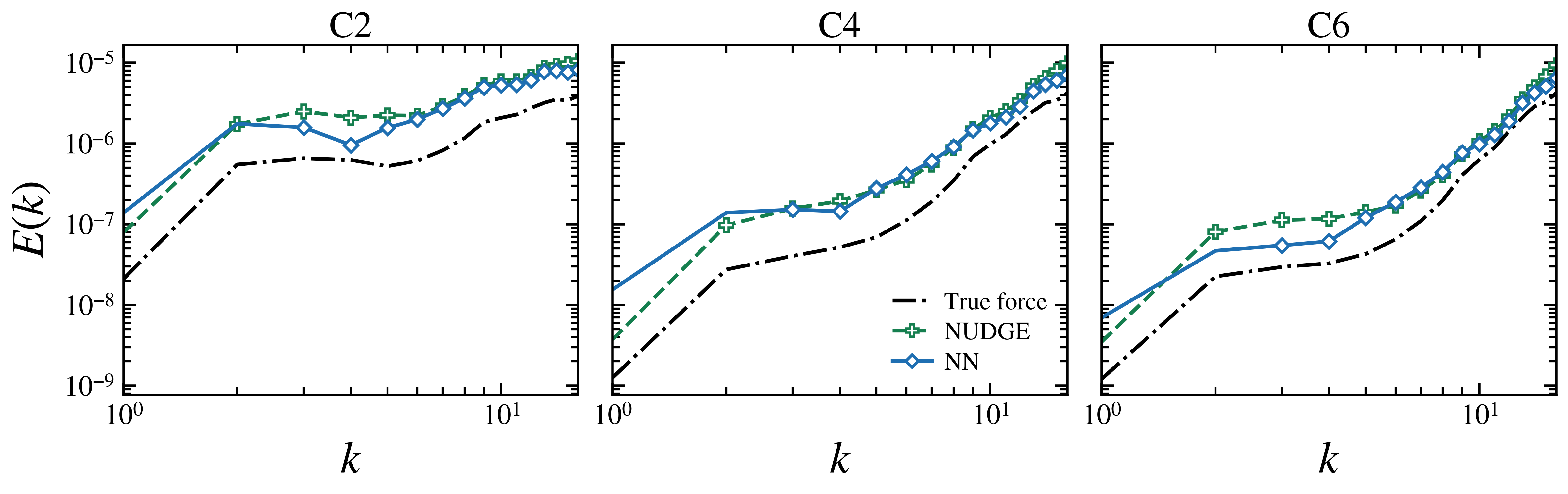}
  \caption{{{Spectrum $E(k)$ of the net forcing compared with that of the true forcing $\boldsymbol{F}_{true}$, for the C2, C4 and C6 schemes in the three-dimensional case.}}}
  \label{fig:true_forcing_spectrum_3d}
\end{figure}

{Despite these differences in the pointwise fields, the spectral comparison in figure~\ref{fig:true_forcing_spectrum_3d} shows the same trend for all three curves across C2, C4 and C6, and here the magnitudes are comparable at the value $\mu=2.5$ used throughout in contrast to the two-dimensional case. Taken together, the two- and three-dimensional comparisons indicate that the recorded correction reproduces the structure of the forcing required by the coarse solver, which is the property the learned closure is intended to inherit.}

\subsection{Generalization across unseen discretizations}\label{sec:generalization_fine_tune}

{A recurring argument (as discussed in \S~\ref{sec:methodology} and \ref{sec:2d_results}) of this work is that closures trained \textit{a-posteriori} tend to overfit to the numerical discretization used during optimization, since a correction obtained by propagating the solution through the solver must account for both the subgrid contribution and the discretization error of that particular solver. The same behavior is observed within the present framework when the nudged targets are drawn from a single scheme; as shown in figure~\ref{fig:single_scheme_2d}, a model trained on VL does not transfer to UPWIND or CD2. The scheme conditioning using FiLM layers addresses this by encoding the training of a fixed set of known discretizations within a single model, by tuning the output features to produce the right amount of forcing to balance the numerical artifacts. Thus, in general, we do not expect this model to transfer to a discretization absent from the training set.}

{The structure of the conditioning nevertheless admits an inexpensive route to calibration. The numerical scheme enters the network only through the FiLM layers, in which a scheme encoding indexes an embedding from which the per-channel scale and shift $(\gamma^{(s)},\beta^{(s)})$ are generated. Because these depend only on the encoding and not on the input field, they are fixed constants once training is complete and are the only scheme-dependent parameters in the model; the Swin backbone is shared across all schemes and receives no scheme information. Hence, an efficient way to adapt the model to new discretizations can be to \emph{fine-tune} the existing model by freezing the existing set of weights and retraining a new set of FiLM parameters to learn the new schemes' forcing discretization dependence.}

{In the trained model these parameters are shared across the blocks of each stage and applied at a single location within each block. Adding a scheme therefore introduces only a single new embedding vector. The number of parameters available for calibration is then set by the embedding dimension alone, independently of the depth or width of the network, and is too small for the calibration to be effective. For the fine-tuning we replace that generator with a table that stores $(\gamma,\beta)$ directly, one row per scheme, and assign a separate pair to every transformer block, applying the modulation at more locations within each block; the network weights are left unchanged. The rows of the previously trained schemes are filled with the values the generator produced, replicated across the blocks of their stage, and the entries introduced at the new locations are set to leave the features unaltered, so that the enlarged parameterization reproduces the original model exactly for those encodings. Only the row corresponding to the new scheme is optimized; the trained rows are held outside the optimized parameter set and cannot change.}

{The schemes used for this test are chosen to lie outside the character of those seen during training. In two dimensions, we use second-order central differencing (CD2), which unlike VL and UPWIND schemes, is not numerically dissipative by construction. In three dimensions, we use a fifth-order Weno scheme (WENO), which is strongly dissipative, whereas C2, C4 and C6 are dispersive and inject spurious content near the grid cutoff; the net effect, however, of these central schemes together with the Smagorinsky closure nevertheless remains dissipative.}

\begin{figure}
  \centering
  \begin{tabular}{@{}c@{\hspace{0.02\textwidth}}c@{}}
    \includegraphics[width=0.47\textwidth]{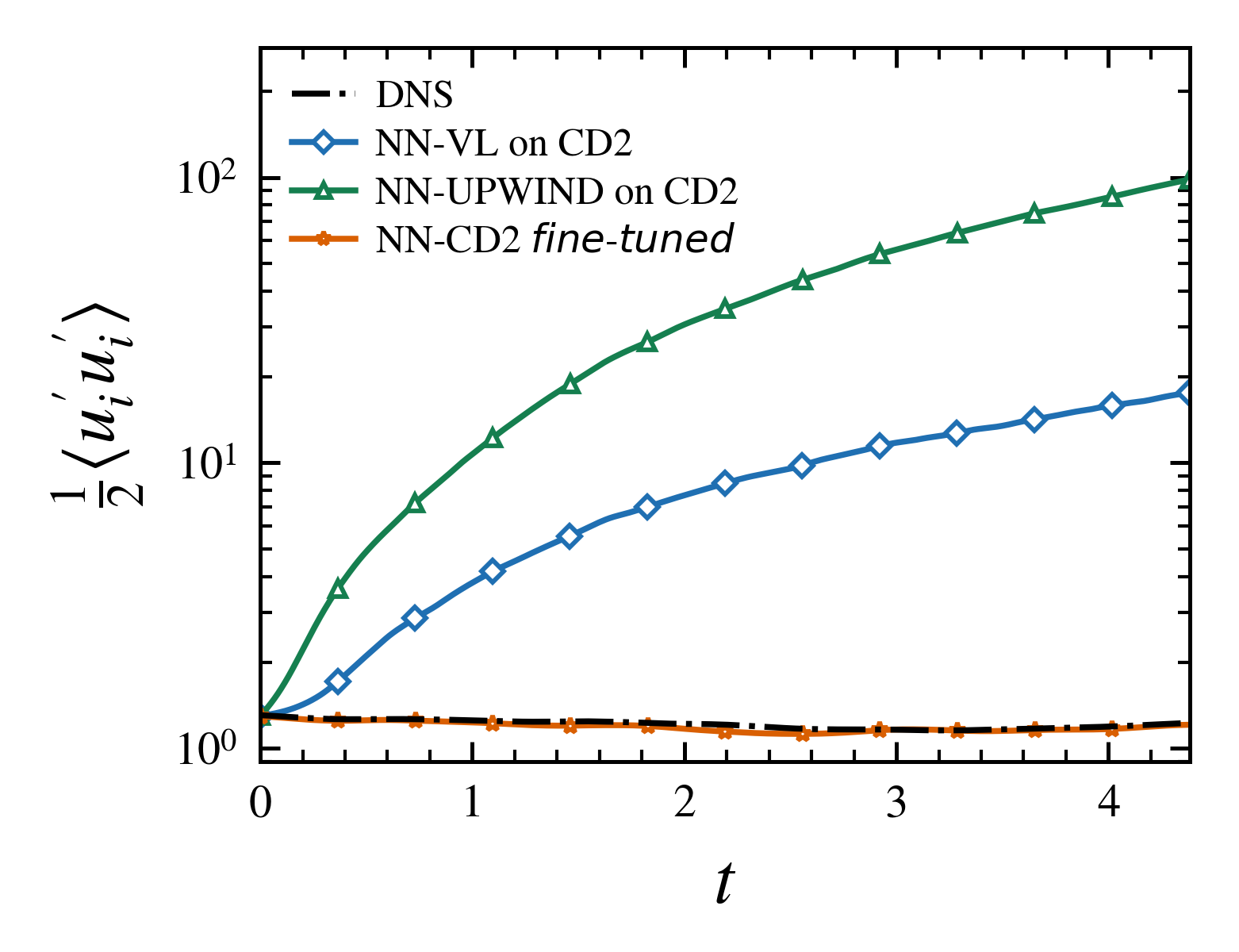} &
    \includegraphics[width=0.47\textwidth]{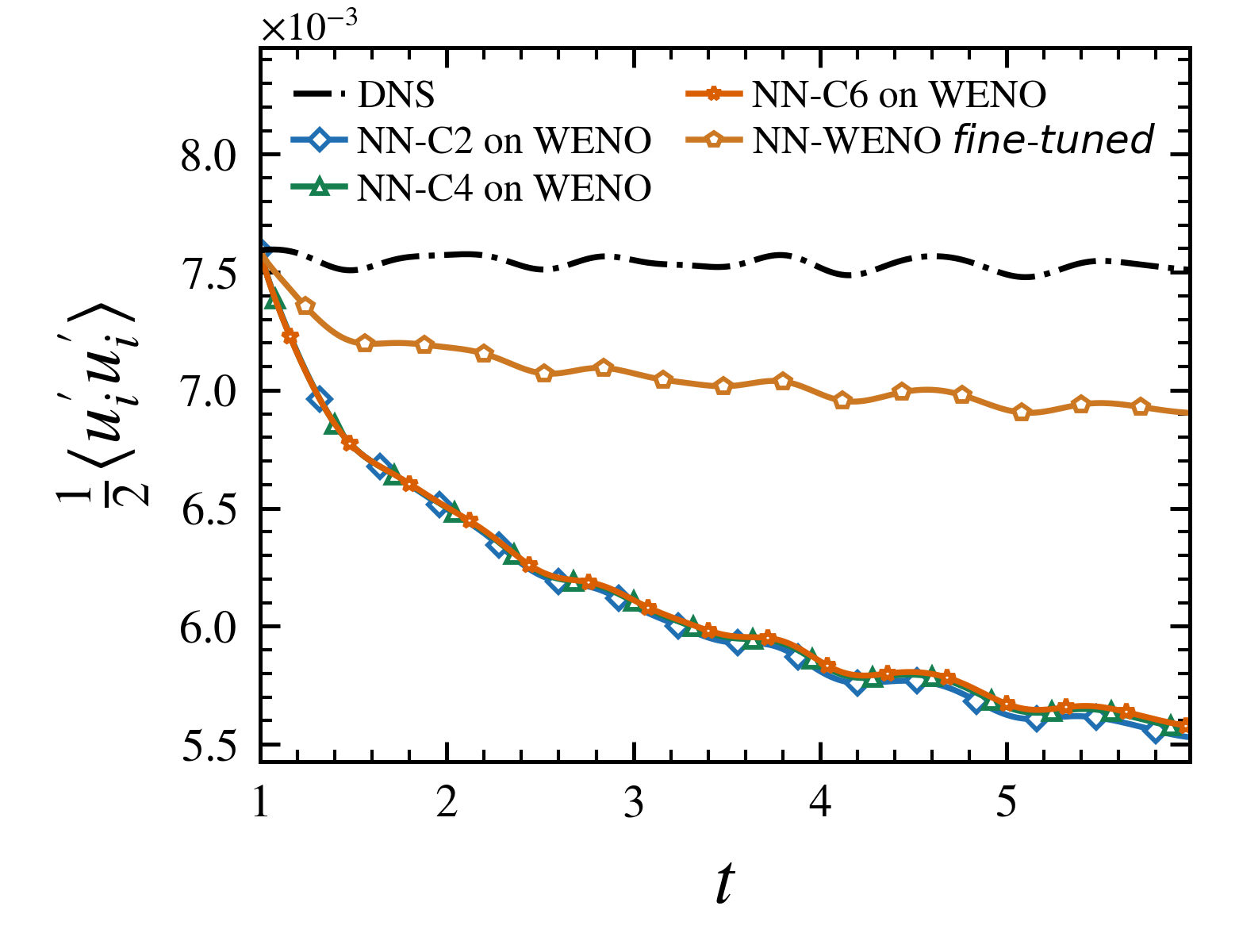} \\
    (a) & (b)
  \end{tabular}
  \caption{
{{TKE evolution, $\tfrac{1}{2}\langle u_i' u_i'\rangle$ versus $t$, for the scheme-conditioned model deployed on a discretization not used during training: (a) CD2 in the two-dimensional and (b) WENO in the three-dimensional case. The model using the old scheme labels (VL, UPWIND and C2, C4, C6) does not generalize to the new scheme whereas the fine-tuned model, trained on a subset of data is able to adapt the FiLM layers to produce the right amount of forcing to adapt to the new scheme. The fine-tuning is done in such a way that the previous FiLM parameters remain the same, so there is no change in the model performance before and after fine-tuning for the older schemes.}}}
  \label{fig:film_fine_tuning}
\end{figure}

{Figure~\ref{fig:film_fine_tuning} compares the TKE evolution obtained when the model is deployed on the new scheme using the previously trained encodings and using the fine-tuned encoding. In the two-dimensional case, the VL and UPWIND forcings both supply considerably more forcing than CD2 requires, having been trained to compensate for numerical dissipation that CD2 does not introduce; the result is a rapid accumulation of energy, reaching roughly two orders of magnitude above the reference value for the UPWIND encoding. The fine-tuned CD2 model instead tracks the reference TKE closely. In three dimensions the C2, C4 and C6 encodings produce considerably less forcing than required due to excessive damping effect of the WENO scheme; the fine-tuned model recovers the greater part of this deficit. The fine-tuning uses $12.5\%$ of the data required for the original training and adjusts $2.25\%$ and $1.3\%$ of the model parameters in the two- and three-dimensional cases respectively, and therefore provides an inexpensive means of calibrating an existing model to a new discretization with limited data.}

{This result also informs the interpretation of what the shared backbone represents. Since a small set of scheme-dependent parameters is sufficient to retune the output of a frozen backbone to a discretization it has not seen, the backbone appears to encode a structure of the correction that is \emph{common} to the discretizations used in training, with the scheme dependence handled by the small set of FiLM parameters. This \emph{shared representation} of the closure remains latent and cannot be cleanly separated from the scheme-specific numerical artifacts; extracting it explicitly, and using it as the basis of a closure that transfers without recalibration, is the subject of ongoing work.}

\section{Conclusion}

We have presented a nudging-based framework for data-driven turbulence closure modeling in which the discrepancy between coarse-grid LES and reference observations is first obtained from synchronized nudged simulations and then learned offline using a neural network. In this way, the learned model inherits information about the corrective term required by the coarse-grid solver without requiring adjoints or backpropagation through the flow solver.

In both two- and three-dimensional test cases, the proposed framework recovers the correction required to reproduce the reference coarse-grid statistics. In the two-dimensional case, the learned model performs well when training and inference are carried out on the same numerical scheme, but does not transfer directly across discretizations, indicating that the learned correction retains a strong dependence on the numerical dynamics of the coarse solver. Explicit scheme conditioning through FiLM layers allows a single model to adapt across multiple schemes by learning the affine feature transformations required to maintain the reference statistics. Similar behavior is observed in the three-dimensional case, where the learned correction recovers the statistical behavior of the reference flow across central discretizations despite the absence of pointwise synchronization during rollout.

The results further suggest that the scheme dependence of the learned correction is closely related to the numerical discretization error introduced by the underlying scheme. Comparisons of scheme-to-scheme differences in the predicted correction with the corresponding differences in the discretized fluxes show similar spatial and spectral trends, supporting the interpretation that the learned model captures not only the SGS correction but also a component of the discretization-induced error. {The recorded correction is also found to reproduce the structure of the \emph{true forcing} required by the coarse solver, to retain the ground truth trajectory, given a true coarse state (here subsampled DNS).}

{We further provide empirical evidence that even the scheme-conditioned model does not extend to a discretization absent from the training set, owing to the differing numerical characteristics of the schemes. Within the present approach, however, the model can be recalibrated to an unseen discretization by retraining only its scheme-dependent parameters, using a small fraction of the original training data.}

{The present study has considered full-state coarse-grid observations in both test cases. This setting provides a convenient testbed for analyzing how the learned correction depends on the numerical discretization, but remains idealized relative to practical applications, where observations are typically much sparser than the coarse grid or may observe only a subset of the flow field variables (typically with measurement noise).} 

Thus, an important next step is to adapt the framework to develop data-driven closures by incorporating experimental observations. {Another key direction is the extension to wall-bounded flows, where the correction is no longer statistically homogeneous but varies with wall-normal position.} {The present framework also does not normalize the network inputs and outputs by the flow scales, or build physical constraints such as Galilean invariance into the network, both of which have been shown to improve the generalization of data-driven closure models~\citep{ling2016reynolds, frezat2021physical, prakash2022invariant}, including to unseen Reynolds numbers and grid resolutions~\citep{kang2023neural}.} The results nevertheless show that nudging provides a practical route to constructing solver-aware offline training targets for turbulence closure learning.

\section*{Acknowledgments}
This work was supported by the ARO cooperative agreement W911NF-25-2-0183 and an ARO Young Investigator Award W911NF-24-1-0315 (PM - Dr. Rob Martin). We also acknowledge support from the DOE Office of Science, Office of Biological and Environmental Research through DOE-FOA-3194 from the Atmospheric System Research program. We also acknowledge computing resources from the Purdue Rosen Center for Advanced Computing. Large-language model based AI-tools were used for the grammar, editing, and consistency checks in this article.

\FloatBarrier
\bibliographystyle{plainnat}   
\bibliography{ref}

@article{germano1991dynamic,
  title={A dynamic subgrid-scale eddy viscosity model},
  author={Germano, Massimo and Piomelli, Ugo and Moin, Parviz and Cabot, William H},
  journal={Physics of fluids a: Fluid dynamics},
  volume={3},
  number={7},
  pages={1760--1765},
  year={1991},
  publisher={American Institute of Physics}
}

@article{zang1993dynamic,
  title={A dynamic mixed subgrid-scale model and its application to turbulent recirculating flows},
  author={Zang, Yan and Street, Robert L and Koseff, Jeffrey R},
  journal={Physics of Fluids A: Fluid Dynamics},
  volume={5},
  number={12},
  pages={3186--3196},
  year={1993},
  publisher={American Institute of Physics}
}

@article{meneveau2000scale,
  title={Scale-invariance and turbulence models for large-eddy simulation},
  author={Meneveau, Charles and Katz, Joseph},
  journal={Annual Review of Fluid Mechanics},
  volume={32},
  number={1},
  pages={1--32},
  year={2000},
  publisher={Annual Reviews 4139 El Camino Way, PO Box 10139, Palo Alto, CA 94303-0139, USA}
}

@article{lilly1992germano,
  author = {Lilly, D. K.},
  title = {A proposed modification of the Germano subgrid-scale closure method},
  journal = {Physics of Fluids A: Fluid Dynamics},
  volume = {4},
  number = {3},
  pages = {633--635},
  year = {1992},
  doi = {10.1063/1.858280}
}

@article{Piomelli1991SubgridscaleBI,
  title={Subgrid-scale backscatter in turbulent and transitional flows},
  author={Ugo Piomelli and William H. Cabot and Parviz Moin and Sangsan Lee},
  journal={Physics of Fluids},
  year={1991},
  volume={3},
  pages={1766-1771},
  url={https://api.semanticscholar.org/CorpusID:54904570}
}

@article{smagorinsky1963general,
  title={General circulation experiments with the primitive equations: I. The basic experiment},
  author={Smagorinsky, Joseph},
  journal={Monthly weather review},
  volume={91},
  number={3},
  pages={99--164},
  year={1963}
}

@article{shankar2023differentiable,
  title={Differentiable physics-enabled closure modeling for Burgers’ turbulence},
  author={Shankar, Varun and Puri, Vedant and Balakrishnan, Ramesh and Maulik, Romit and Viswanathan, Venkatasubramanian},
  journal={Machine Learning: Science and Technology},
  volume={4},
  number={1},
  pages={015017},
  year={2023},
  publisher={IOP Publishing}
}

@article{maulik2019subgrid,
  title={Subgrid modelling for two-dimensional turbulence using neural networks},
  author={Maulik, Romit and San, Omer and Rasheed, Adil and Vedula, Prakash},
  journal={Journal of Fluid Mechanics},
  volume={858},
  pages={122--144},
  year={2019},
  publisher={Cambridge University Press}
}

@article{chakraborty2024note,
  title={A note on the error analysis of data-driven closure models for large eddy simulations of turbulence},
  author={Chakraborty, Dibyajyoti and Barwey, Shivam and Zhang, Hong and Maulik, Romit},
  journal={arXiv preprint arXiv:2405.17612},
  year={2024}
}

@article{maulik2018data,
  title={Data-driven deconvolution for large eddy simulations of Kraichnan turbulence},
  author={Maulik, Romit and San, Omer and Rasheed, Adil and Vedula, Prakash},
  journal={Physics of Fluids},
  volume={30},
  number={12},
  year={2018},
  publisher={AIP Publishing}
}

@article{Duraisamy_2019,
   title={Turbulence Modeling in the Age of Data},
   volume={51},
   ISSN={1545-4479},
   url={http://dx.doi.org/10.1146/annurev-fluid-010518-040547},
   DOI={10.1146/annurev-fluid-010518-040547},
   number={1},
   journal={Annual Review of Fluid Mechanics},
   publisher={Annual Reviews},
   author={Duraisamy, Karthik and Iaccarino, Gianluca and Xiao, Heng},
   year={2019},
   month=jan, pages={357–377} }

@article{beck2019deep,
  title={Deep neural networks for data-driven LES closure models},
  author={Beck, Andrea and Flad, David and Munz, Claus-Dieter},
  journal={Journal of Computational Physics},
  volume={398},
  pages={108910},
  year={2019},
  publisher={Elsevier}
}

@article{List_Chen_Thuerey_2022, title={Learned turbulence modelling with differentiable fluid solvers: physics-based loss functions and optimisation horizons}, volume={949}, DOI={10.1017/jfm.2022.738}, journal={Journal of Fluid Mechanics}, author={List, Björn and Chen, Li-Wei and Thuerey, Nils}, year={2022}, pages={A25}}

@article{Sirignano_MacArt_2023, title={Deep learning closure models for large-eddy simulation of flows around bluff bodies}, volume={966}, DOI={10.1017/jfm.2023.446}, journal={Journal of Fluid Mechanics}, author={Sirignano, Justin and MacArt, Jonathan F.}, year={2023}, pages={A26}}

@misc{sanderse2024scientificmachinelearningclosure,
      title={Scientific machine learning for closure models in multiscale problems: a review}, 
      author={Benjamin Sanderse and Panos Stinis and Romit Maulik and Shady E. Ahmed},
      year={2024},
      eprint={2403.02913},
      archivePrefix={arXiv},
      primaryClass={math.NA},
      url={https://arxiv.org/abs/2403.02913}, 
}

@article{kim2026generalizable,
  title={Generalizable data-driven turbulence closure modeling on unstructured grids with differentiable physics},
  author={Kim, Hojin and Shankar, Varun and Viswanathan, Venkatasubramanian and Maulik, Romit},
  journal={Computers \& Fluids},
  pages={107200},
  year={2026},
  publisher={Elsevier}
}

@article{mohan2024you,
  title={What You See is Not What You Get: Neural Partial Differential Equations and The Illusion of Learning},
  author={Mohan, Arvind and Chattopadhyay, Ashesh and Miller, Jonah},
  journal={arXiv preprint arXiv:2411.15101},
  year={2024}
}

@article{beck2023toward,
  title={Toward discretization-consistent closure schemes for large eddy simulation using reinforcement learning},
  author={Beck, Andrea and Kurz, Marius},
  journal={Physics of Fluids},
  volume={35},
  number={12},
  year={2023},
  publisher={AIP Publishing}
}

@article{PhysRevFluids.6.050504,
  title = {Perspectives on machine learning-augmented Reynolds-averaged and large eddy simulation models of turbulence},
  author = {Duraisamy, Karthik},
  journal = {Phys. Rev. Fluids},
  volume = {6},
  issue = {5},
  pages = {050504},
  numpages = {16},
  year = {2021},
  month = {May},
  publisher = {American Physical Society},
  doi = {10.1103/PhysRevFluids.6.050504},
  url = {https://link.aps.org/doi/10.1103/PhysRevFluids.6.050504}
}

@article{um2020solver,
  title={Solver-in-the-loop: Learning from differentiable physics to interact with iterative pde-solvers},
  author={Um, Kiwon and Brand, Robert and Fei, Yun Raymond and Holl, Philipp and Thuerey, Nils},
  journal={Advances in neural information processing systems},
  volume={33},
  pages={6111--6122},
  year={2020}
}

@article{chakraborty2026binned,
  title={Binned spectral power loss for improved prediction of chaotic systems},
  author={Chakraborty, Dibyajyoti and Mohan, Arvind T and Maulik, Romit},
  journal={Journal of Computational Physics},
  pages={114866},
  year={2026},
  publisher={Elsevier}
}

@article{shankar2025differentiable,
  title={Differentiable turbulence: Closure as a partial differential equation constrained optimization},
  author={Shankar, Varun and Chakraborty, Dibyajyoti and Viswanathan, Venkatasubramanian and Maulik, Romit},
  journal={Physical Review Fluids},
  volume={10},
  number={2},
  pages={024605},
  year={2025},
  publisher={APS}
}

@article{Guan_2022,
   title={Stable a posteriori LES of 2D turbulence using convolutional neural networks: Backscattering analysis and generalization to higher Re via transfer learning},
   volume={458},
   ISSN={0021-9991},
   url={http://dx.doi.org/10.1016/j.jcp.2022.111090},
   DOI={10.1016/j.jcp.2022.111090},
   journal={Journal of Computational Physics},
   publisher={Elsevier BV},
   author={Guan, Yifei and Chattopadhyay, Ashesh and Subel, Adam and Hassanzadeh, Pedram},
   year={2022},
   month=jun, pages={111090} }

@article{Guan_2023,
   title={Learning physics-constrained subgrid-scale closures in the small-data regime for stable and accurate LES},
   volume={443},
   ISSN={0167-2789},
   url={http://dx.doi.org/10.1016/j.physd.2022.133568},
   DOI={10.1016/j.physd.2022.133568},
   journal={Physica D: Nonlinear Phenomena},
   publisher={Elsevier BV},
   author={Guan, Yifei and Subel, Adam and Chattopadhyay, Ashesh and Hassanzadeh, Pedram},
   year={2023},
   month=Jan, pages={133568} }

@article{Jakhar_2024,
   title={Learning Closed‐Form Equations for Subgrid‐Scale Closures From High‐Fidelity Data: Promises and Challenges},
   volume={16},
   ISSN={1942-2466},
   url={http://dx.doi.org/10.1029/2023MS003874},
   DOI={10.1029/2023ms003874},
   number={7},
   journal={Journal of Advances in Modeling Earth Systems},
   publisher={American Geophysical Union (AGU)},
   author={Jakhar, Karan and Guan, Yifei and Mojgani, Rambod and Chattopadhyay, Ashesh and Hassanzadeh, Pedram},
   year={2024},
   month=jul}

@article{Subel_2023,
   title={Explaining the physics of transfer learning in data-driven turbulence modeling},
   volume={2},
   ISSN={2752-6542},
   url={http://dx.doi.org/10.1093/pnasnexus/pgad015},
   DOI={10.1093/pnasnexus/pgad015},
   number={3},
   journal={PNAS Nexus},
   publisher={Oxford University Press (OUP)},
   author={Subel, Adam and Guan, Yifei and Chattopadhyay, Ashesh and Hassanzadeh, Pedram},
   editor={Yortsos, Yannis},
   year={2023},
   month=Jan }

@inproceedings{ling2025numerically,
  title={Numerically consistent data-driven subgrid-scale model via data assimilation and machine learning},
  author={Ling, Yuenong and Lozano-Duran, Adrian},
  booktitle={AIAA SCITECH 2025 Forum},
  pages={1280},
  year={2025}
}

@article{kochkov2021machine,
  title={Machine learning--accelerated computational fluid dynamics},
  author={Kochkov, Dmitrii and Smith, Jamie A and Alieva, Ayya and Wang, Qing and Brenner, Michael P and Hoyer, Stephan},
  journal={Proceedings of the National Academy of Sciences},
  volume={118},
  number={21},
  pages={e2101784118},
  year={2021},
  publisher={National Academy of Sciences}
}

@article{Bezgin2025,
   author = {Deniz A. Bezgin and Aaron B. Buhendwa and Nikolaus A. Adams},
   doi = {10.1016/j.cpc.2024.109433},
   issn = {00104655},
   journal = {Computer Physics Communications},
   month = {3},
   pages = {109433},
   title = {JAX-Fluids 2.0: Towards HPC for differentiable CFD of compressible two-phase flows},
   volume = {308},
   url = {https://linkinghub.elsevier.com/retrieve/pii/S0010465524003564},
   year = {2025},
}

@article{Bezgin2023,
   author = {Deniz A. Bezgin and Aaron B. Buhendwa and Nikolaus A. Adams},
   doi = {10.1016/j.cpc.2022.108527},
   issn = {00104655},
   journal = {Computer Physics Communications},
   month = {1},
   pages = {108527},
   title = {JAX-Fluids: A fully-differentiable high-order computational fluid dynamics solver for compressible two-phase flows},
   volume = {282},
   url = {https://linkinghub.elsevier.com/retrieve/pii/S0010465522002466},
   year = {2023},
}

@article{cao2022continuous,
  title={Continuous data assimilation for the 3D Ladyzhenskaya model: analysis and computations},
  author={Cao, Yu and Giorgini, Andrea and Jolly, Michael and Pakzad, Ali},
  journal={Nonlinear Analysis: Real World Applications},
  volume={68},
  pages={103659},
  year={2022},
  publisher={Elsevier}
}

@article{Azouani_2013,
   title={Continuous Data Assimilation Using General Interpolant Observables},
   volume={24},
   ISSN={1432-1467},
   url={http://dx.doi.org/10.1007/s00332-013-9189-y},
   DOI={10.1007/s00332-013-9189-y},
   number={2},
   journal={Journal of Nonlinear Science},
   publisher={Springer Science and Business Media LLC},
   author={Azouani, Abderrahim and Olson, Eric and Titi, Edriss S.},
   year={2013},
   month=nov, pages={277–304} }

@article{larios2025data,
  title={Data Assimilation in Large Eddy Simulation: Addressing Model-Observation Mismatch from Navier-Stokes Data},
  author={Larios, Adam and Pakzad, Ali and White, Nicholas},
  journal={arXiv preprint arXiv:2508.07492},
  year={2025}
}

@article{zerfas2019continuous,
  title={Continuous data assimilation reduced order models of fluid flow},
  author={Zerfas, Camille and Rebholz, Leo G and Schneier, Michael and Iliescu, Traian},
  journal={Computer Methods in Applied Mechanics and Engineering},
  volume={357},
  pages={112596},
  year={2019},
  publisher={Elsevier}
}

@misc{wu2025interpolateddiscrepancydataassimilation,
      title={Interpolated Discrepancy Data Assimilation for PDEs with Sparse Observations}, 
      author={Tong Wu and Humberto Godinez and Vitaliy Gyrya and James M. Hyman},
      year={2025},
      eprint={2510.24944},
      archivePrefix={arXiv},
      primaryClass={math.NA},
      url={https://arxiv.org/abs/2510.24944}, 
}

@article{clark2020synchronization,
  title={Synchronization to big data: Nudging the Navier-Stokes equations for data assimilation of turbulent flows},
  author={Clark Di Leoni, Patricio and Mazzino, Andrea and Biferale, Luca},
  journal={Physical Review X},
  volume={10},
  number={1},
  pages={011023},
  year={2020},
  publisher={APS}
}

@article{buzzicotti2020synchronizing,
  title={Synchronizing subgrid scale models of turbulence to data},
  author={Buzzicotti, Michele and Clark Di Leoni, Patricio},
  journal={Physics of Fluids},
  volume={32},
  number={12},
  year={2020},
  publisher={AIP Publishing}
}

@article{Clark_Di_Leoni_2018,
   title={Inferring flow parameters and turbulent configuration with physics-informed data assimilation and spectral nudging},
   volume={3},
   ISSN={2469-990X},
   url={http://dx.doi.org/10.1103/PhysRevFluids.3.104604},
   DOI={10.1103/physrevfluids.3.104604},
   number={10},
   journal={Physical Review Fluids},
   publisher={American Physical Society (APS)},
   author={Clark Di Leoni, Patricio and Mazzino, Andrea and Biferale, Luca},
   year={2018},
   month=oct}

@article{li2022synchronizing,
  title={Synchronizing large eddy simulations with direct numerical simulations via data assimilation},
  author={Li, Jian and Tian, Mengdan and Li, Yi},
  journal={Physics of Fluids},
  volume={34},
  number={6},
  year={2022},
  publisher={AIP Publishing}
}

@article{Bessaih_2015,
   title={Continuous data assimilation with stochastically noisy data},
   volume={28},
   ISSN={1361-6544},
   url={http://dx.doi.org/10.1088/0951-7715/28/3/729},
   DOI={10.1088/0951-7715/28/3/729},
   number={3},
   journal={Nonlinearity},
   publisher={IOP Publishing},
   author={Bessaih, Hakima and Olson, Eric and Titi, Edriss S},
   year={2015},
   month=feb, pages={729–753} }

@article{Zauner_Mons_Marquet_Leclaire_2022, title={Nudging-based data assimilation of the turbulent flow around a square cylinder}, volume={937}, DOI={10.1017/jfm.2022.133}, journal={Journal of Fluid Mechanics}, author={Zauner, M. and Mons, V. and Marquet, O. and Leclaire, B.}, year={2022}, pages={A38}}

@article{wang2023ensemble,
  title={Ensemble data assimilation-based mixed subgrid-scale model for large-eddy simulations},
  author={Wang, Yunpeng and Yuan, Zelong and Wang, Jianchun},
  journal={Physics of Fluids},
  volume={35},
  number={8},
  year={2023},
  publisher={AIP Publishing}
}

@article{ephrati2025continuous,
  title={Continuous data assimilation closure for modeling statistically steady turbulence in large-eddy simulation},
  author={Ephrati, Sagy R and Franken, Arnout and Luesink, Erwin and Cifani, Paolo and Geurts, Bernard J},
  journal={Physical Review Fluids},
  volume={10},
  number={1},
  pages={013801},
  year={2025},
  publisher={APS}
}

@inproceedings{
loshchilov2018decoupled,
title={Decoupled Weight Decay Regularization},
author={Ilya Loshchilov and Frank Hutter},
booktitle={International Conference on Learning Representations},
year={2019},
url={https://openreview.net/forum?id=Bkg6RiCqY7},
}

@inproceedings{liu2021swin,
  title={Swin transformer: Hierarchical vision transformer using shifted windows},
  author={Liu, Ze and Lin, Yutong and Cao, Yue and Hu, Han and Wei, Yixuan and Zhang, Zheng and Lin, Stephen and Guo, Baining},
  booktitle={Proceedings of the IEEE/CVF international conference on computer vision},
  pages={10012--10022},
  year={2021}
}

@inproceedings{perez2018film,
  title={Film: Visual reasoning with a general conditioning layer},
  author={Perez, Ethan and Strub, Florian and De Vries, Harm and Dumoulin, Vincent and Courville, Aaron},
  booktitle={Proceedings of the AAAI conference on artificial intelligence},
  volume={32},
  year={2018}
}

@inproceedings{rahaman2019spectral,
  title={On the spectral bias of neural networks},
  author={Rahaman, Nasim and Baratin, Aristide and Arpit, Devansh and Draxler, Felix and Lin, Min and Hamprecht, Fred and Bengio, Yoshua and Courville, Aaron},
  booktitle={International conference on machine learning},
  pages={5301--5310},
  year={2019},
  organization={PMLR}
}

@article{rosales2005linear,
  title={Linear forcing in numerical simulations of isotropic turbulence: Physical space implementations and convergence properties},
  author={Rosales, Carlos and Meneveau, Charles},
  journal={Physics of fluids},
  volume={17},
  number={9},
  year={2005},
  publisher={AIP Publishing}
}

@article{kennedy2008reduced,
  title={Reduced aliasing formulations of the convective terms within the Navier--Stokes equations for a compressible fluid},
  author={Kennedy, Christopher A and Gruber, Andrea},
  journal={Journal of Computational Physics},
  volume={227},
  number={3},
  pages={1676--1700},
  year={2008},
  publisher={Elsevier}
}

@article{Lu_2021,
   title={Learning nonlinear operators via DeepONet based on the universal approximation theorem of operators},
   volume={3},
   ISSN={2522-5839},
   url={http://dx.doi.org/10.1038/s42256-021-00302-5},
   DOI={10.1038/s42256-021-00302-5},
   number={3},
   journal={Nature Machine Intelligence},
   publisher={Springer Science and Business Media LLC},
   author={Lu, Lu and Jin, Pengzhan and Pang, Guofei and Zhang, Zhongqiang and Karniadakis, George Em},
   year={2021},
   month=Mar, pages={218–229} }

@inproceedings{
li2021fourier,
title={Fourier Neural Operator for Parametric Partial Differential Equations},
author={Zongyi Li and Nikola Borislavov Kovachki and Kamyar Azizzadenesheli and Burigede liu and Kaushik Bhattacharya and Andrew Stuart and Anima Anandkumar},
booktitle={International Conference on Learning Representations},
year={2021},
url={https://openreview.net/forum?id=c8P9NQVtmnO}
}

@article{ling2016reynolds,
  title={Reynolds averaged turbulence modelling using deep neural networks with embedded invariance},
  author={Ling, Julia and Kurzawski, Andrew and Templeton, Jeremy},
  journal={Journal of Fluid Mechanics},
  volume={807},
  pages={155--166},
  year={2016},
  publisher={Cambridge University Press}
}

@article{frezat2021physical,
  title={Physical invariance in neural networks for subgrid-scale scalar flux modeling},
  author={Frezat, Hugo and Balarac, Guillaume and Le Sommer, Julien and Fablet, Ronan and Lguensat, Redouane},
  journal={Physical Review Fluids},
  volume={6},
  number={2},
  pages={024607},
  year={2021},
  publisher={APS}
}

@article{pirozzoli2010generalized,
  title={Generalized conservative approximations of split convective derivative operators},
  author={Pirozzoli, Sergio},
  journal={Journal of Computational Physics},
  volume={229},
  number={19},
  pages={7180--7190},
  year={2010},
  publisher={Elsevier}
}

@article{hoekstra2024reduced,
  title={Reduced data-driven turbulence closure for capturing long-term statistics},
  author={Hoekstra, Rik and Crommelin, Daan and Edeling, Wouter},
  journal={Computers \& Fluids},
  volume={285},
  pages={106469},
  year={2024},
  publisher={Elsevier}
}

@article{ephrati2025probabilistic,
  title={Probabilistic data-driven turbulence closure modeling by assimilating statistics},
  author={Ephrati, Sagy R},
  journal={Journal of Computational Physics},
  pages={114234},
  year={2025},
  publisher={Elsevier}
}

@article{ghosal1996analysis,
  title={An analysis of numerical errors in large-eddy simulations of turbulence},
  author={Ghosal, Sandip},
  journal={Journal of Computational Physics},
  volume={125},
  number={1},
  pages={187--206},
  year={1996},
  publisher={Elsevier}
}

@article{chow2003further,
  title={A further study of numerical errors in large-eddy simulations},
  author={Chow, Fotini Katopodes and Moin, Parviz},
  journal={Journal of Computational Physics},
  volume={184},
  number={2},
  pages={366--380},
  year={2003},
  publisher={Elsevier}
}

@article{park2021toward,
  title={Toward neural-network-based large eddy simulation: Application to turbulent channel flow},
  author={Park, Jonghwan and Choi, Haecheon},
  journal={Journal of Fluid Mechanics},
  volume={914},
  pages={A16},
  year={2021},
  publisher={Cambridge University Press}
}

@article{cho2024recursive,
  title={A recursive neural-network-based subgrid-scale model for large eddy simulation: application to homogeneous isotropic turbulence},
  author={Cho, Chonghyuk and Park, Jonghwan and Choi, Haecheon},
  journal={Journal of Fluid Mechanics},
  volume={1000},
  pages={A76},
  year={2024},
  publisher={Cambridge University Press}
}

@article{novati2021automating,
  title={Automating turbulence modelling by multi-agent reinforcement learning},
  author={Novati, Guido and De Laroussilhe, Hugues Lascombes and Koumoutsakos, Petros},
  journal={Nature Machine Intelligence},
  volume={3},
  number={1},
  pages={87--96},
  year={2021},
  publisher={Nature Publishing Group UK, London}
}

@article{subel2021data,
  title={Data-driven subgrid-scale modeling of forced Burgers turbulence using deep learning with generalization to higher Reynolds numbers via transfer learning},
  author={Subel, Adam and Chattopadhyay, Ashesh and Guan, Yifei and Hassanzadeh, Pedram},
  journal={Physics of Fluids},
  volume={33},
  number={3},
  year={2021},
  publisher={AIP Publishing}
}

@article{bernardini2023streams,
  title={STREAmS-2.0: Supersonic turbulent accelerated Navier-Stokes solver version 2.0},
  author={Bernardini, Matteo and Modesti, Davide and Salvadore, Francesco and Sathyanarayana, Srikanth and Della Posta, Giacomo and Pirozzoli, Sergio},
  journal={Computer Physics Communications},
  volume={285},
  pages={108644},
  year={2023},
  publisher={Elsevier}
}

@article{wang2010hybrid,
  title={A hybrid numerical simulation of isotropic compressible turbulence},
  author={Wang, Jianchun and Wang, L-P and Xiao, Zuoli and Shi, Y and Chen, Shiyi},
  journal={Journal of Computational Physics},
  volume={229},
  number={13},
  pages={5257--5279},
  year={2010},
  publisher={Elsevier}
}

@inproceedings{sanchez2020learning,
  title={Learning to simulate complex physics with graph networks},
  author={Sanchez-Gonzalez, Alvaro and Godwin, Jonathan and Pfaff, Tobias and Ying, Rex and Leskovec, Jure and Battaglia, Peter},
  booktitle={International conference on machine learning},
  pages={8459--8468},
  year={2020},
  organization={Pmlr}
}

@article{stachenfeld2021learned,
  title={Learned coarse models for efficient turbulence simulation},
  author={Stachenfeld, Kimberly and Fielding, Drummond B and Kochkov, Dmitrii and Cranmer, Miles and Pfaff, Tobias and Godwin, Jonathan and Cui, Can and Ho, Shirley and Battaglia, Peter and Sanchez-Gonzalez, Alvaro},
  journal={arXiv preprint arXiv:2112.15275},
  year={2021}
}

@article{MoralesBrotons2024ExponentialMA,
  title={Exponential Moving Average of Weights in Deep Learning: Dynamics and Benefits},
  author={Daniel Morales-Brotons and Thijs Vogels and Hadrien Hendrikx},
  journal={Trans. Mach. Learn. Res.},
  year={2024},
  volume={2024},
  url={https://api.semanticscholar.org/CorpusID:270355881}
}

@article{prakash2022invariant,
  title={Invariant data-driven subgrid stress modeling in the strain-rate eigenframe for large eddy simulation},
  author={Prakash, Aviral and Jansen, Kenneth E and Evans, John A},
  journal={Computer Methods in Applied Mechanics and Engineering},
  volume={399},
  pages={115457},
  year={2022},
  publisher={Elsevier}
}

@article{kang2023neural,
  title={Neural-network-based mixed subgrid-scale model for turbulent flow},
  author={Kang, Myeongseok and Jeon, Youngmin and You, Donghyun},
  journal={Journal of Fluid Mechanics},
  volume={962},
  pages={A38},
  year={2023},
  publisher={Cambridge University Press}
}

\FloatBarrier
\begin{appendices}

\section{Notation}\label{sec:appendix_A}
\begin{table}[H]
\centering
\begin{tabular}{ll}
\hline
$\boldsymbol{u}$ & DNS (reference) velocity field\\
$\boldsymbol{v}$ & coarse-grid LES (assimilated) velocity field\\
$\boldsymbol{u}_{obs}$ & subsampled DNS on the coarse grid (observations)\\
$\boldsymbol{d}=\boldsymbol{u}_{obs}-\boldsymbol{v}$ & discrepancy field / unscaled nudging target\\
$\boldsymbol{f}_{nudge},\ \mu$ & nudging correction (forcing), nudging coefficient\\
$\mathcal{N}[\cdot],\ \mathcal{S}[\cdot]$ & Navier-Stokes operator, Smagorinsky closure operator\\
$\langle\,\cdot\,\rangle$, $\langle\,\cdot\,\rangle_t$ & spatial and temporal average\\
$\boldsymbol{f}_{sgs},\ \boldsymbol{\tau}_{sgs}$ & subgrid (SGS) force and stress\\
$\mathcal{M}_\theta$ & neural network model with parameters $\theta$\\
$s$ & numerical-scheme encoding\\
VL, UPWIND, CD2 & 2D convective schemes\\
C2, C4, C6, WENO & 3D 2nd/4th/6th-order central and Weno convective schemes\\
$\mathcal{C}(\cdot)$ & discretized convective operator\\
$\boldsymbol{F}_{p},\ F_{e}$ & momentum and energy nudging source terms\\
$\tfrac{1}{2}\langle u_i' u_i'\rangle,\ E(k)$ & turbulent kinetic energy, energy spectrum\\
$\varepsilon$ & relative $L_2$ error\\
$\boldsymbol{F}_{true}$ & single-step true forcing measure for the coarse grid solver\\
$\Phi_{c}$ & coarse grid solver without a closure term\\
$\gamma^{(s)},\ \beta^{(s)}$ & FiLM scale and shift for scheme encoding $s$\\
{$Re_\lambda$} & {Taylor-microscale Reynolds number}\\
\hline
\end{tabular}
\vspace{3mm}
\caption{Summary of notation.}
\label{tab:notation}
\end{table}

\section{Dataset generation}\label{sec:appendix_B}
\subsection{Two-dimensional turbulence with Kolmogorov forcing}\label{sec:appendix_data_2d_turbulence}
As a first test case, we consider forced two-dimensional turbulence on the
doubly periodic domain $[0,2\pi]^2$, governed by
\begin{equation}
\partial_t \boldsymbol{u}+\nabla\cdot(\boldsymbol{u}\otimes\boldsymbol{u})
=-\frac{1}{\rho}\nabla p+\frac{1}{Re}\nabla^2\boldsymbol{u}+\boldsymbol{f},
\qquad \nabla\cdot\boldsymbol{u}=0,
\end{equation}
where $\boldsymbol{u}$ is the velocity, $p$ the pressure, $Re$ is the Reynolds number and $\rho$ the density. We use Kolmogorov forcing with linear drag,
\begin{equation}
\boldsymbol{f}=A\sin(kx)\hat{\boldsymbol{e}}_y-r\boldsymbol{u}.
\end{equation}
{Here $\boldsymbol{f}$ is the external body forcing that drives
the turbulence, and is distinct from the nudging term $\boldsymbol{f}_{nudge}$
in~\eqref{eq:nudge_force}}.

{During training-data generation, the coarse-grid LES is evolved
with the explicit nudging forcing,}
\begin{equation}
\partial_t \boldsymbol{v}+\nabla\cdot(\boldsymbol{v}\otimes\boldsymbol{v})
=-\frac{1}{\rho}\nabla p+\frac{1}{Re}\nabla^2\boldsymbol{v}
+\nabla\cdot\boldsymbol{\tau}_{smag}+\boldsymbol{f}
+\underbrace{\mu\,(\boldsymbol{u}_{obs}-\boldsymbol{v})}_{\boldsymbol{f}_{nudge}},
\qquad \nabla\cdot\boldsymbol{v}=0,
\end{equation}
where the Smagorinsky stress is
\begin{equation}
\boldsymbol{\tau}_{smag}=-2\nu_t\boldsymbol{S},
\qquad
\nu_t=(C_s\Delta)^2|\boldsymbol{S}|,
\qquad
\boldsymbol{S}=\tfrac{1}{2}\left(\nabla\boldsymbol{v}+\nabla\boldsymbol{v}^{T}\right),
\end{equation}
with $C_s$ the Smagorinsky coefficient and $\Delta$ the coarse-grid filter width.
{Setting $\boldsymbol{f}_{nudge}=0$ recovers the baseline
Smagorinsky LES (SMAG). Here $\boldsymbol{u}_{obs}$ is the DNS velocity field subsampled onto the coarse grid, and the recorded discrepancy
$\boldsymbol{d}=(\boldsymbol{u}_{obs}-\boldsymbol{v})$ is the unscaled nudging term stored at every step.}

The dataset is generated using the finite-volume solver JAX-CFD~\citep{kochkov2021machine}, with the DNS on a $1024^2$ grid and the coarse-grid LES on a $64^2$ grid; {the viscous term uses a second-order central five-point stencil and incompressibility is enforced by a pressure-projection step solved by fast diagonalization. These are identical in the DNS and LES, and only the convective discretization is varied.} Multiple DNS trajectories are generated from distinct initial conditions, each advanced until a statistically stationary state is reached before the nudged simulations are performed. The DNS uses a second-order Van Leer discretization of the convective term; the LES uses $C_s=0.2$ for both the baseline and nudged runs, with nudging coefficient $\mu=25$. All simulations use $\rho=1$, $Re=1000$ (equivalently, kinematic viscosity $\nu=10^{-3}$), $A=2$, $k=8$, $r=0.1$ and $\Delta t\approx4.4\times10^{-4}$, obtained with $\mathrm{CFL}=0.5$.

The training and test datasets are then constructed from 15 independent trajectories, each generated for second-order Van Leer (VL) and first-order upwind (UPWIND). The first 12 trajectories are used for training and the remaining 3 for testing. For each trajectory, samples are extracted from the full $10,000$ simulation steps, yielding 600 samples per trajectory per scheme. This gives a total of 14,400 training samples and 3,600 test samples. Each sample consists of a coarse velocity field, the corresponding nudging-closure target, and a scheme label identifying the discretization branch.

\subsection{Three-dimensional forced homogeneous isotropic turbulence}\label{sec:appendix_data_3d_turbulence}

The dataset for the three-dimensional case is generated using the compressible finite-volume solver JAX-Fluids \citep{Bezgin2023, Bezgin2025}. The flow state is represented in terms of the primitive variables
\begin{equation}
\mathbf{W}=[\rho,u,v,w,p]^T,
\end{equation}
where $\rho$ is the density, $(u,v,w)$ are the three velocity components, and $p$ is the pressure. The corresponding conservative variables are
\begin{equation}
\mathbf{U}=[\rho,\rho u,\rho v,\rho w,E]^T,
\end{equation}
where the total energy per unit volume is
\begin{equation}
E=\rho e+\frac{1}{2}\rho \boldsymbol{u}\cdot\boldsymbol{u},
\end{equation}
with $e$ denoting the internal energy per unit mass and $\boldsymbol{u}=(u,v,w)^T$ the velocity vector.

The governing equations are solved in conservative form as
\begin{equation}
\partial_t \mathbf{U}+\nabla\cdot\mathbf{F}^{c}(\mathbf{U})
=\nabla\cdot\mathbf{F}^{v}(\mathbf{U},\nabla\mathbf{U})+\mathbf{S},
\end{equation}
where $\mathbf{F}^{c}$ and $\mathbf{F}^{v}$ denote the convective and viscous fluxes, respectively, and $\mathbf{S}$ is the source term. Although the solver is compressible, we restrict attention to a low-Mach-number regime so that the dynamics remain close to the incompressible limit.

{To sustain the turbulence, we apply a linear forcing~\citep{rosales2005linear} which, in primitive form, is written as
\begin{equation}
\boldsymbol{f}=A\boldsymbol{u}',
\end{equation}
with $A=0.02$, where $\boldsymbol{u}'$ is the velocity fluctuation obtained by subtracting the mean flow. Although forced homogeneous isotropic turbulence has zero mean in principle, we subtract the mean as a safeguard, since numerical simulations usually carry a small mean bias and forcing the full velocity amplifies it exponentially over long runs. This is relevant in our case, as the coarse-grid simulations are initialized by subsampling the DNS, which introduces a small mean, and the neural network prediction (during inference) may also carry a small bias. For the DNS (initialized by machine precision zero mean), the mean remains negligible and both forms are equivalent.} {In the conservative system this enters through the
source term as a momentum source $\rho\boldsymbol{f}$ and the corresponding energy
source $\rho\boldsymbol{f}\cdot\boldsymbol{u}$, i.e.
$\mathbf{S}_{forcing}=[\,0,\;\rho\boldsymbol{f},\;\rho\boldsymbol{f}\cdot\boldsymbol{u}\,]^T$}. The flow is initialized using an exponential energy spectrum with peak wavenumber
$k_0=2$, {initial Reynolds number (box-scale)
$Re_0=u'_{rms,0}L/\nu\approx 726$}, reference density $\rho_{ref}=1$,
initial Mach number $Ma=0.08$, kinematic viscosity $\nu=10^{-3}$.
{The working fluid is an ideal gas with ratio of specific heats
$\gamma=1.4$, governed by the equation of state $p=\rho R T$. The reference
temperature is set to unity, $T_{ref}=1$, and the reference sound speed
is $c_{ref}=2.5$, giving a specific gas constant
$R=c_{ref}^2/(\gamma T_{ref})=4.4643$ and reference pressure $p_{ref}=4.4643$ from the equation of state.} {Given the low Mach number, heat conduction is neglected in the DNS and the nudging simulations, so the energy equation retains only the convective and viscous fluxes. In the long coarse-grid rollouts, which have no subgrid-scale model for the heat flux, heat conduction (Prandtl number $0.72$) is included to prevent the accumulation of temperature and density fluctuations, with negligible effect on the velocity statistics.}

The DNS is performed on a \(128^3\) grid over a triply periodic box of $[0,2\pi]^3$, while the corresponding coarse-grid simulations are carried out on a \(32^3\) grid. Time advancement is performed using a third-order Runge-Kutta scheme with constant time step $\Delta t=0.001$. The density remains nearly constant in time throughout. {In the statistically stationary state the Taylor-microscale Reynolds number is $Re_\lambda\approx36$, with $\Delta_{DNS}/\eta=1.16$ and $k_{max}\eta=2.70$, where $\eta$ is the Kolmogorov length scale, so that the DNS is fully resolved.}

Under these conditions, the flow is not sufficiently developed to exhibit a clear inertial-range scaling, but it still serves as an effective test case for evaluating our framework. The coarse-grid simulations use the same forcing and time stepping, together with a Smagorinsky closure with $C_s=0.2$.

As in the two-dimensional case, each DNS trajectory is first advanced until a statistically stationary state is reached. {During training-data generation, the nudging source ($\mu=2.5$) is added to
$\mathbf{S}$ through the momentum and energy source terms $\boldsymbol{F}_p$ and $F_e$ introduced in \S~\ref{sec:3d_results}, with $\rho_{obs}$ and $\boldsymbol{u}_{obs}$ the DNS density and velocity subsampled onto the $32^3$ coarse grid. Setting the nudging source to zero recovers the baseline SMAG LES, and the recorded discrepancy is stored at every step as the supervised training
target.}

The fields are then restricted to the \(32^3\) grid and stored at every time step to provide the coarse-grid observations used in the nudging simulations. {For the DNS, the convective flux uses the two-point kinetic-energy-preserving split form~\citep{kennedy2008reduced,pirozzoli2010generalized}, which limits aliasing errors and provides nonlinear stability without explicit dealiasing or filtering; the viscous fluxes use a fourth-order central stencil. For the coarse-grid nudging simulations, the convective discretization is varied across the unsplit second-, fourth-, and sixth-order central schemes, while the viscous discretization and the remaining components of the numerical setup are kept unchanged.}

For dataset generation, the DNS is evolved until a statistically stationary state is reached at approximately \(t=130\). Data are then stored from $t=135$ to $t=220$ in 17 blocks of 5000 time steps, each spanning 5 time units. For training, only the first 12 blocks are used. Within each block, we sample every second field-target pair for preparing the dataset. The same procedure is repeated for all three coarse-grid schemes. Validation uses the next three blocks, while the remaining blocks are excluded from training.

For testing, an additional DNS trajectory is generated using the same configuration but a different initial condition. The results in figure~\ref{fig:inference_stats_3d} demonstrate that the model generalizes {to a flow generated from a different initial condition and evolution trajectory in the same flow configuration}.

\subsection{Choice of nudging strength}\label{sec:appendix_nudge_strength}

\begin{figure}
  \centering
  \begin{tabular}{@{}c@{\hspace{0.02\textwidth}}c@{}}
    \includegraphics[width=0.45\textwidth]{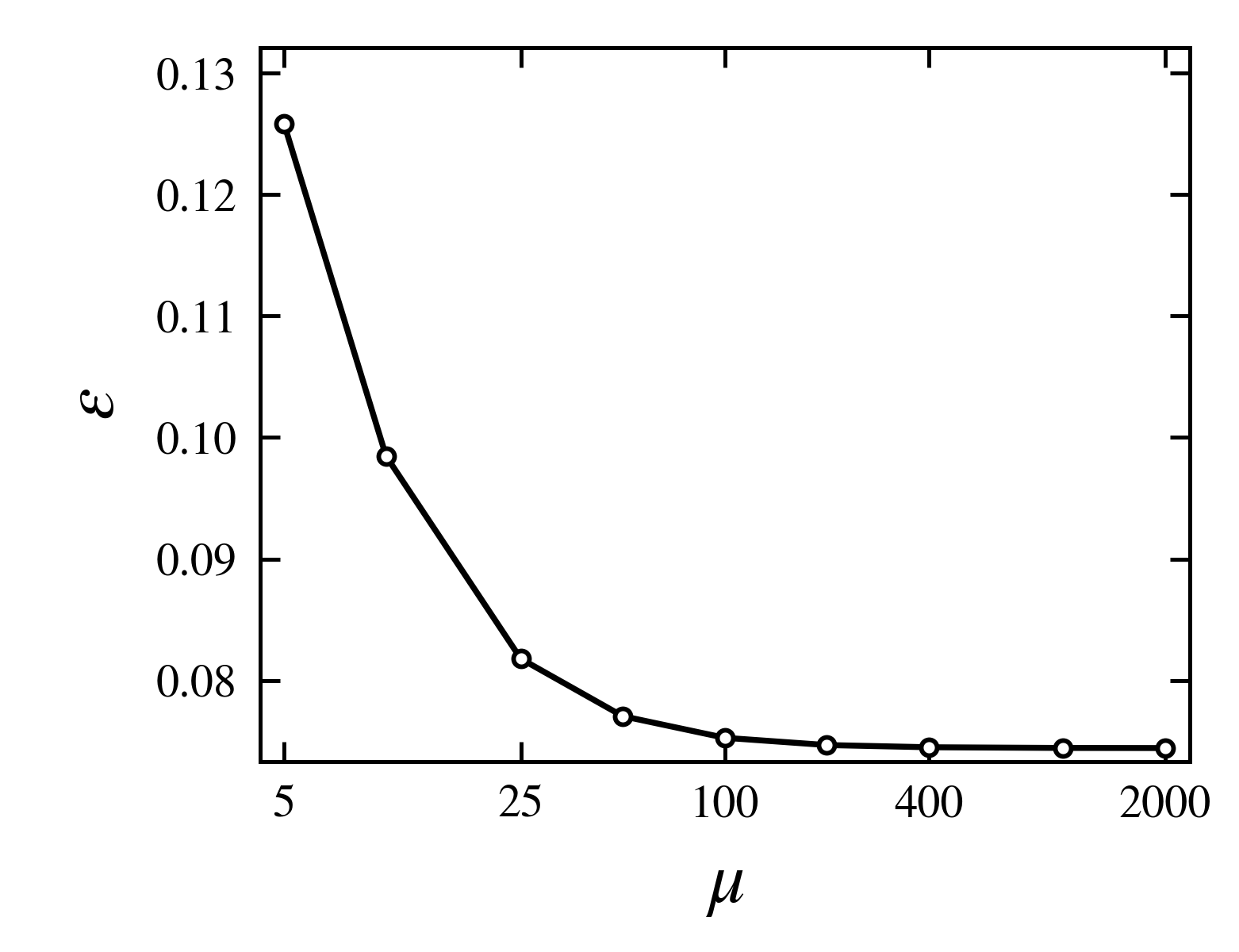} &
    \includegraphics[width=0.47\textwidth]{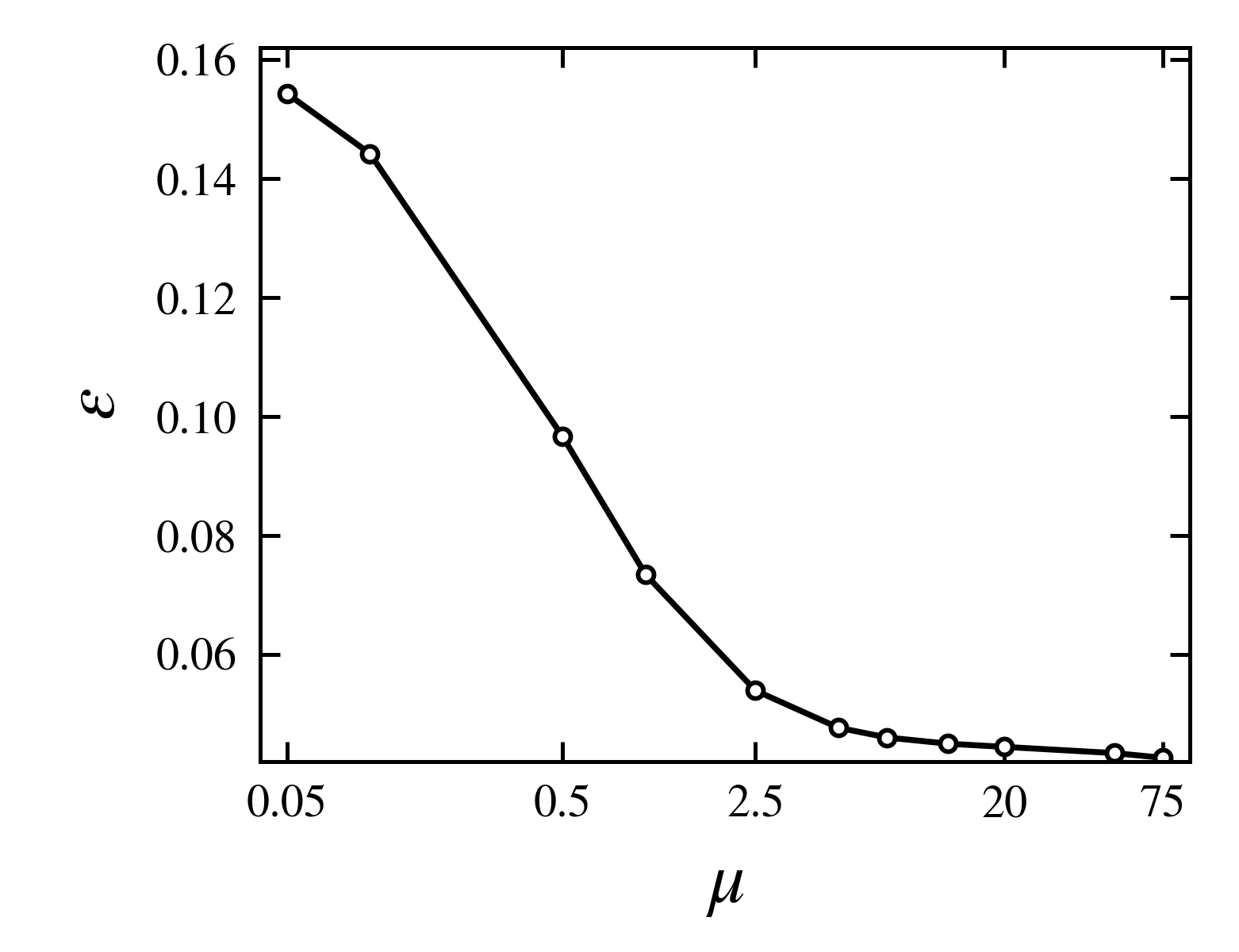} \\
    (a) & (b)
  \end{tabular}
  \caption{
{{Residual floor, measured by the relative $L_2$ error $\varepsilon$, of the nudged coarse-grid simulation as a function of the nudging coefficient $\mu$, for (a) the two-dimensional case (VL) and (b) the three-dimensional case (C4). The selected values, $\mu=25$ in 2D and $\mu=2.5$ in 3D, lie at the elbow of the respective curves.}}}
  \label{fig:mu_sweep_residual_floor}
\end{figure}

{The nudging coefficient $\mu$ represents the strength of the feedback term and thereby determines the degree of synchronization attained by the assimilated system. Since the assimilated system is itself a coarse-grid LES, the discrepancy does not vanish but settles to a residual floor, measured here by the relative $L_2$ error $\varepsilon$ defined in \S~\ref{sec:2d_results}. For both the two- and three-dimensional cases, the value of $\mu$ is fixed by performing a sweep and determining an \emph{optimal} value for learning the residual. Figure~\ref{fig:mu_sweep_residual_floor} reports this residual floor for different values of $\mu$. All remaining parameters are held at the values given previously, and the nudging is performed with the Van Leer (VL) and fourth-order central (C4) convective discretizations.}

{In both cases the residual decreases monotonically with $\mu$ and saturates towards an asymptotic floor, so that the curves exhibit a clear ``elbow''. We select this value of nudging strength, giving $\mu=25$ in two dimensions and $\mu=2.5$ in three dimensions. At this value we obtain a sufficient level of synchronization, with further increases in $\mu$ yielding diminishing returns.}

{Another factor considered when choosing $\mu$ is the \emph{learnability} of the residual itself. While increasing $\mu$ leads to better pointwise synchronization with the ground truth, it also means that the residual content becomes increasingly dominated by small-scale, low-amplitude fluctuations. Owing to the spectral bias of neural networks~\citep{rahaman2019spectral, chakraborty2026binned}, the learning problem becomes increasingly difficult, even with the binned spectral power (BSP) loss. Hence, the value chosen for this study represents a compromise between the degree of synchronization and the learnability of the recorded target.}

\section{Training details}\label{sec:appendix_C}

{For all training runs, we use a Swin Transformer architecture \citep{liu2021swin}. Self-attention is well suited to the non-local dependence of the correction, and the Swin variant evaluates it within local windows whose partitioning shifts between successive layers, which keeps the cost tractable on dense two- and three-dimensional fields while still connecting neighboring regions. Here the token grid and embedding width are held fixed across stages, so the representation remains at the coarse-grid resolution throughout, as suits a field-to-field regression. The framework is nonetheless independent of this choice, and any sufficiently expressive model, such as a convolutional network or a neural operator~\citep{li2021fourier, Lu_2021}, could be used in its place.} The numerical scheme is encoded through FiLM layers applied within each transformer block, which perform feature-wise affine transformations conditioned on the scheme label. {The scale and shift parameters are generated per stage and shared by the blocks within that stage.} In this way, the model is allowed to modulate its internal representation according to the target discretization. The network is trained to predict the coarse-grid unscaled nudging forcing term. We use mean-squared error (MSE) as the primary loss in all cases,

\begin{equation}
\mathcal{L}_{MSE}
=
\frac{1}{B\,C\,|\Omega|}
\sum_{b=1}^{B}
\sum_{c=1}^{C}
\sum_{\mathbf{x}\in\Omega}
\left(
\hat{F}_{b,c}(\mathbf{x})
-
F_{b,c}(\mathbf{x})
\right)^2,
\end{equation}

where $\hat{F}$ is the predicted component and $F$ is the actual nudging term. For the two-dimensional case, training is performed using the AdamW \citep{loshchilov2018decoupled} optimizer with learning rate $10^{-3}$ and batch size $32$. Only the MSE loss is used. The model uses embedding dimension $48$ and stage depths $(2,2,4,2)$, corresponding to $10$ transformer blocks in total, and has $341,578$ trainable parameters.

For the three-dimensional case, we add a binned spectral power (BSP) loss \citep{chakraborty2026binned} in addition to the MSE loss in order to improve agreement in the spectrum of the target field.
The binned spectral loss is defined as
\begin{equation}
\mathcal{L}_{BSP}
=
\frac{1}{B C N_K}
\sum_{b=1}^{B}
\sum_{c=1}^{C}
\sum_{m=1}^{N_K}
\left[
1-
\frac{
E^{pred}_{b,c}(k_m)+\epsilon
}{
E^{true}_{b,c}(k_m)+\epsilon
}
\right]^2.
\end{equation}
\begin{equation}
E_c(k_m)
=
\frac{1}{N_m}
\sum_{\mathbf{k}\in\mathcal{K}_m}
\frac{1}{2}
\left|
\widehat{F}_c(\mathbf{k})
\right|^2,
\qquad
\widehat{F}_c(\mathbf{k}) = \mathcal{F}[F_c](\mathbf{k}).
\end{equation}

where $F_c$ denotes the $c$th output channel, $\widehat{F}_c$ its
Fourier transform, $\hat{F}_c$ the predicted $c$th output channel in physical
space, $E^{pred}$ and $E^{true}$ the binned
spectra of the predicted and target fields, $\epsilon$ a small positive
constant, $\mathcal{K}_m$ the set of Fourier modes in the $m$th wavenumber bin,
$N_m$ the number of modes in that bin, $B$ the batch size, $C$ the number of
output channels, $N_K$ the total number of spectral bins, and
$|\Omega| = N_x N_y N_z$ the number of spatial grid points. The total loss is given by,

\begin{equation}
\mathcal{L} = \lambda_1\mathcal{L}_{MSE} + \lambda_2 \mathcal{L}_{BSP},
\end{equation}

with $\lambda_1 = 100$ and $\lambda_2= 10^{-6}$.{ The disparity between the two weights balances the very different scales of the two terms rather than reflecting their relative importance; the binned spectral loss is initially much larger than the mean-squared error, so the weighted spectral term dominates the early epochs and shapes the predicted spectrum before decaying to become negligible. Larger values of $\lambda_2$ improve the spectral agreement further but degrade the pointwise accuracy and eventually destabilize training, so the term is retained as a soft regularizer rather than a primary objective.}

For the three-dimensional case, training is performed using AdamW with initial learning rate $10^{-3}$, weight decay $10^{-4}$ and per-device batch size 1. The model uses embedding dimension 96, stage depths $(2,2,6,2)$, window size $(4,4,4)$ and attention heads $(3,6,12,24)$, corresponding to 12 transformer blocks in total. Scheme dependence is encoded through a learned 32-dimensional scheme embedding with stage-wise FiLM generators, and the full model has $1,462,839$ trainable parameters. {The final model weights are taken as an exponential moving average of the weights at the end of training, a standard practice for obtaining a more robust model~\citep{MoralesBrotons2024ExponentialMA}.}

\section{Generalizability across different temporal schemes}\label{sec:appendix_D}

\begin{figure}[H]
  \centering
  \begin{tabular}{@{}c@{\hspace{0.02\textwidth}}c@{}}
    \includegraphics[width=0.47\textwidth]{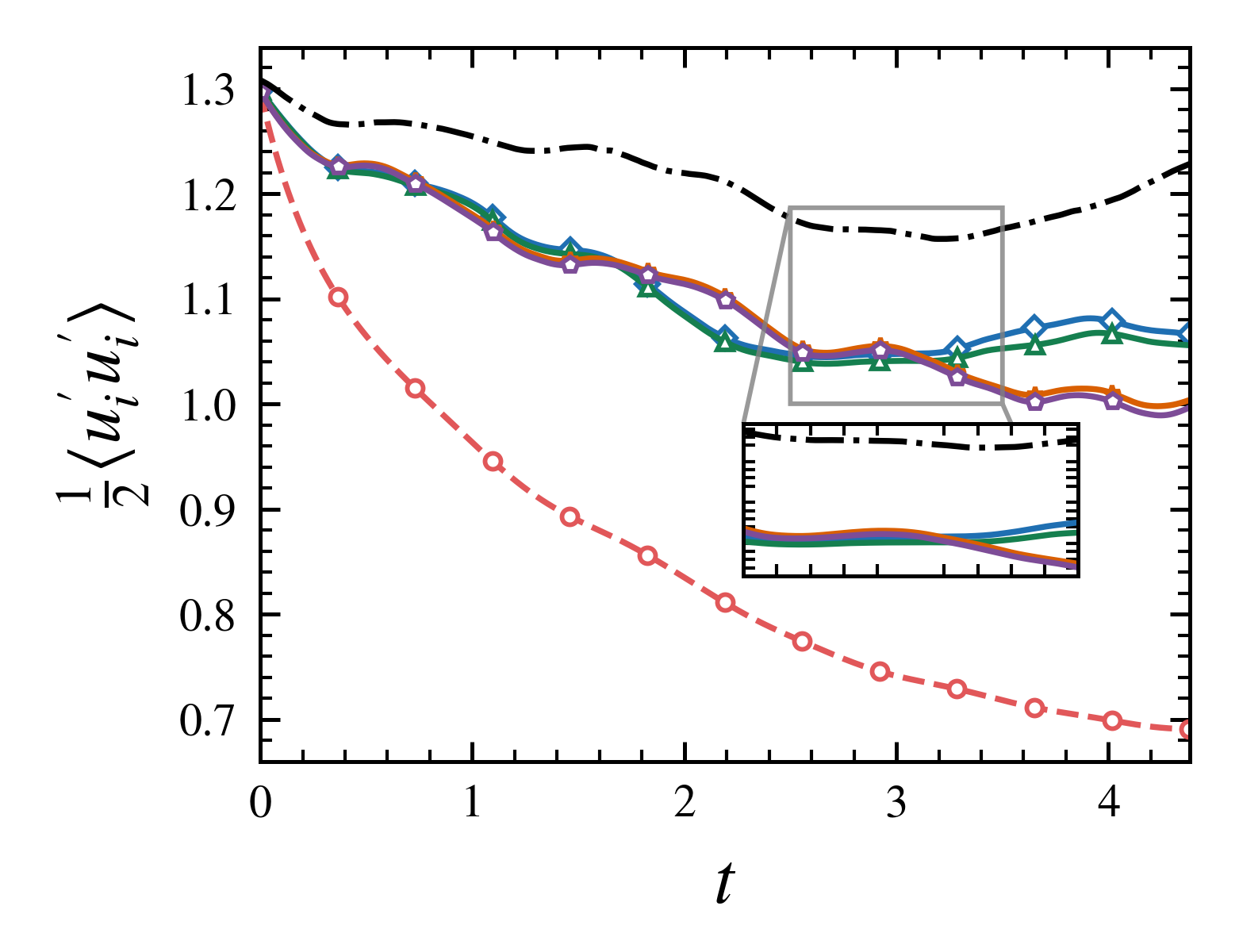} &
    \includegraphics[width=0.47\textwidth]{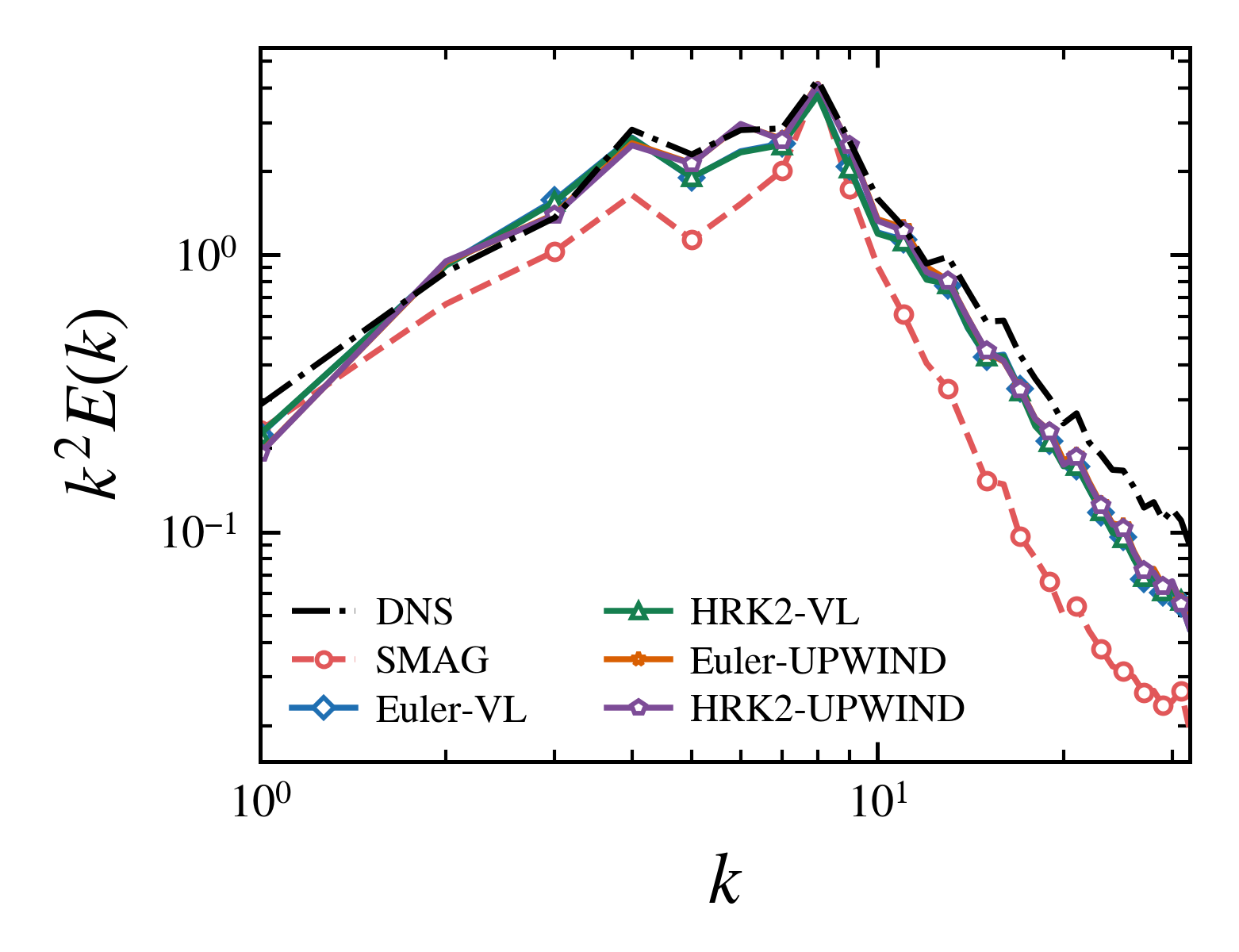} \\
    (a) & (b) \\[0.5em]
    \includegraphics[width=0.47\textwidth]{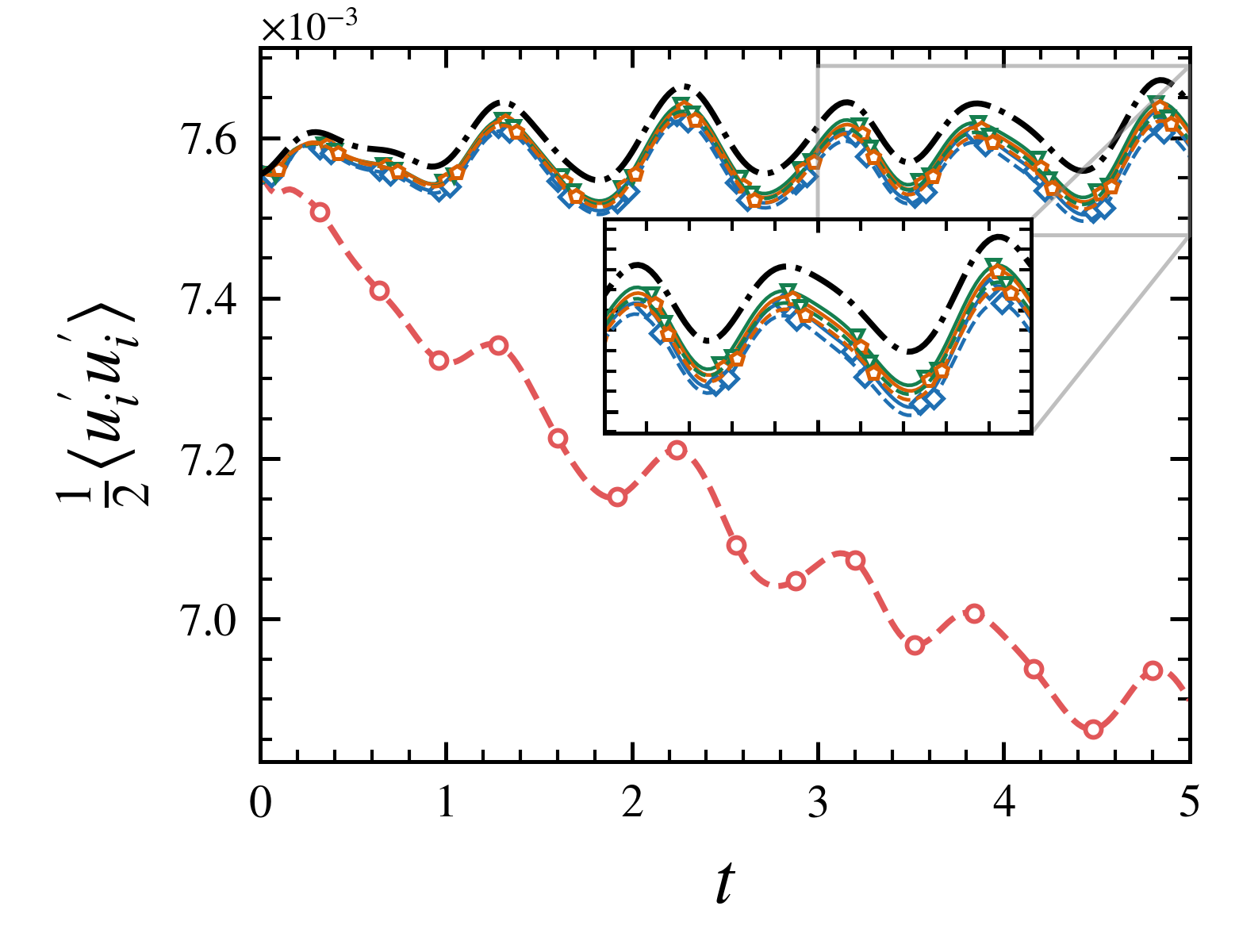} &
    \includegraphics[width=0.46\textwidth]{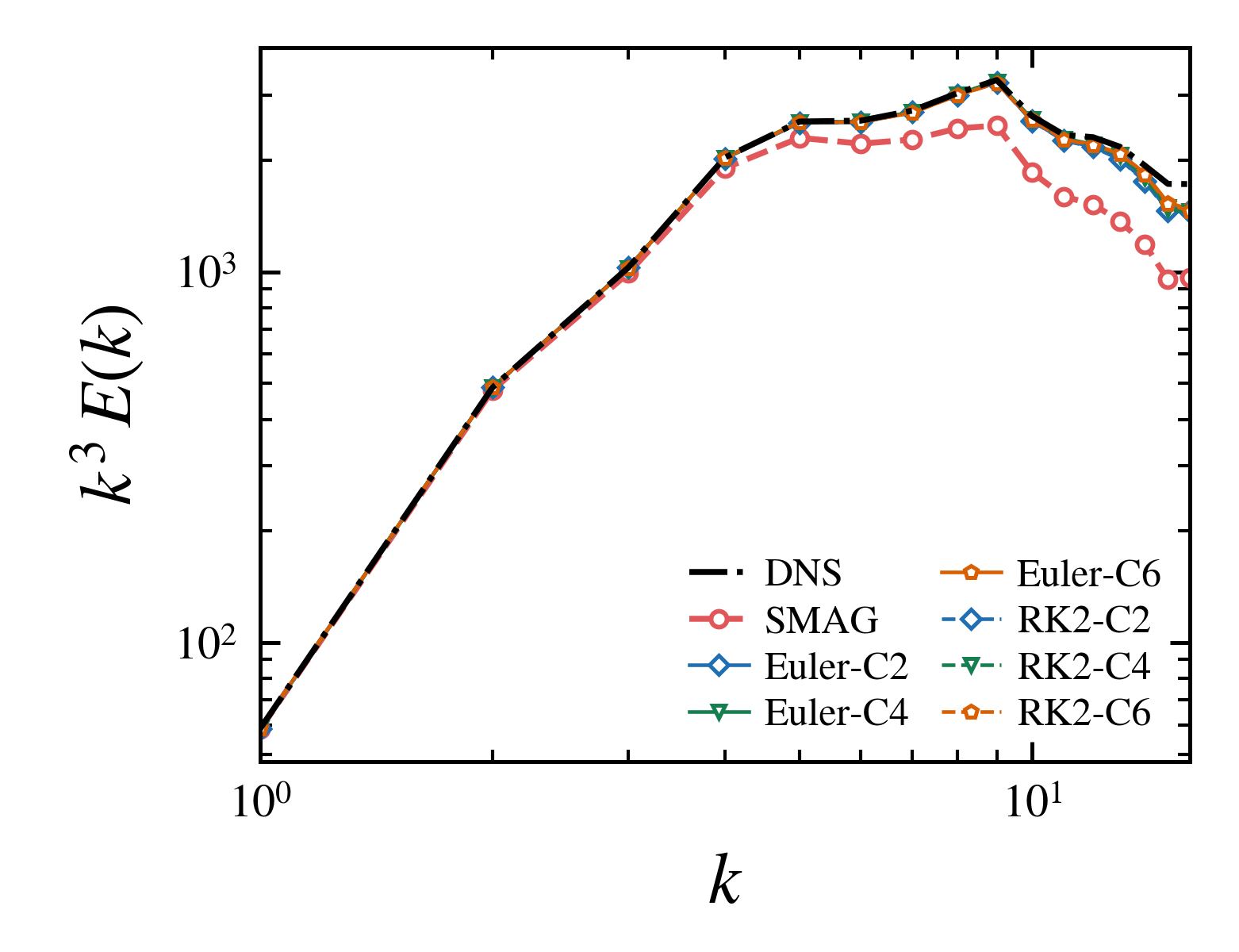} \\
    (c) & (d)
  \end{tabular}
  \caption{
{Generalization across time-stepping schemes not used during training: TKE and scaled time-averaged energy spectrum for the two-dimensional (a,b) and three-dimensional (c,d) cases. Both cases use DNS and SMAG baselines. In 2D, the model is deployed on Heun second-order Runge-Kutta (HRK2) and Forward-Euler (Euler) for the two trained schemes VL and UPWIND. In the 3D case, the model is deployed on the three convective schemes (C2, C4 and C6) for Euler, RK2 schemes. The model remains stable in these tests and retains good agreement with the TKE and spectral reference statistics for temporal schemes not used during training.}}
  \label{fig:diff_time_int}
\end{figure}

We test whether the learned correction generalizes (see figure~\ref{fig:diff_time_int}) across the temporal discretizations utilized in this study. In contrast to the pronounced dependence observed for the spatial scheme, the sensitivity to the time-stepping method is comparatively weak. For the two-dimensional case, the model is trained on RK4 (on all schemes) and tested on Forward-Euler and Heun-RK2, whereas the three-dimensional model is trained on RK3 and tested on Forward-Euler and RK2.

The close-up comparisons also show that the generalization is not exact: lower-order time integrators exhibit slightly larger departures from the reference evolution, whereas the higher-order schemes retain closer agreement. The stability and accuracy of the time-integration schemes also depend on the chosen time step and the underlying flow field. Overall, these results suggest that the learned correction depends much more strongly on the spatial discretization than on the time-stepping scheme, although a residual dependence on the temporal integrator is still present.

\section{Extension to higher Reynolds numbers}\label{sec:appendix_highRe}

{Since the focus of this study is the dependence of closures on the numerical discretization, all the test cases previously considered are at a single Reynolds number. Here, we discuss the performance of the framework at a higher Reynolds number. In general, a data-driven closure trained at one Reynolds number is not expected to \emph{transfer} directly to a different one due to the differences in the flow dynamics that remain unseen during training. Some studies~\citep{park2021toward, kochkov2021machine} show generalization to different Reynolds numbers, but are limited to cases where the LES grid size relative to the smaller scales of the flow remains close to that seen during training. The most common approaches to such generalization are either incorporating multiple Reynolds numbers during training, which improves generalization to unseen regimes~\citep{novati2021automating, cho2024recursive}, or fine-tuning a pre-trained model at a new Reynolds number~\citep{subel2021data, Guan_2022}. We examine both the direct deployment of the pre-trained model (zero-shot testing) and after fine-tuning it on nudging data at the higher Reynolds number.}

\begin{figure}[!hbtp]
  \centering
  \begin{tabular}{@{}c@{}}
    \includegraphics[width=0.49\textwidth]{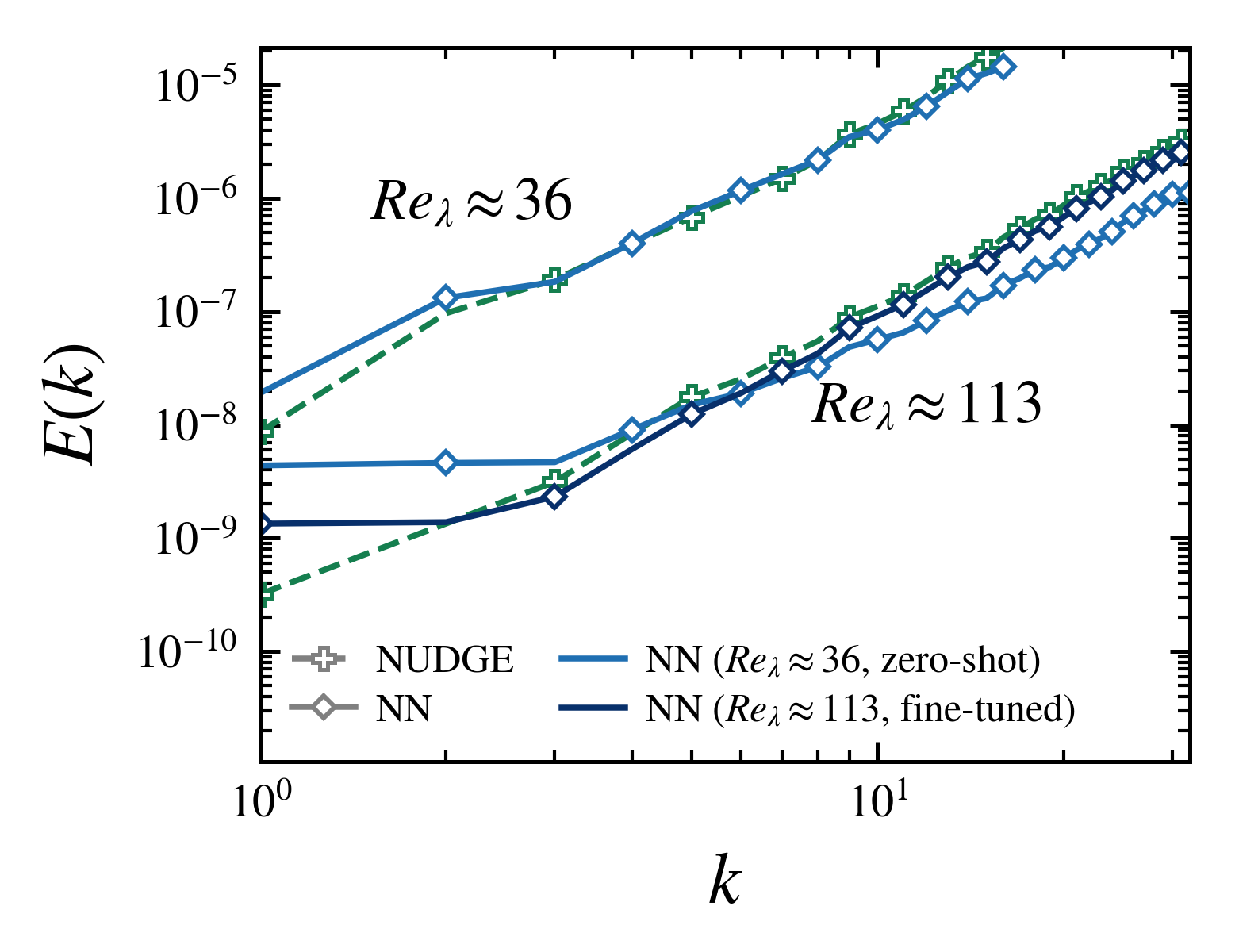} \\
    (a) \\[0.5em]
    \includegraphics[width=0.96\textwidth]{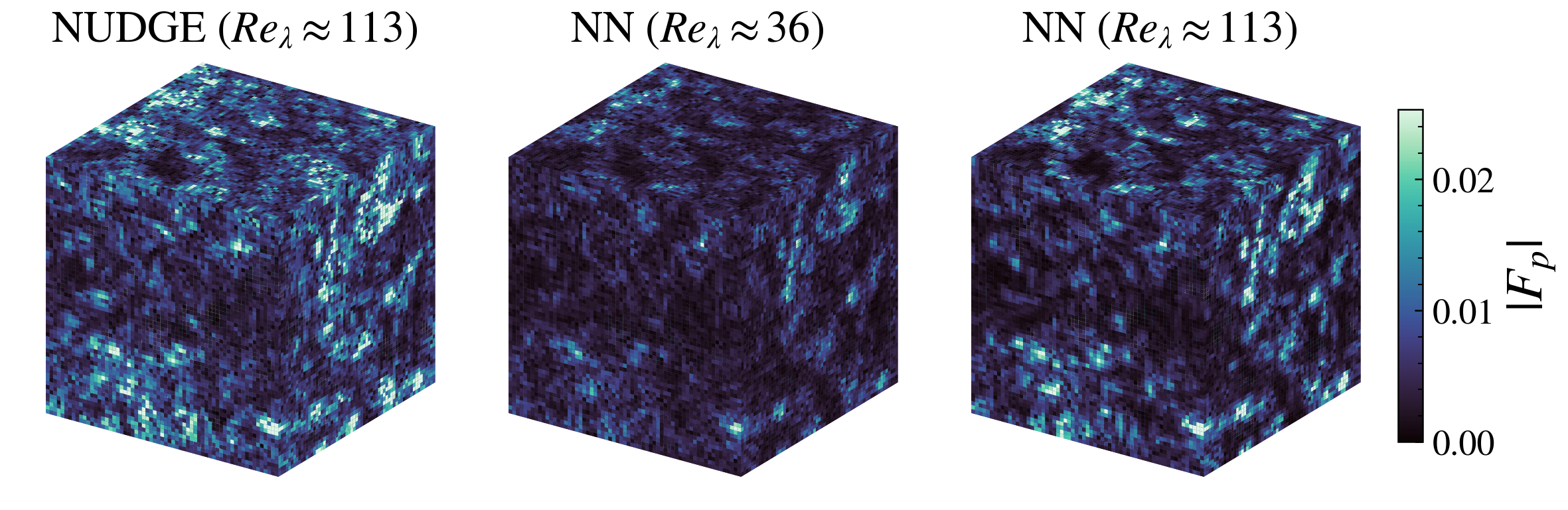} \\
    (b)
  \end{tabular}
  \caption{{\textit{A-priori} comparison of the nudging correction at $Re_\lambda \approx 113$. (a) Forcing term spectrum of the momentum correction for both the Reynolds number cases. (b) Momentum forcing magnitude $|F_p|$ on a nudged state at $Re_\lambda \approx 113$. Here, NN ($Re_{\lambda} \approx 36$, zero-shot) refers to the model from the previous lower Reynolds number case tested on the higher Reynolds number without any fine-tuning on the new flow; and the NN ($Re_\lambda \approx 113$, fine-tuned) represents the previous model fine-tuned on higher Reynolds number.}}
  \label{fig:highRe_closure}
\end{figure}

{We consider forced homogeneous isotropic turbulence at a higher Reynolds number, $Re_\lambda \approx 113$ (compared with $Re_\lambda \approx 36$ used in the main study). The DNS is performed using the compressible finite-difference solver STREAmS~\citep{bernardini2023streams} on a $256^3$ periodic grid ($\Delta_{DNS}/\eta \approx 1.2$), with a sixth-order kinetic-energy-preserving central scheme for the convective terms, sixth-order central differences for the viscous terms, and a third-order Runge-Kutta scheme at a CFL number of $0.3$. We use a large-scale forcing similar to that of~\citet{wang2010hybrid}, which keeps the energy of the solenoidal velocity in the two lowest wavenumber shells fixed, with the injected energy removed uniformly from the internal energy. The turbulent Mach number is approximately $0.08$ (comparable to the previous case), so that the flow remains close to the incompressible limit. The DNS fields are subsampled onto a $64^3$ grid to form the observations, retaining the coarsening ratio of four as before. The nudged LES is then performed with JAX-Fluids using the C4 scheme, with the same forcing as the DNS, following the procedure of Appendix~\ref{sec:appendix_data_3d_turbulence} (with the same $\mu = 2.5$).}

\begin{figure}
  \centering
  \begin{tabular}{@{}c@{\hspace{0.02\textwidth}}c@{}}
    \includegraphics[width=0.49\textwidth]{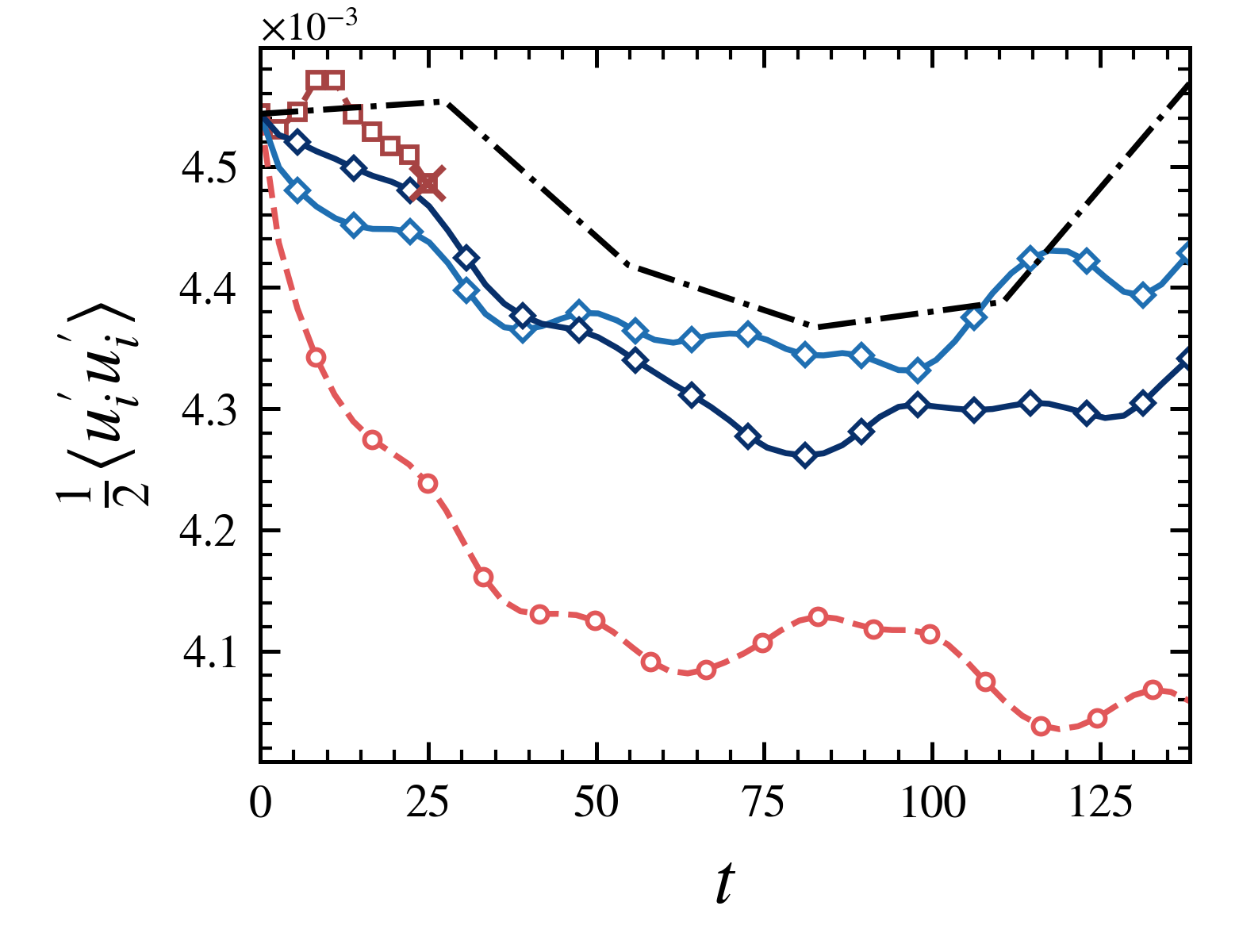} &
    \includegraphics[width=0.49\textwidth]{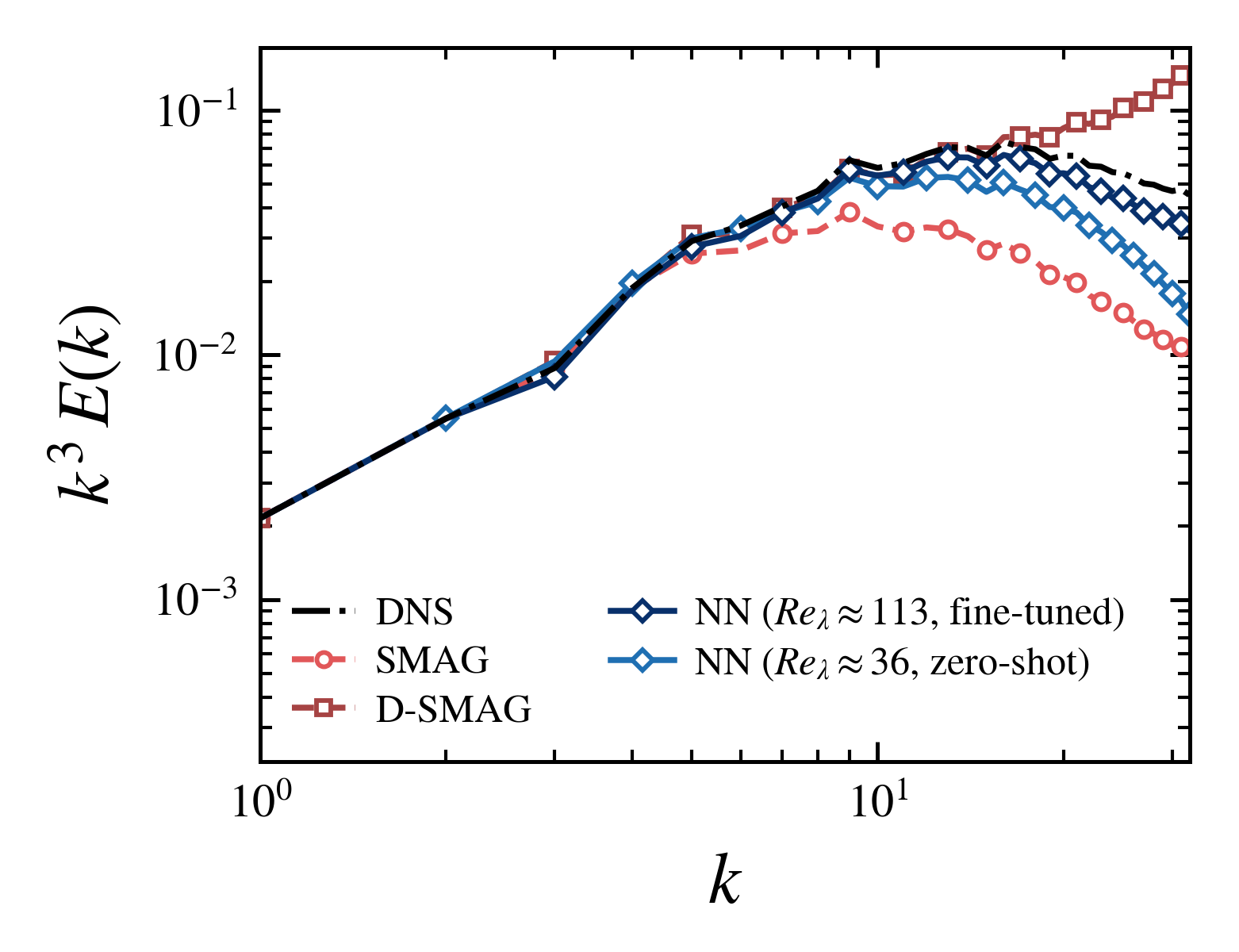} \\
    (a) & (b)
  \end{tabular}
  \caption{{Results for the model trained at $Re_\lambda \approx 36$ and the fine-tuned model at $Re_\lambda \approx 113$. (a) TKE evolution and (b) scaled time-averaged energy spectrum. SMAG remains overly dissipative at the higher wavenumbers, whereas D-SMAG blows up due to the accumulation of energy near the cutoff. The fine-tuned model follows the reference statistics and remains stable throughout. The rollout spans around $5$ DNS eddy-turnover times ($L_I/u_{rms}$, with $L_I$ the integral length scale) and starts from an initial condition unseen during fine-tuning.}}
  \label{fig:highRe_stats}
\end{figure}

{The pre-trained model of the main study, with the scheme conditioning fixed to C4, is fine-tuned on the nudging data at this Reynolds number using the mean-squared-error loss alone, which, in this case, performed better than in combination with the BSP loss. Further, during training, high-frequency noise is added to a fraction of the input fields and the recorded correction term is adjusted accordingly, an approach that has been shown to improve deployment stability in similar settings~\citep{sanchez2020learning, stachenfeld2021learned}. As expected, this training on perturbed inputs leaves the model \emph{slightly more conservative} in its supply of energy relative to the nudging term, which can be attributed to the model learning to remove spurious content at the smaller scales.}

{We first compare the closures \emph{a-priori} on unseen nudged states at $Re_\lambda \approx 113$ (see figure~\ref{fig:highRe_closure}). Structurally, the nudging term retains the form it has in the previous case, i.e. it predominantly behaves as a small-scale correction with negligible correction in the low-$k$ region (as before for the C4 scheme), and hence depends little on the large-scale forcing term; which differs between the two cases. Since the model takes unnormalized fields, deploying it without fine-tuning is possible here because the velocity fields and the nudging correction at this Reynolds number remain comparable in magnitude to those of the main study; in general, normalizing the inputs (say by $u_{rms}$) or using invariant inputs would be needed. Without the fine-tuning, the $Re_\lambda \approx 36$ model captures the spatial structure of the correction (figure~\ref{fig:highRe_closure}), but under-predicts its magnitude, mainly near the cutoff, indicating a difference in the small-scale content injected back by the nudging term at this Reynolds number. The fine-tuned model recovers the correction at the scales where it acts, remaining only slightly below the nudging term.}

{Both the fine-tuned model and the $Re_\lambda \approx 36$ model are then deployed in a coarse-grid rollout from an initial condition outside the training data, and compared with SMAG and D-SMAG in figure~\ref{fig:highRe_stats}. Both learned closures remain stable here and closely track the reference TKE, despite the differences in the forcing and the time step (here $\Delta t \approx 5.2 \times 10^{-3}$, compared with $10^{-3}$ in the main study). Since the TKE is dominated by the large, energy-containing scales, the two models differ mainly near the cutoff, where the correction acts. Consistent with the \emph{a-priori} comparison, the $Re_\lambda \approx 36$ model under-supplies the correction and remains more dissipative at the higher wavenumbers, whereas the fine-tuned model recovers most of the energy there. Similar to the previous case, SMAG is overly dissipative at the higher wavenumbers; however, its TKE does not recover and remains below the reference as the forcing here keeps the large-scale energy fixed. As before, D-SMAG accumulates excess energy near the cutoff, which at this Reynolds number causes the simulation to become unstable early in the rollout.}

{These results show that the nudging-based framework extends to a higher Reynolds number and a different flow configuration, where the pre-trained closure transfers stably and fine-tuning recovers most of the small-scale correction that the direct transfer misses. Since the nudging term retains a similar structure across the two Reynolds numbers, a single model trained on a few Reynolds numbers, with the Reynolds number as an additional input in the same way as the FiLM layers condition the model on the numerical scheme, could generalize across a range of Reynolds numbers.}

\section{Nudging using partial and discrete observations}\label{sec:appendix_E}
In many practical applications, observations are available for only a subset of the state variables. To demonstrate how the present framework extends to such settings, we consider partial nudging using only the \(u\)-velocity component in the two-dimensional case. In the three-dimensional case, we use the observed sets \((\rho,u)\) and \((\rho,u,v)\), while \((\rho,u,v,w)\) serves as the baseline full-observation case. No separate density forcing is introduced in 3D; instead, density enters only through the construction of the momentum nudging term. For this analysis, the numerical schemes are kept the same as in the main experiments, namely Van Leer in 2D and C4 in 3D.

\begin{figure}[H]
  \centering
  \begin{tabular}{@{}c@{\hspace{0.02\textwidth}}c@{}}
    \includegraphics[width=0.46\textwidth]{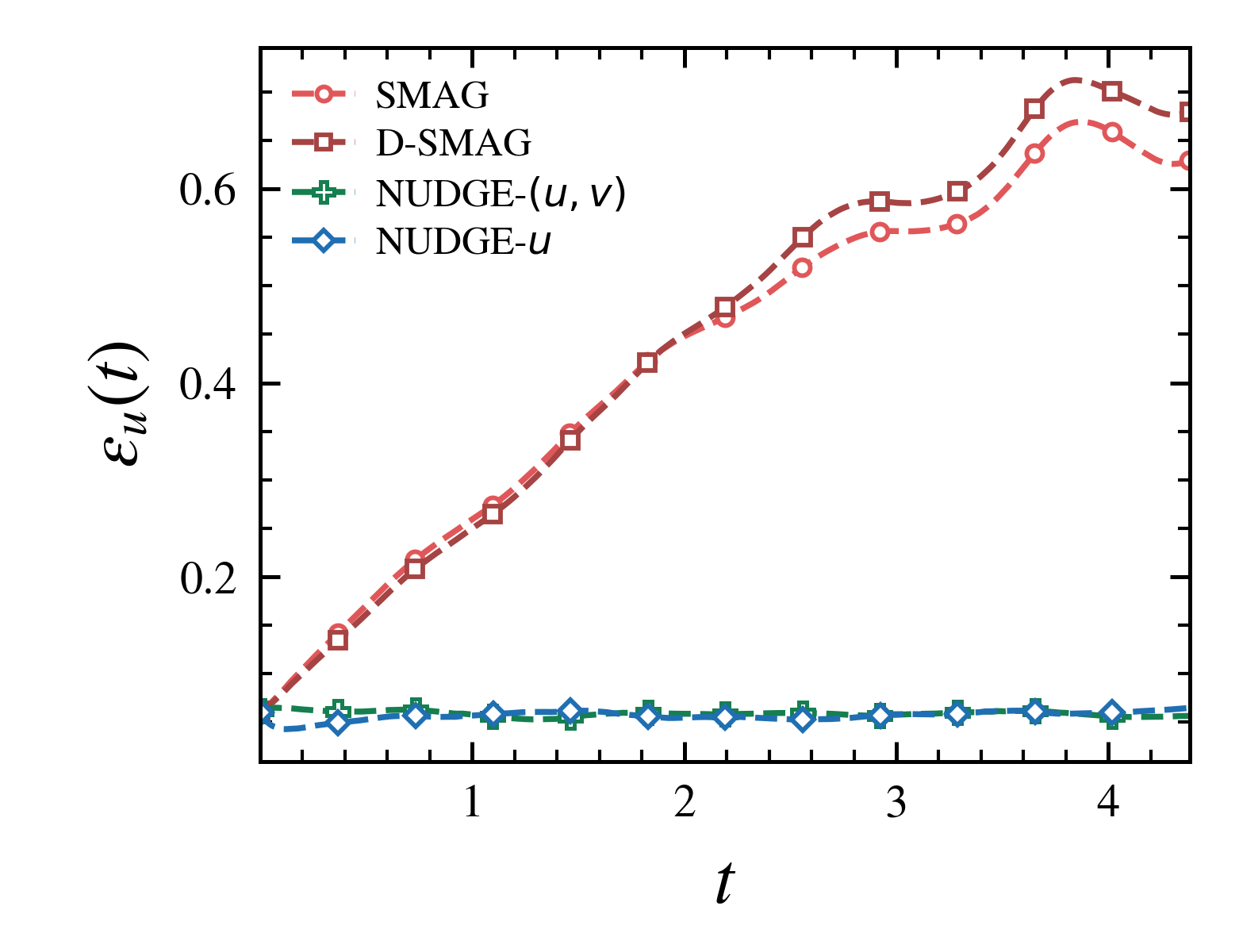} &
    \includegraphics[width=0.47\textwidth]{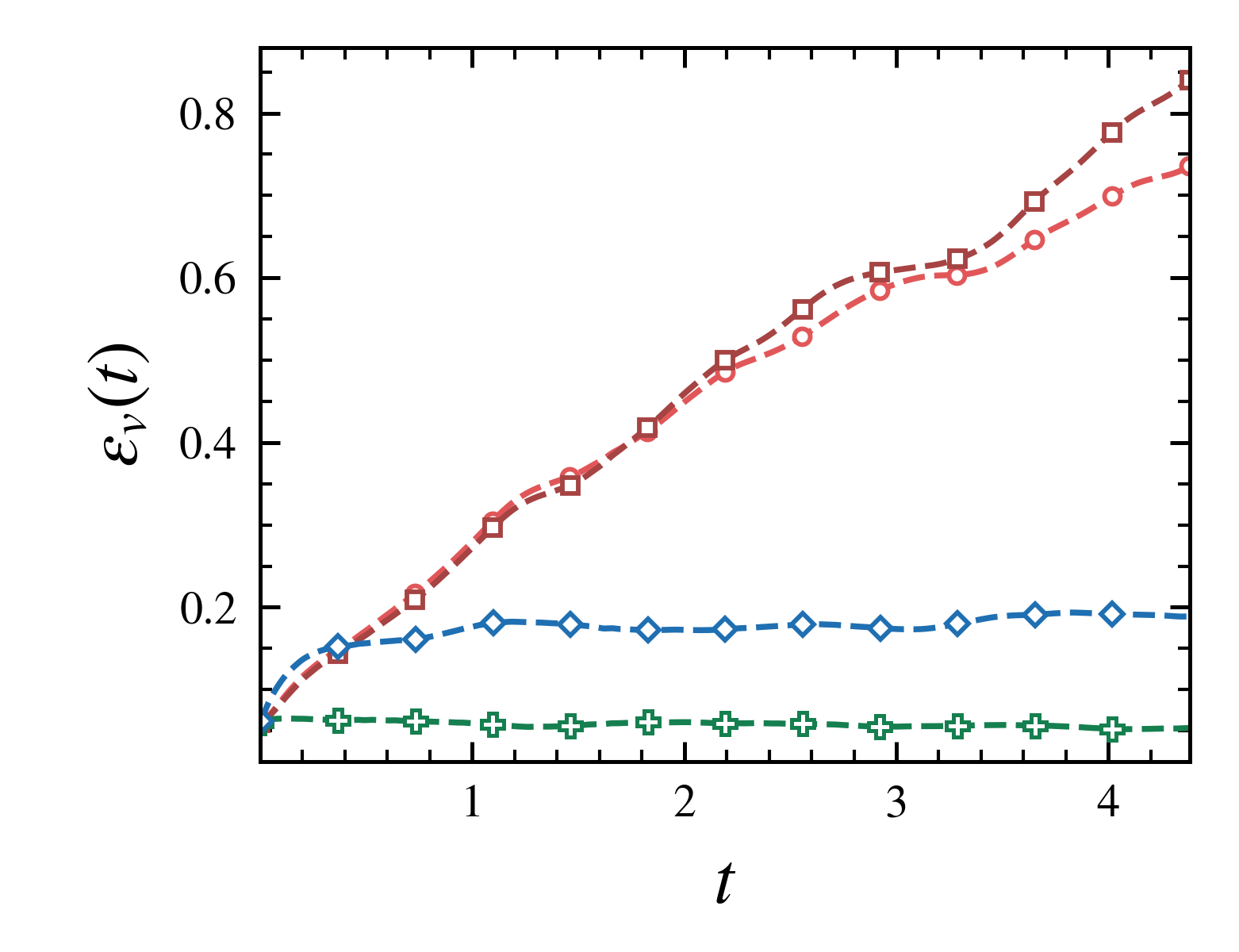} \\
    (a) & (b)
  \end{tabular}

  \vspace{0.5em}

  \begin{tabular}{@{}c@{}}
    \includegraphics[width=0.47\textwidth]{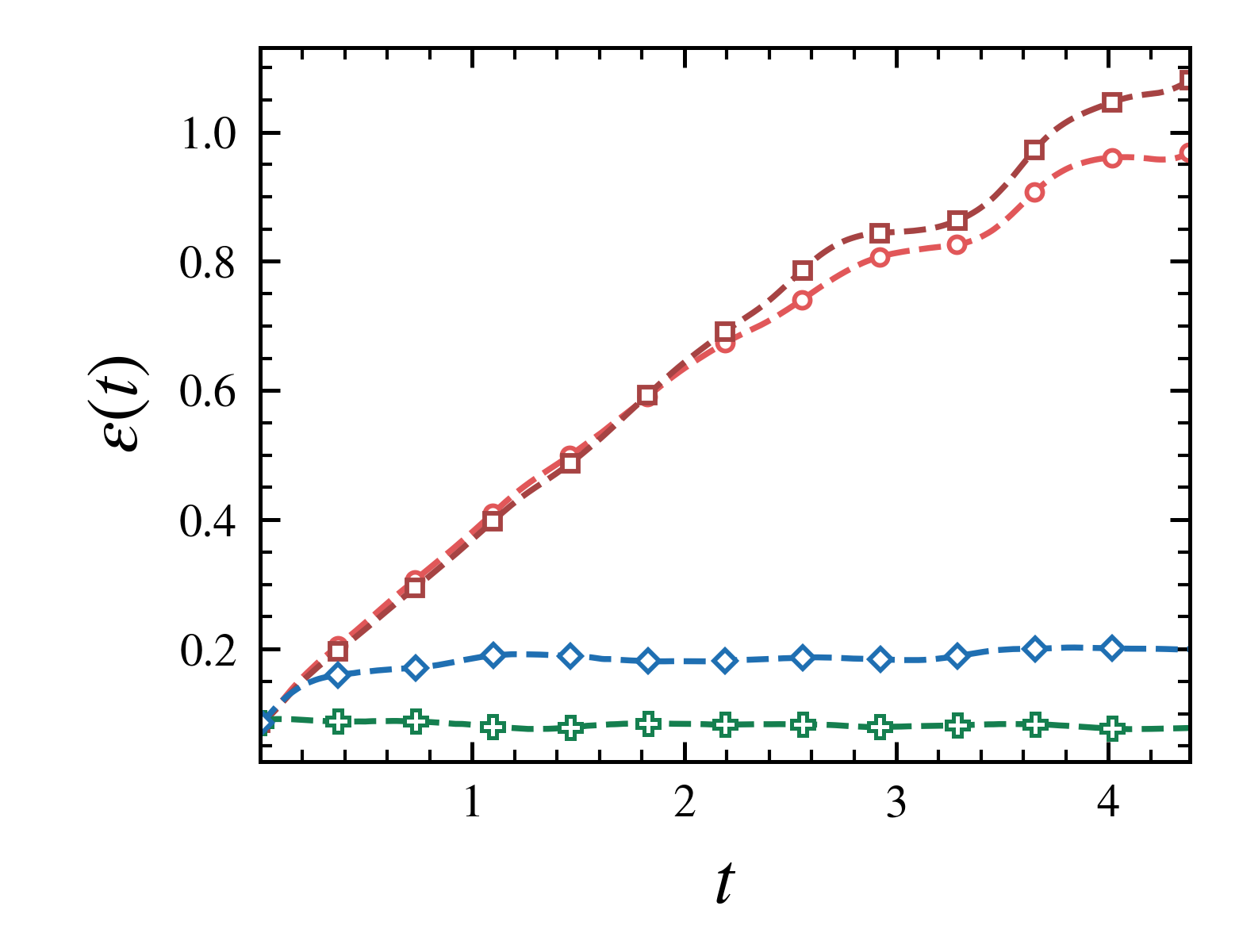} \\
    (c)
  \end{tabular}
  \caption{
{Error comparison for partial nudging using a subset of observed variables, two-dimensional case. (a,b) Relative $L_2$ error of the $x$ and $y$ velocity components, $u$ and $v$, $\varepsilon_u(t)=\|u-u_{obs}\|_2/\|\boldsymbol{u}_0\|_2$ and $\varepsilon_v(t)=\|v-v_{obs}\|_2/\|\boldsymbol{u}_0\|_2$, and (c) the
full-field relative error $\varepsilon(t)=\|\boldsymbol{v}-\boldsymbol{u}_{obs}\|_2/\|\boldsymbol{u}_0\|_2$, all normalized by the initial norm of the full reference velocity field $\|\boldsymbol{u}_0\|_2$. SMAG and D-SMAG are the baselines; NUDGE-$(u,v)$ is the full-observation case discussed in the main text, while NUDGE-$u$ nudges only the $x$-component velocity field. Even under partial observation the error settles to a residual level below SMAG and D-SMAG.}}
  \label{fig:error_2d_partial_nudging}
\end{figure}

\begin{figure}
  \centering
  \begin{tabular}{@{}c@{\hspace{0.02\textwidth}}c@{}}
    \includegraphics[width=0.47\textwidth]{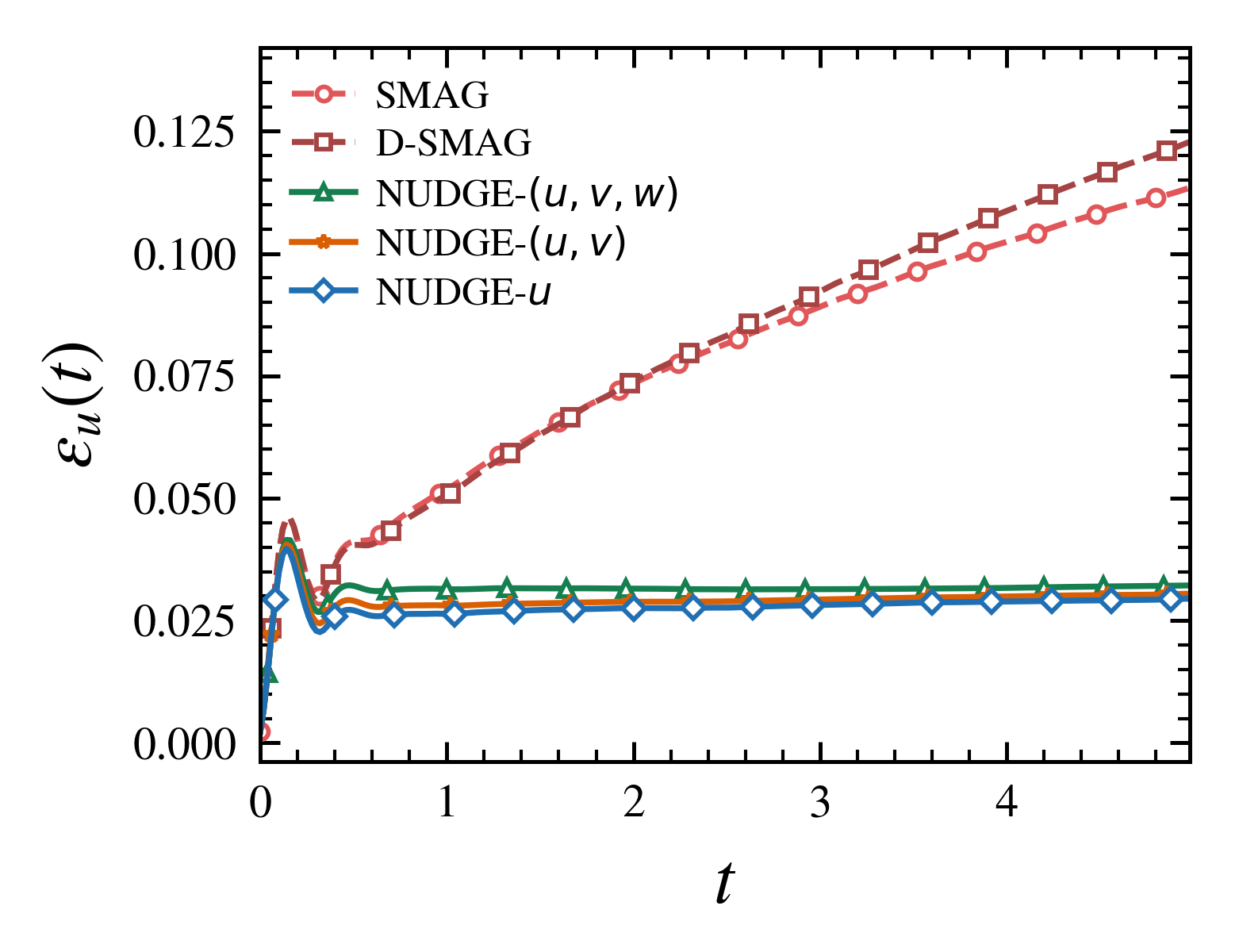} &
    \includegraphics[width=0.47\textwidth]{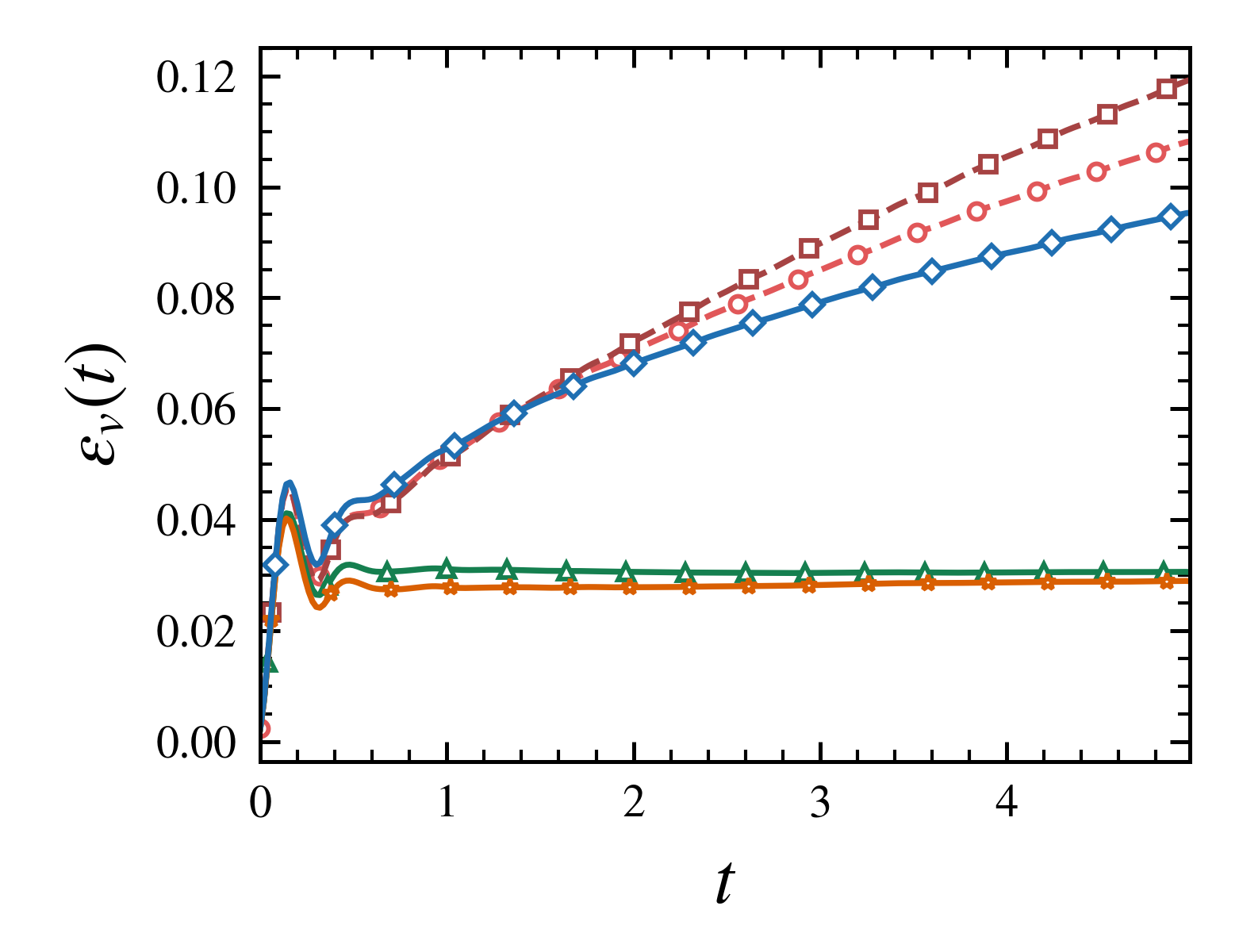} \\
    (a) & (b) \\[0.5em]
    \includegraphics[width=0.47\textwidth]{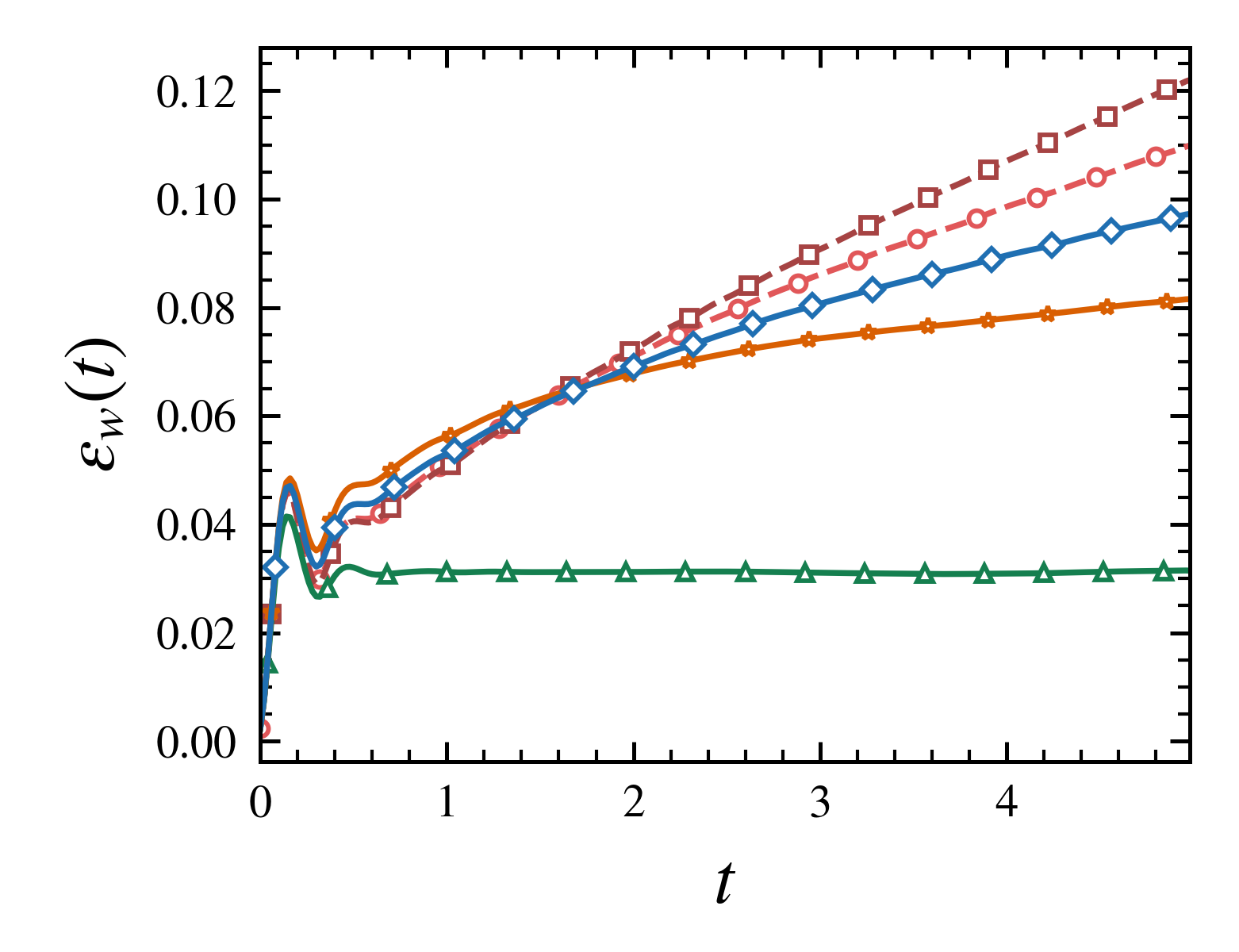} &
    \includegraphics[width=0.47\textwidth]{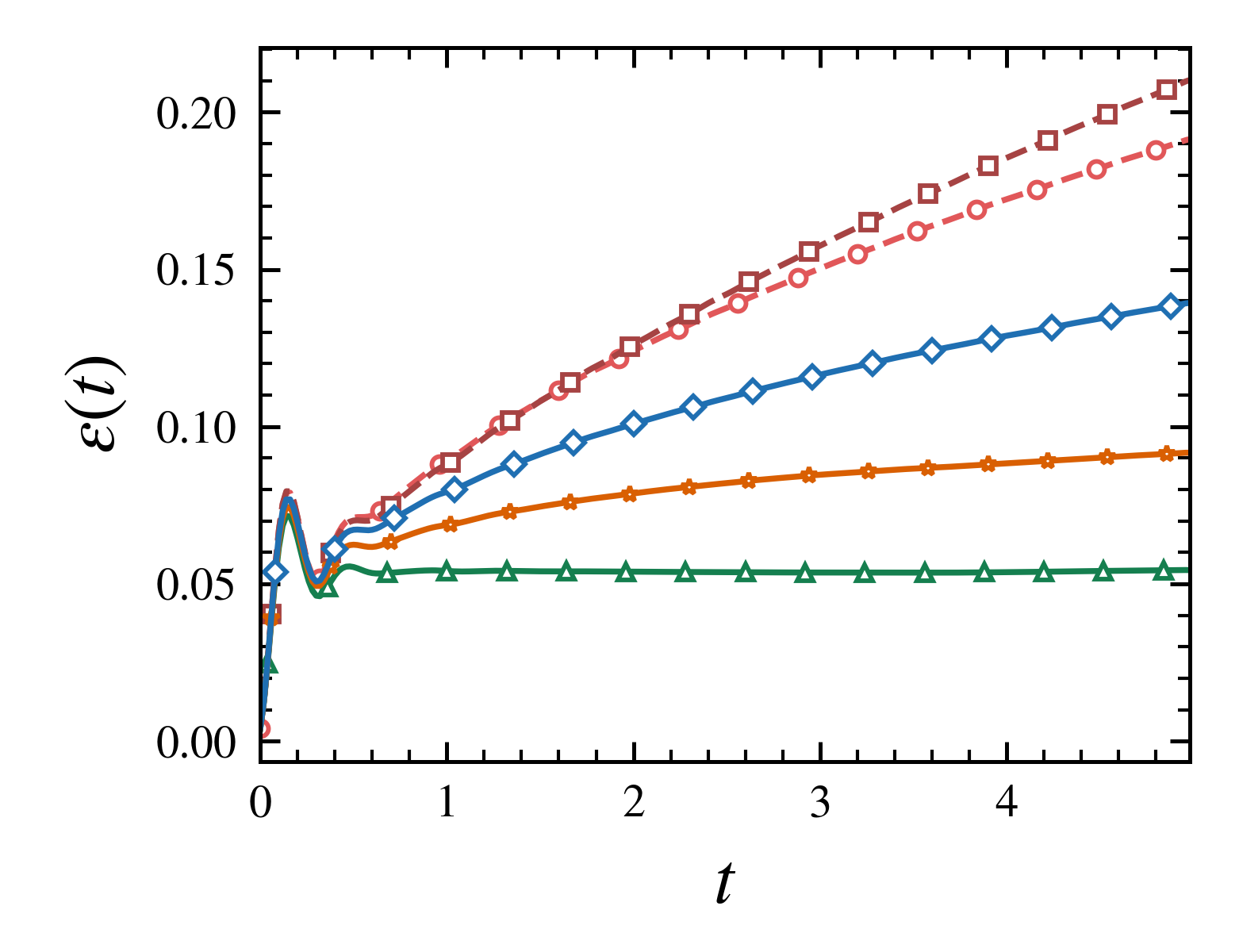} \\
    (c) & (d)
  \end{tabular}
  \caption{
{Error comparison for partial nudging, three-dimensional case. (a,b,c) Relative $L_2$ error $\varepsilon_u(t)$, $\varepsilon_v(t)$, $\varepsilon_w(t)$ of the $u$, $v$ and $w$ velocity components and (d) the full-field relative error $\varepsilon(t)$; the metrics and curves are defined similar to figure~\ref{fig:error_2d_partial_nudging}.}}
  \label{fig:error_3d_partial_nudging}
\end{figure}

\begin{figure}
  \centering
  \begin{tabular}{@{}c@{\hspace{0.02\textwidth}}c@{}}
    \includegraphics[width=0.47\textwidth]{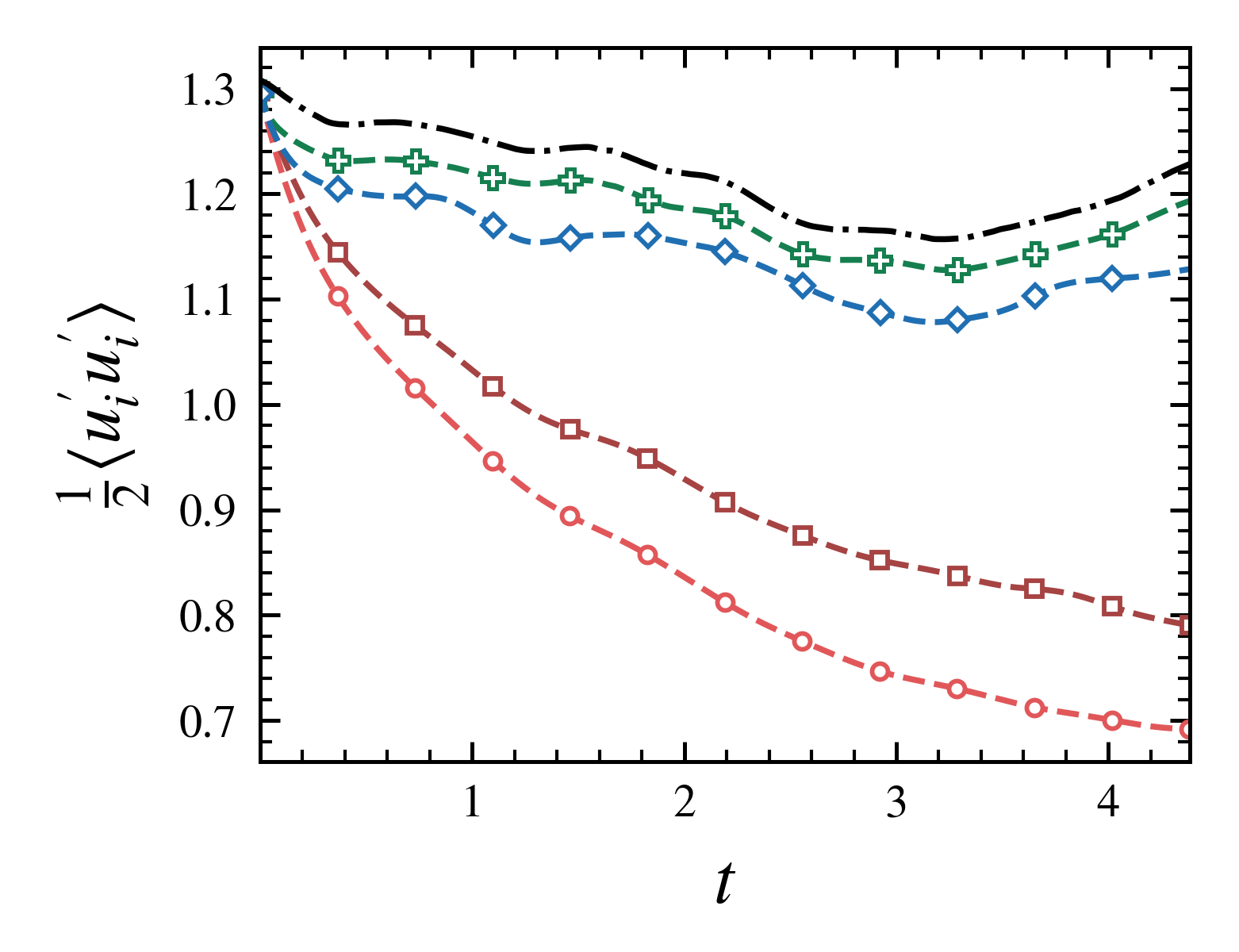} &
    \includegraphics[width=0.47\textwidth]{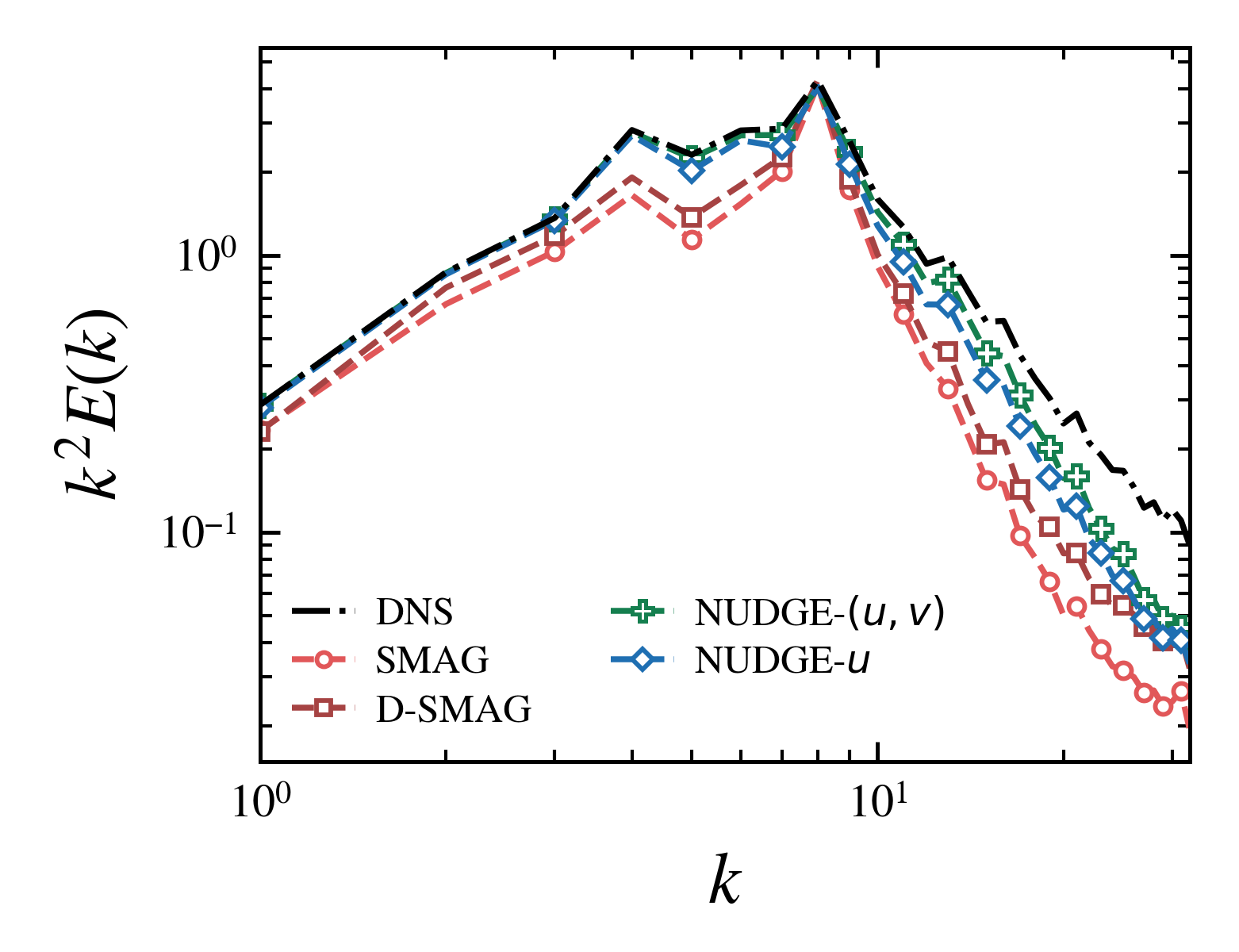} \\
    (a) & (b) \\[0.5em]
    \includegraphics[width=0.47\textwidth]{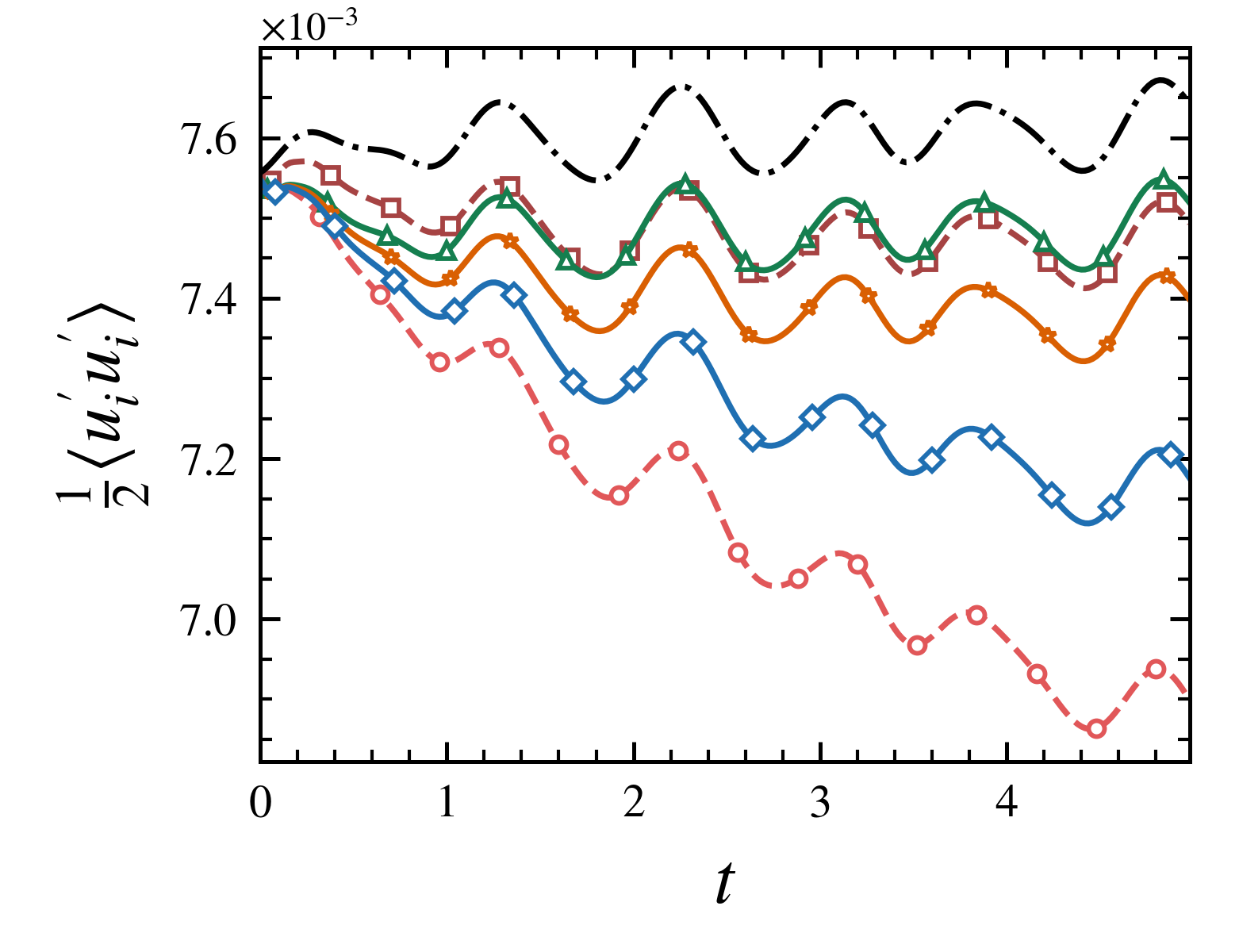} &
    \includegraphics[width=0.47\textwidth]{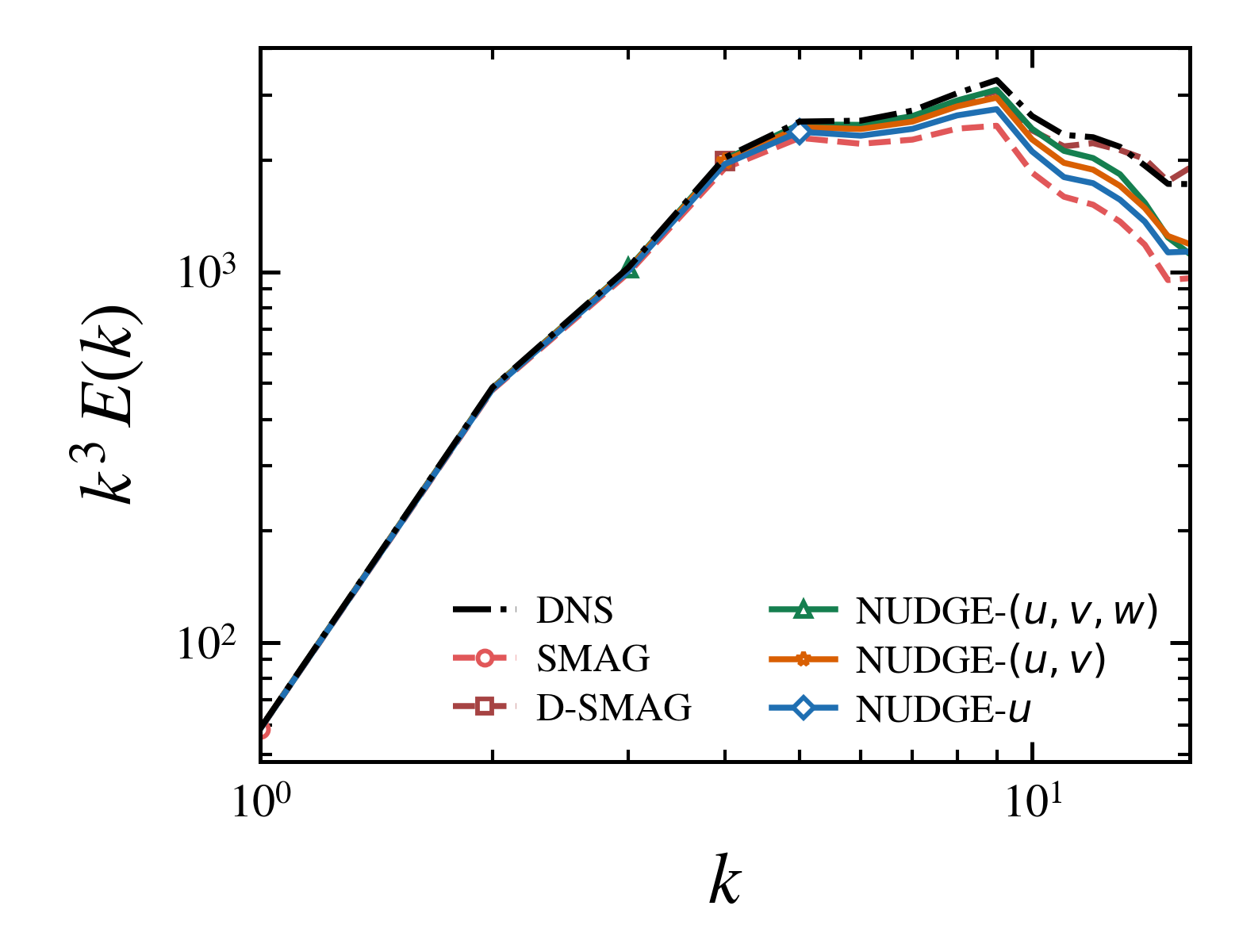} \\
    (c) & (d)
  \end{tabular}
  \caption{
{TKE (a,c) and scaled time-averaged energy spectrum (b,d) for partial nudging, for the two-dimensional (a,b) and three-dimensional (c,d) cases. The methods follow figures~\ref{fig:error_2d_partial_nudging} (2D) and~\ref{fig:error_3d_partial_nudging} (3D), with DNS added as the reference.}}
  \label{fig:stats_partial_nudging}
\end{figure}

Figures~\ref{fig:error_2d_partial_nudging} and~\ref{fig:error_3d_partial_nudging} compare the total and component-wise relative $L_2$ error with the coarse DNS for the partial-nudging cases, along with the baseline case in which all variables are nudged. The nudging coefficient is kept the same as in the full-state setting. As expected, the total relative error is smallest when all available variables are nudged, and increases as the observations are restricted to a subset of the fields. The component-wise errors further show that the error in a given field remains closest to the baseline only when that field is directly nudged. This is consistent with the pointwise nature of the present formulation, since no direct constraint is imposed on fields that are not included in the observed set. Even so, partial nudging still yields substantially lower errors than SMAG and D-SMAG. The TKE evolution similarly indicates that the additional corrective forcing from nudging retains the reference statistics more accurately than SMAG and D-SMAG, although the level of agreement depends strongly on the number of observed variables, as shown in figure~\ref{fig:stats_partial_nudging}.

\begin{figure}
  \centering
  \begin{tabular}{@{}c@{\hspace{0.02\textwidth}}c@{}}
    \includegraphics[width=0.47\textwidth]{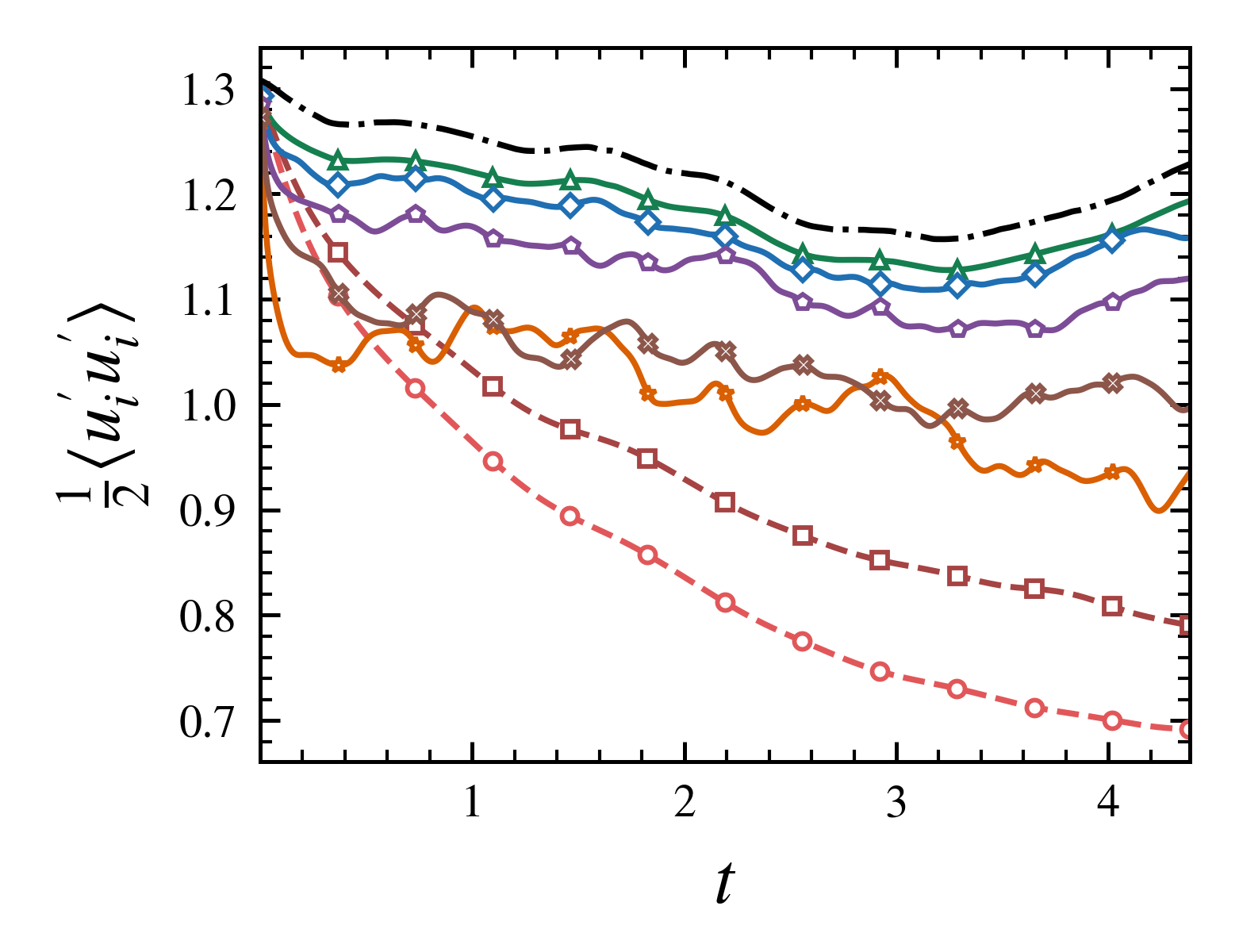} &
    \includegraphics[width=0.47\textwidth]{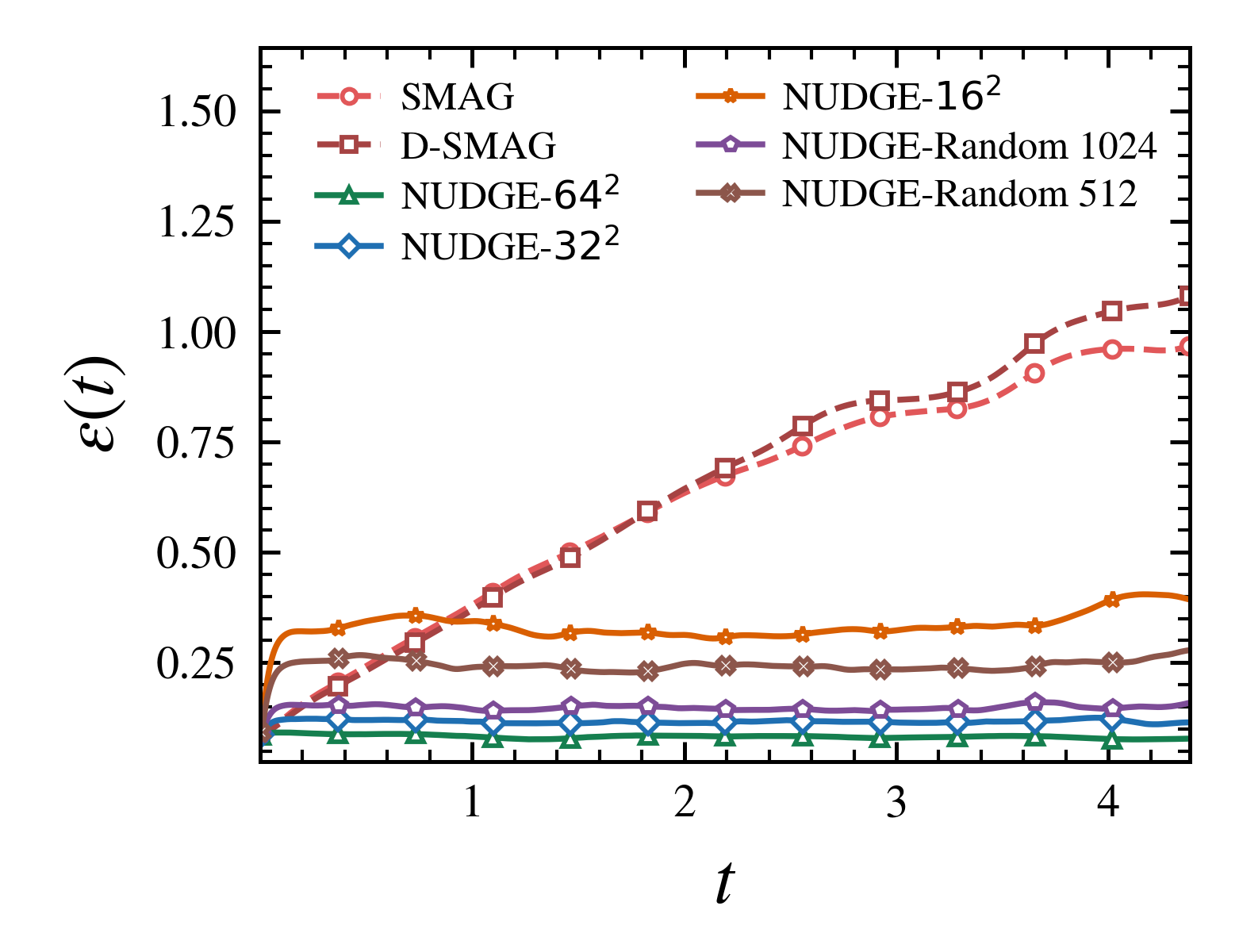} \\
    (a) & (b) \\[0.5em]
    \includegraphics[width=0.47\textwidth]{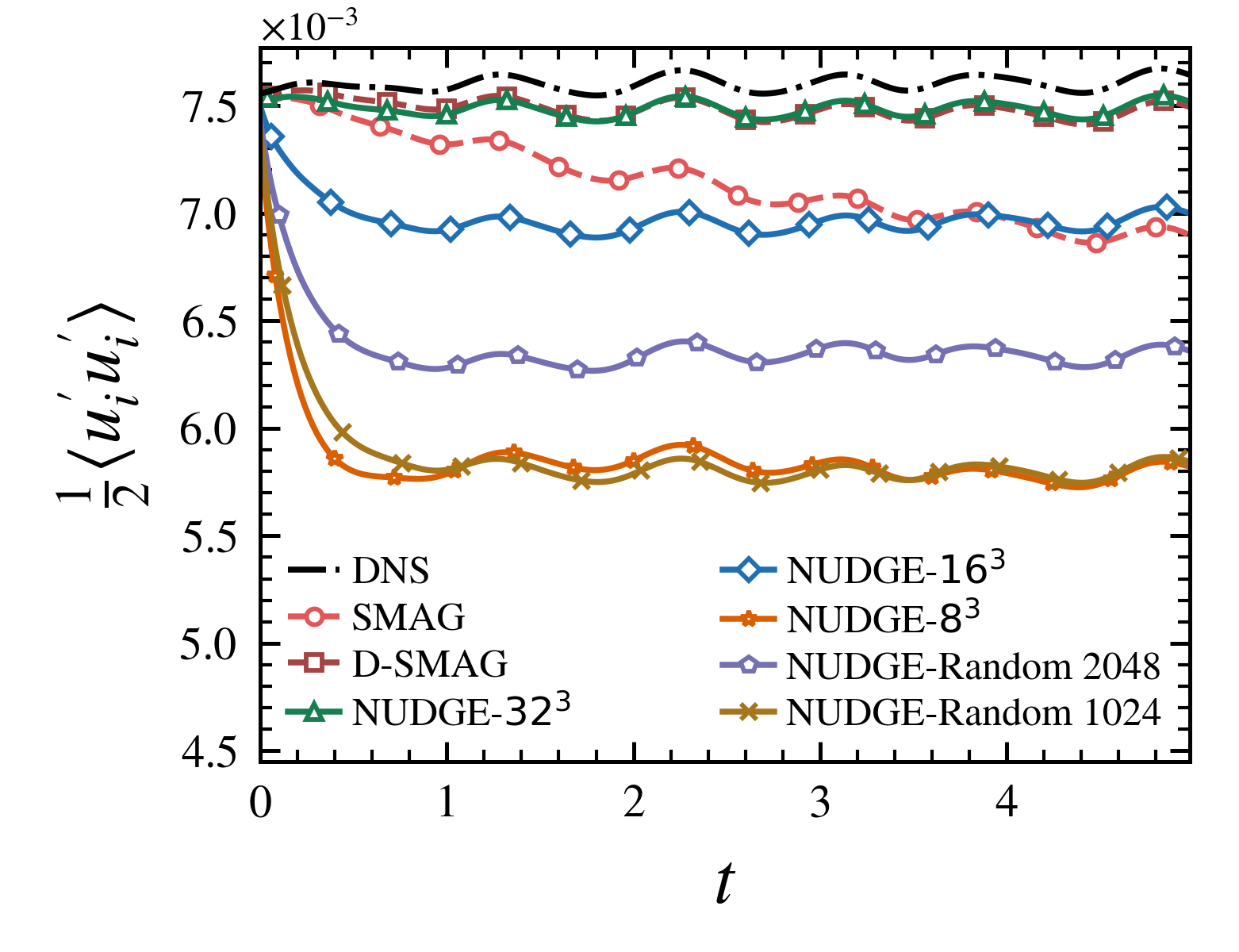} &
    \includegraphics[width=0.47\textwidth]{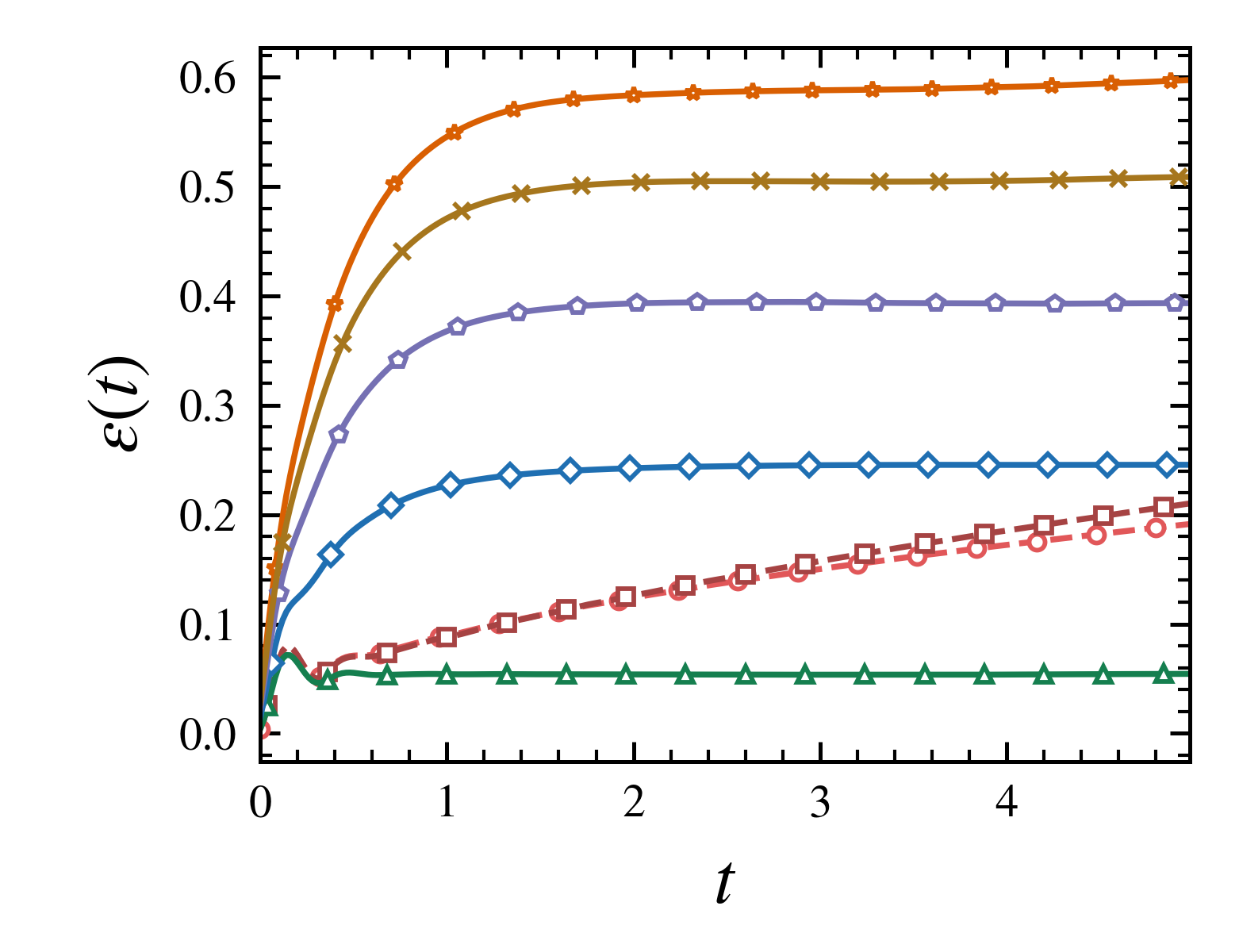} \\
    (c) & (d)
  \end{tabular}
  \caption{{TKE (a,c) and relative $L_2$ error $\varepsilon(t)$ (b,d) for nudging using observations sparser than the coarse grid, for the two-dimensional (a,b) and three-dimensional (c,d) cases. Baselines are DNS, SMAG and D-SMAG. The nudged cases are labeled by their observation set. NUDGE-$64^2$, NUDGE-$32^2$ and NUDGE-$16^2$ in 2D, and NUDGE-$32^3$, NUDGE-$16^3$ and NUDGE-$8^3$ in 3D, use observations regularly subsampled from the ground truth onto a grid of the stated size; NUDGE-Random~$N$ uses $N$ observation points placed at random within the coarse grid ($N=1024$ and $512$ in 2D; $N=2048$ and $1024$ in 3D). In every case the observations are interpolated onto the coarse grid before the nudging term is formed. NUDGE-$64^2$ (2D) and NUDGE-$32^3$ (3D) sample every coarse-grid point and therefore serve as the full observation reference.}}
  \label{fig:stats_discrete_nudging}
\end{figure}

We also consider a more challenging setting in which the measurements are made even sparser, either through coarse subsampling of the reference data or through a set of observation points that are randomly distributed in space and fixed in time. Figure~\ref{fig:stats_discrete_nudging} shows the TKE of the nudged system together with the relative $L_2$ error for the different measurement configurations. In these cases, all field variables are included, and the observations are interpolated onto the coarse grid using radial basis function (RBF) interpolation. In the presence of model mismatch associated with coarse-grid evolution, the degree of synchronization attained by the nudged system depends critically on both the observation spacing and the nudging coefficient. In both the two- and three-dimensional cases, the residual error increases as the observations become sparser.

The TKE exhibits a similar trend (see figure~\ref{fig:stats_discrete_nudging}). This may be attributed to the smoothing of the measurement field caused by sparse observations, which reduces the high-wavenumber content available to the nudged system. As a result, when the observations are very sparse, the simulation is driven towards an overly smoothed representation of the reference flow leading to worse agreement with the DNS. This effect is more pronounced in the three-dimensional case, where the SMAG and D-SMAG baselines outperform nudging using such sparse observations. To examine this dependence more systematically in 3D, we next vary the nudging coefficient and assess its effect on the relative $L_2$ error and TKE.

\begin{figure}
  \centering
  \begin{tabular}{@{}c@{\hspace{0.02\textwidth}}c@{}}
    \includegraphics[width=0.47\textwidth]{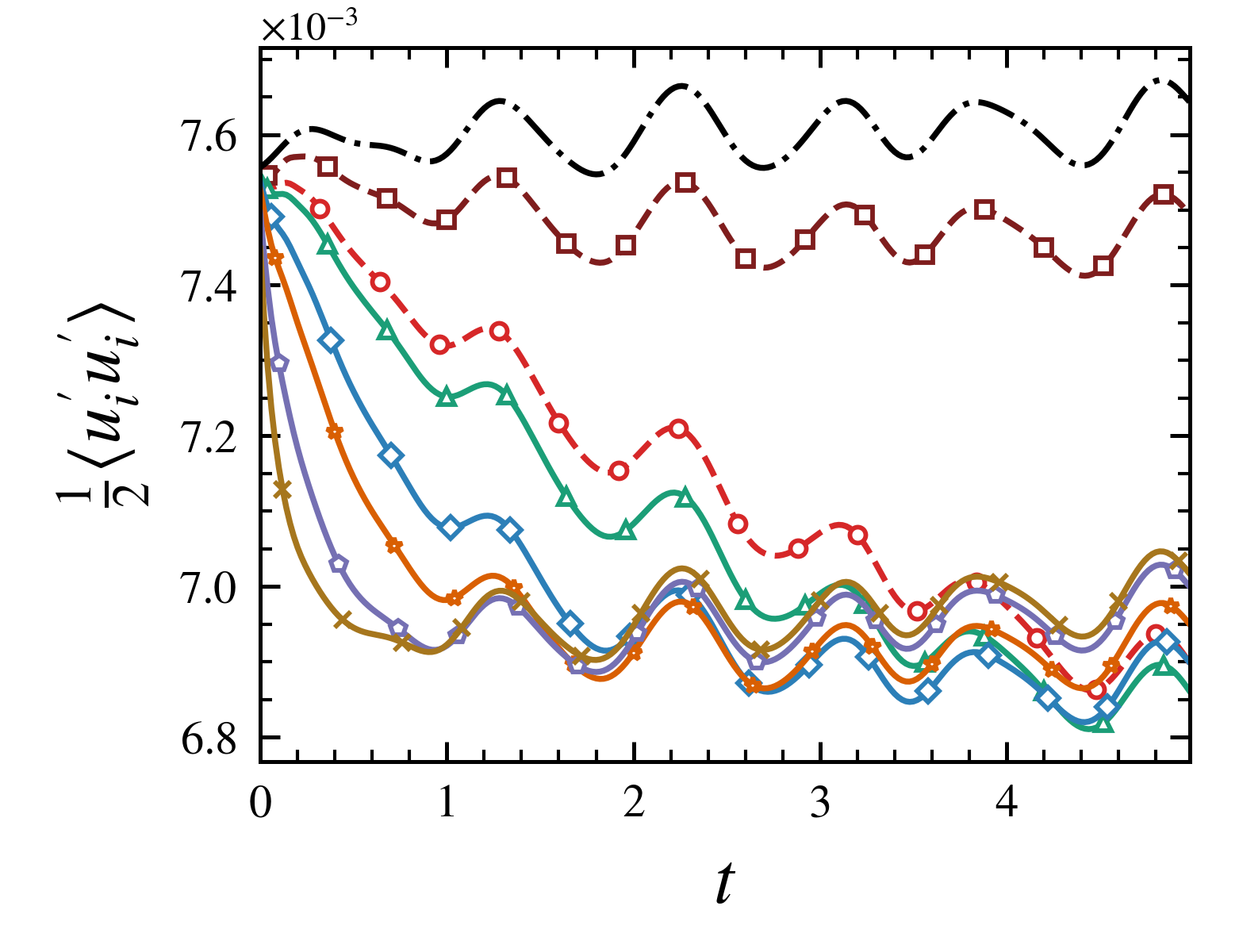} &
    \includegraphics[width=0.47\textwidth]{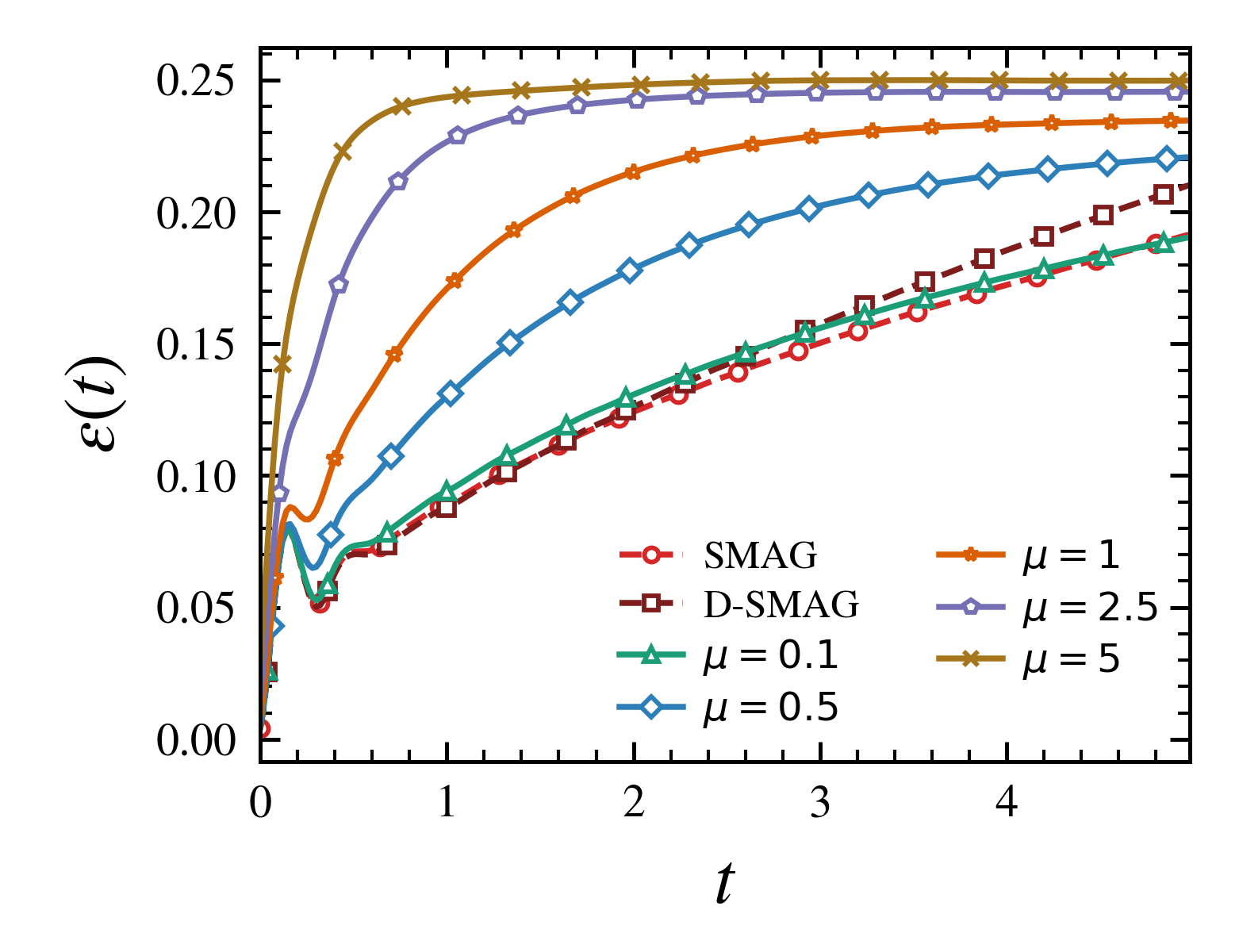} \\
    (a) & (b)
  \end{tabular}
  \caption{
{Dependence of the three-dimensional nudging result on the nudging coefficient $\mu$ for a representative sparse-observation configuration ($16^3$ regular subsampling). (a) TKE and (b) relative $L_2$ error.}}
  \label{fig:3d_mu_sweep}
\end{figure}

Figure~\ref{fig:3d_mu_sweep} shows that the residual error depends strongly on~$\mu$, taking values between the SMAG error level and a larger saturated value as~$\mu$ increases. For small~$\mu$, the forcing is too weak to produce meaningful synchronization, and the solution remains close to the underlying SMAG-LES. For large~$\mu$, the system is driven too strongly towards the interpolated measurements, which again increases the error. These results indicate that an appropriate choice of nudging strength is essential, and further highlight the strong dependence of nudging performance on the observation spacing.

\end{appendices}\clearpage

\end{document}